  \pdfoutput=1
  \documentclass[graphicx,amssymb,amsmath,enumerate,11pt]{report}
  \usepackage{fancyhdr}
  \fancyheadoffset[]{0.1cm}
  \usepackage{epsfig}
  \usepackage{axodraw}
  \usepackage{url}
  \newcommand\mchapter[2]{\chapter*{#1}
  \vskip -0.5cm \noindent {\it \LARGE #2}
  \addcontentsline{toc}{chapter}{#1\\{\normalsize\it #2}}} 
\usepackage{color}
\usepackage{multirow}
\usepackage{subfig}
\usepackage{amssymb}
\usepackage[utf8x]{inputenc}

\newcommand{\lsun}{{\rm L_{\odot}}}
\newcommand{\rsun}{{\rm R_{\odot}}}
\newcommand{\msun}{{\rm M_{\odot}}}
\newcommand{\zxsun}{\left(Z/X\right)_\odot}
\newcommand{\rcz}{R_{\rm CZ}}
\newcommand{\ys}{Y_{\rm S}}
\newcommand{\zs}{Z_{\rm S}}
\newcommand{\yc}{Y_{\rm C}}
\newcommand{\zc}{Z_{\rm C}}
\newcommand{\yini}{Y_{\rm ini}}
\newcommand{\zini}{Z_{\rm ini}}
\newcommand{\xini}{X_{\rm ini}}

\begin{document}

 \rhead{\bfseries Solar Neutrinos}

 \mchapter{Solar Neutrinos}
 {Authors:\ V. Antonelli$^a$, L. Miramonti$^a$, C. Pe\~na-Garay$^b$ and A. Serenelli$^c$}
 \label{ch-14:mycontribution}

\vspace{0.5cm}

\begin{center}
$^a$ {\it Dipartimento di Fisica, Universit\'a degli Studi di Milano and INFN Milano, Via Celoria 16,\\ 
I-20133 Milano, Italy} \\ [6pt]
$^b$ {\it Instituto de Fisica Corpuscular, CSIC-UVEG, Valencia E-46071, Spain}\\ [6pt]
$^c$ {\it Instituto de Ciencias del Espacio (CSIC-IEEC), Facultad de Ciencias, Campus UAB, Bellaterra, 08193, Spain}
\end{center}

\vspace{3cm}

\begin{center}
{\bf Abstract}
\end{center}
The study of solar neutrinos has given since ever a fundamental contribution both to 
astroparticle and to elementary particle physics, offering an ideal test of solar models
and offering at the same time relevant indications on the fundamental interactions among 
particles. After reviewing the striking results of the last two decades, which were determinant 
to solve the long standing solar neutrino puzzle and refine the Standard Solar Model, 
we focus our attention on the more recent results in this field and on the experiments 
presently running or planned for the near future. The main focus at the moment is 
to improve the knowledge of the mass and mixing pattern and especially to study in detail
the lowest energy part of the spectrum, which represents most of solar neutrino spectrum
but is still a partially unexplored realm. We discuss this research project and the 
way in which present and future experiments could contribute to make the 
theoretical frawemork more complete and stable, understanding the origin of some
``anomalies'' that seem to emerge from the data and contributing to answer some present 
questions, like the exact mechanism of the vacuum to matter transition and the 
solution of the so called solar metallicity problem. 

\section{Motivations for the solar neutrino study}
The analysis  of neutrinos emitted in  the fusion processes inside  the Sun is
one of most  significant examples of the relevant role played  by the study of
neutrino properties  in elementary particle  physics and astrophysics and in
creating a link between these two sectors. The pioneering work in the 
sixties \cite{14-Davis1968cp} had the main goal of
understanding  better the  way in  which  our star  shines and  to test  solar
models.  But, the surprising  result of  an apparent  deficit in  the electron
neutrino flux  reaching the detector marked  the raise of the  so called solar
neutrino  puzzle, and  opened a  whole  new field  of research  that has  been
central in elementary particle physics for many decades.

The  experimental results  obtained using  different techniques  in  more than
thirty years  and the parallel  theoretical advancements confirmed at  the end
the  validity  of  Pontecorvo's  revolutionary idea  of  neutrino  oscillation
\cite{14-Pontecorvo},  proving  in a  crystal  clear  way  that neutrinos  are
massive and  oscillating particles. This is  one of the first  pieces of clear
evidence  of  the  need  to  go  beyond  the  Standard  Model  of  electroweak
interactions  and the attempt  to accommodate  the experimental  results about
neutrino  masses and  mixing  is a  test  every theory  ``beyond the  Standard
Model'' has to pass. Therefore, it is clear why these results have had a great
impact on elementary particle physics  and also on cosmological models. At the
same time, the  possibility of measuring directly at  least some components of
the solar neutrino spectrum and of  recovering in an indirect way the value of
total solar neutrino flux have been fundamental for the progressive refinement
of the Standard  Solar Model (SSM), which has evolved during these  years and 
is now in a general good agreement with the solar neutrino experiments.

Despite the fundamental steps forward made in the last decades, many questions
are still open about the real  nature and the main properties of neutrinos and
the exact  mixing mechanism,  e.g. are neutrinos  Majorana or  Dirac fermions,
determination of mass hierarchy  and exact mass values, accurate determination
of the mixing angles, presence of CP violation. The solar neutrino experiments
presently running or  planned for the future can contribute  to solve at least
some of these puzzles.  The new frontier in this field is the study of the low
energy part of the solar neutrino spectrum which represents the great majority
of  the  spectrum, and  is  still  an almost  unexplored  realm.  Some of  the
challenges ahead are: reducing  significantly the indetermination on {\it pep}
and CNO neutrinos and attaching the {\it pp} solar neutrino measurement.
This would be essential to test  the stability and consistency of the standard
explanation of the oscillation  mechanism, confirming or definitely disproving
the presence of discrepancies between theory and experiments, which has lately
stimulated a  flourishing of models  introducing the so called  ``Non Standard
Interactions'' (Section~\ref{14-sec:status-mixing}).
%\ref{14-sec:NSI}).  
Once more,  these results  would be  of great  interest to
improve the knowledge both  of elementary particle properties and interactions
and  of  the  astrophysical  models  of  the Sun.  They  could  help  also  to
discriminate between different  versions of the solar models,  for instance of
the so-called  ``solar abundance problem'', and to  deepen the comparison with
the  results  coming  from  other  studies  of  solar  properties,  e.g.  from
helioseismology.   This  research project  would  of  course  imply a  further
improvement of the already known  detection techniques and the introduction of
new ones (see, for instance,  the section \ref{14-sec:future}). Also from this
point  of view, solar  neutrino physics  will continue  to give  a stimulating
contribution both to elementary particle physics and to astrophysics.

\section{Brief history and solution of the solar neutrino problem}
\subsection{From Homestake to Super-Kamiokande}\

The  first  experiment built  to  detect solar  neutrinos  took  place in  the
Homestake  gold  mine in  South  Dakota  \cite{14-Davis1968cp}.  The  detector
consisted of  a large  tank containing 615  tons of  liquid perchloroethylene,
chosen because it is rich in chlorine and the experiment operated continuously
from 1970 until 1994.  Neutrinos were detected via the reaction:
\begin{equation}
        \nu_{e} + {\rm ^{37}Cl} \rightarrow {\rm ^{37}Ar} + e^{-} \, .
        \label{eq:homestake}
\end{equation}
The energy threshold of this reaction, $E_{th} = 814 \: \rm{keV}$, allowed the
detection of $^{7}$Be  and $^{8}$B (and a small signal from  the CNO and
{\it pep}) but not that of  {\it $pp$} neutrinos, because of their low maximal
energy of  $0.42 \:  \rm{MeV}$.  The radioactive  $^{37}Ar$ isotopes  decay by
electron capture with a $\tau_{1/2}$ of about 35 days into ${\rm ^{37}Cl^{*}}$:
\begin{equation}
        {\rm ^{37}Ar} + e^{-} \rightarrow {\rm ^{37}Cl^{*}} + \nu_{e} \, .  \label{eq:argon37}
\end{equation}
Once a month, after bubbling helium through the tank, the $^{37}$Ar atoms were
extracted and counted. The number of atoms created was only about $5$ atoms of
$^{37}$Ar  per  month in  615  tons  ${\rm C_{2}Cl_{4}}$.   The number  of  detected
neutrinos  was about  1/3  lower than  expected  by the  Solar Standard  Model. 
This  discrepancy is the essence  of the Solar  Neutrino Problem, which
has been for many years an important puzzle among physicists.

There  were three  possible explanations  to the  Solar Neutrino  Problem. The
first one  was to consider that  Homestake could be wrong,  i.e. the Homestake
detector could be inefficient and, in  this case, its reactions would not have
been cpredicted correctly. After all, to detect a handful of atoms per week in
more than  600 tons of material  is not an easy  task.  The second  one was to
consider that the SSM was not correct, but as helioseismology\footnote{The science 
that studies the interior of the Sun by looking at its vibration modes.}  started 
to  provide  independent  tests of  solar  models the  SSM 
  passed all tests. Indeed, non-standard solar models constructed
ad-hoc  to  resolve the  Solar  Neutrino  Problem  seemed very  unlikely  when
scrutinized  under the  light  of  helioseismology.  The  third  one, and  the
strangest hypothesis, was to consider  that something happens to the neutrinos
while traveling from the core of the Sun to the Earth.

The first real time solar neutrino detector, Kamiokande, was built in Japan in
1982-83  \cite{14-Kamiokande89}. It  consisted of  a large  water \v{C}erenkov
detector with a total mass of 3000  tons of pure water.  In real time neutrino
experiments  scientists  study the  bluish  light  produced  by the  electrons
scattered by an impinging neutrino according to the following equation:
\begin{equation}
        \nu_{x} + e^{-} \rightarrow \nu_{x} + e^{-} \, .
        \label{eq:ES}
\end{equation}
In the  Kamiokande detector  light is recorded  by 1000  photomultiplier tubes
(PMT) and the energy threshold of  the reaction is $E_{th} = 7.5\; {\rm MeV}$;
therefore only $^{8}$B and $hep$  neutrinos are detected.  At the beginning of
the '90s  a much larger version  of the detector  was built, Super-Kamiokande,
where the active  mass was 50000 tons  of pure water viewed by  11200 PMTs. In
Super-Kamiokande  the energy  threshold was  lowered to  $E_{th} =  5.5\; {\rm
  MeV}$ \cite{14-SK99}.

Radiochemical experiments  integrate in  time and in  energy because  they are
slow and  need time  to produce  measurable results. This  causes the  loss of
information about  single individual energy values. In  real time experiments,
instead,  it is  possible to  obtain single  values and  therefore  a spectrum
energy  to  distinguish the  different  neutrino contributions.   Furthermore,
given  that  the  scattered  electron  maintains the  same  direction  of  the
impinging  neutrino, it is  possible to  infer the  direction of  the incoming
neutrino and therefore  to point at its source. This  proved that the detected
neutrinos actually  came from  the Sun. The  number of detected  neutrinos was
about  1/2 lower  than  the number  of  expected ones,  aggravating the  Solar
Neutrino Problem.

Until 1990 there  were no observations of the initial  reaction in the nuclear
fusion  chain, i.e.  the detection  of ${\it  pp}$ neutrinos,  which  are less
model-dependent and hence more significant  to test the hypothesis that fusion
of hydrogen powers the Sun.  Two radiochemical experiments were built in order
to detect solar ${\it pp}$ neutrinos, both employing the reaction:
\begin{equation}
        \nu_{e} + {\rm ^{71}Ga} \rightarrow {\rm ^{71}Ge} + e^{-} \, .
        \label{eq:gallex-sage}
\end{equation}
which has a threshold of $E_{th} = 233\; {\rm keV}$.

In the Gallex experiment, located  at the Gran Sasso underground laboratory in
Italy,      30     tons     of      natural     gallium      were     employed
\cite{14-Hamp99,14-Alt05},\  while in  the soviet-american  experiment (SAGE),
located in the  Baksan underground laboratory, there were  50 tons of metallic
gallium  \cite{14-Abd99}.   Calibration  tests  with  an  artificial  neutrino
source, $^{51}$Cr, confirmed the efficiency  of both detectors. Once again the
measured neutrino  signal was  smaller than predicted  by the  SSM ($\approx\;
60\%$).

All experiments  detected fewer  neutrinos than expected  from the  SSM. Table
\ref{tab:observed  vs expected  ratio in  the four  experiments  (before SNO)}
summarizes the observed vs expected ratio for all experiments.

\begin{table}[h]
\centering

\vspace{0.5truecm}

\begin{tabular}{c c}
\hline \hline
Homestake & $0.34 \pm 0.03$ \\
Super-K & $0.46 \pm 0.02$ \\
SAGE & $0.59 \pm 0.06$ \\
Gallex and GNO & $0.58 \pm 0.05$ \\
\hline \hline
%\vspace{0.5truecm}
\end{tabular}
\caption{Observed vs expected  ratio in the four experiments  (before SNO, see
  later).}
\label{tab:observed vs expected ratio in the four experiments (before SNO)}
\end{table}
%%%%%%%%%%%%%%%%%%%%%%%%%%%%%%%%%%%%%%%%%%%%%%%

\vspace{1cm}

\subsection{The advent of SNO and Kamland: the solution of the Solar Neutrino 
Problem}\

The real breakthrough  in solar neutrino physics was due to  the advent of the
SNO  (Sudbury Neutrino  Observatory)  experiment. It  had  the peculiarity  to
measure simultaneously,  by means of a deuterium  \v{C}erenkov detector, three
different interaction channels for  neutrinos: the neutral current (NC: $\nu_X
+ d  \to \nu_X + p^+ +  n$), receiving contributions from  all active flavors,
the elastic scattering (ES: $\nu_X + e^- \to \, \nu_X + e^- $) and the charged
current (CC:  $\nu_e +  d \to e^-  + p^+  + p^+$), that  is sensitive  only to
electronic neutrinos.   In this way it has  been possible to prove  in a clear
and  direct  way that  the  measured  total neutrino  flux  was  in very  good
agreement with the SSM predictions, but only a fraction of these neutrinos had
conserved its flavor during their way  from the production point in the Sun to
the detector.

The first  SNO data \cite{14-SNOES}, including elastic  scattering and charged
current  analysis,  published  in  2001,  confirmed the  results  obtained  by
previous    solar   neutrino    experiments,   mainly    by   Super-Kamiokande
\cite{14-SK2001},  providing a significant  evidence (at  the $3.3  \, \sigma$
level) of  the presence of a  non-electronic active neutrino  component in the
solar flux.  For the first time it  was possible to indicate  the Large Mixing
Angle (LMA)  as the preferred solution  of the solar neutrino  puzzle, even if
different alternative  possibilities (and  in particular the  low probability,
low mass -LOW-  solution) were still surviving \cite  {14-Global2001}.  In the
following years, the SNO experiment measured also the neutral current channel,
using  different techniques.  The data  of these  different ``phases''  of the
experiment   are  usually   reported   as  SNO   I   \cite{14-SNOI},  SNO   II
\cite{14-SNOII}  (characterized  by  the  addition  of  salt  to  improve  the
efficiency of  neutral current detection)  and SNO III  \cite{14-SNOIII} (with
the use of helium chamber proportional counters).

The  year 2002  is very  often  denoted as  the ``annus  mirabilis'' of  solar
neutrino physics:  on April  the first SNO  results including  neutral current
detection \cite{14-SNOI,14-Ahmad2002ka} marked a  turning point in the history
of the  solar neutrino  problem, in  October the Nobel  prize for  physics was
awarded  to  R.  Davis  Jr.  \cite{14-Davis2002fb} and  M.Koshiba  (for  their
pioneering work on  the detection of cosmic neutrinos) and  on December of the
same year the first results of the Kamiokande Liquid scintillator AntiNeutrino
Detector  (KamLAND)  \cite{14-firstKL}  offered  the first  clear  terrestrial
confirmation of the validity of the oscillation solution to the solar neutrino
problem.

The total $^8$B neutrino  flux,  
$\phi_{NC}=  5.09^{+0.44}_{-0.43}(\rm stat)^{+0.46}_{-0.43}(syst)  \times 10^6
\,   cm^{-2}\   s^{-1}$,  
measured by  SNO  with  neutral  currents was  in   very   good   agreement  
with   the   SSM
\cite{14-bahcall:2001}.  Assuming the standard  shape for the component of the
solar neutrino  flux (undistorted spectrum hypothesis),  the SNO collaboration
recovered also a value of the non-electronic component of the flux
%$\phi_{\mu \tau}}= 3.41^{+0.45}_{-0.45}({\rm stat})^{+0.48}_{-0.45}
%({\rm syst}) \times 10^6 \, cm^{-2} s^{-1}$.
which was $5.3 \, \sigma$ different from zero, providing a direct proof of the
validity  of the  oscillation hypothesis.   These data  were also  decisive to
indicate the LMA region as the solution to the solar neutrino puzzle.
%A global analyis including the results from SNO phase I and the data from all
%the previous experiments on solar neutrinos brings to the following
%conclusions\cite{14-afterSNO2002} ... .

%An independent confirmation was given by the first data coming KamLAND,
%even if it is a reactor antineutrino experiment.
Looking at  the oscillation probability\footnote{For  instance, in a  simple 2
  flavor  analysis,  the  flavor   transition  probability  is  given  by  the
  expression    \hbox{$P_{12}    =    \sin{^2   (2    \theta_{12})}    \sin{^2
      \left(\frac{\Delta  m_{12}^2  ({\rm eV^2})  \,  L({\rm  km})}{4 E  ({\rm
          GeV})}\right)}$}, where  $\theta_{12} $ is the  mixing angle between
  the  two  flavors,  $\Delta  m^2_{\rm  ij}\equiv  m^2_{\rm  1}-m^2_{2}$  the
  difference of  the masses  squared, $L$ the  distance traveled, and  $E$ the
  neutrino  energy.}, it  is apparent  that the  reactor experiments  that run
before KamLAND, and used neutrino energy beams  of the order of the MeV with a
baseline of the  order of 1 km,  could test only values of  $\Delta m^2$ above
$10^{-3}$  $\rm{eV}^2$.   The KamLAND  experiment,  instead,  with an  average
baseline of about 180 km, was ideal to probe the LMA region, which corresponds
to values  of $\Delta  \rm{m}^2$ of the  order $10^{-5}-10^{-4}  \, \rm{eV}^2$
\cite{14-SK_KLpotentialities}.   The KamLAND experiment  studied the  ratio of
the number of inverse $\beta$  decay events (due to reactor $\bar{\nu}_e$ with
an  energy threshold  of 3.4  MeV) to  the expected  number of  events without
disappearance  and also  the spectrum  shape \cite{14-firstKL}.   The observed
deficit  of events  was  inconsistent with  the  expected rate  in absence  of
oscillation at the $99.95 \%$ confidence level.

Since one would  expect a negligible reduction of  the $\bar{\nu}_e$ flux from
the  SMA, LOW  and  vacuum solar  neutrino  solutions, the  LMA  was the  only
oscillation solution compatible with KamLAND results and CPT invariance.  This
evidence were further reinforced by the data published by the collaboration in
the  following   years  (with  greater  statistical   precisions  and  reduced
systematic  errors), which  showed also  a  spectral distortion  in very  good
agreement  with   the  oscillation  solution  \cite{14-KL2004,14-KLfollowing2,
  14-KLfollowing3}.  KamLAND data also restricted  the allowed LMA region in a
significant way.  The preferred values for $\Delta m^2_{12}$ and $\theta_{12}$
are slightly  higher than the ones  corresponding to the best  fit solution of
the solar  neutrino experiments,  but this small  tension can be  explained by
taking into  account the experimental uncertainties.  Moreover, the difference
on the $\Delta  m^2_{12}$ parameter has been reduced by  the more recent solar
neutrino data.

%\section{Solar models: neutrino production, fluxes and densities}

\section{Standard Solar Model}
\label{14-section:SSM}

SSMs have to be understood, primarily, as a framework within which solar models can 
be constructed and clear predictions can be made with respect to the properties of 
the solar interior, including the production of solar neutrinos. The defining 
characteristics are simple: the SSM is the result of the evolution of a 1~$\msun$ 
star since its formation and, the evolutionary models have to include the 
physical ingredients considered {\it standard} in stellar structure and evolution 
models (here, {\it standard} also implies trying to keep to a minimum the number of 
free tunable parameters -knobs- in the model). SSMs are therefore progressively 
refined as our understanding of stellar physics progresses.

In practice, a  SSM is  constructed  as  follows. An initial chemically homogeneous 
model of a 1~$\msun$ stellar model on the \hbox{pre-main} sequence is constructed 
with a composition determined by a guess (educated one) for the initial mass 
fractions of hydrogen $\xini$, 
helium $\yini$, and metals $\zini$ ($\xini+\yini+\zini=1$); additionally, a third 
free parameter has to be specified, the mixing length parameter $\alpha_{\rm MLT}$
of convection. This model is then evolved up  to the
solar   system   age   \hbox{$\tau_\odot=  4.57$~Gyr}   \cite{14-bahcall:1982,
  14-bahcall:1989}.  At this age the  model is required to match the present-day 
  solar
luminosity   $\lsun$   and   radius   $\rsun$,   as  well   as   the   surface
metal-to-hydrogen  abundance ratio $\zxsun$. The  initial and  final surface
\hbox{metal-to-hydrogen} ratios differ by about  10 to 15\% due to the effects
of gravitational  settling. In general, the SSM constructed with the first set  of 
guesses for
$\alpha_{\rm MLT}$,  $\yini$, and  $\zini$ will not  lead to  a satisfactory
agreement with the surface constraints, and an iterative procedure is used to refine 
the free parameters until the right surface conditions are achieved at $\tau_\odot$.
In general, surface conditions are matched to one part in $10^5$ or $10^6$ within two 
or three iterations. It is  important to keep  in mind
that the  SSM is  not just  a snapshot aimed  at representing  the present-day
structure of the Sun, but actually the result of taking into account all its previous 
history. There  are alternative  ways to  construct a  model  of the
present-day solar structure using,  for example, helioseismic constraints. 
These kind of models are  constructed `ad-hoc' to match helioseismic data and
have, therefore, limited predictive power. 

The internal  structure of a SSM depends  on the values adopted  for the three
constraints  mentioned above and,  of course,  on the  physical inputs  of the
models such  as the radiative  opacities, cross sections of  nuclear reactions
and others.   Next we describe  the changes/updates that have  occurred during
the last decade that impact predictions of solar models.

\subsection{Input physics and parameters}

\subsubsection{Solar Surface Composition}\label{14-sec:solarcompo}

The constraint imposed by the surface metallicity of the Sun or, more precisely, the 
surface metal-to-hydrogen ratio $\zxsun$, is critical in the construction of 
solar models. The reason is that, aside from the 10 to 15\% change in this value due 
to the action of gravitational settling, $\zxsun$ determines almost directly the 
metallicity of solar models. As for any other star, the metal content in the Sun 
has a fundamental role in its structure through its contribution to the radiative 
opacity $\kappa$, which determines,  in turn,  the temperature gradient  in the  
radiative solar interior. It is important, in fact, that the abundance of 
individual metals are accurately determined, because different 
elements contribute to the radiative opacities in different regions of the Sun. 

The abundance of metals in the  solar surface has to be determined or inferred
from  a  variety of  sources:  photospheric  abundances from  solar
spectra, chemical  analysis of primitive  meteorites, emission lines  from the
solar  corona, composition  of the  solar wind  \cite{14-lodders:2009}.  While
meteoritic abundances are  the most precisely determined, at  least 2/3 of the
solar metallicity  is composed by  the volatile elements  C, N, and O  and can
only be determined from analysis of the solar spectrum. 

Over  the   last  decade,  the  development   of  three-dimensional  radiation
hydrodynamic (3D RHD)  models of the solar atmosphere  has prompted a thorough
revision of the solar composition determined from the solar spectrum. These 3D
RHD models of the solar atmosphere  capture the dynamics of convection and its
interaction with the  radiation field, and are able  to reproduce features such
as the solar granulation  pattern, observed limb-darkening, asymmetries in the
shapes  of spectral lines  \cite{14-beeck:2012}.  The  structure of  the solar
model atmospheres derived by different  groups are nicely consistent with each
other, adding to the credibility of the models. Newly derived spectroscopic 
abundances rely on the 3D atmosphere model, or more appropriately on a one-
dimensional model obtained from a suitably averaged 3D model,
 as the background on top of  which detailed radiative
transfer and line formation calculations are performed {\it a posteriori}. 
It is this second step that leads, finally, to
the  determination of  the  abundances  of the  different  elements. The  most
thorough  and consistent  determination of  the solar  photospheric abundances
based on  3D model atmospheres has been presented by Asplund and collaborators
\cite{14-asplund:2009,  14-asplund:2005}, although  revision  on key  elements
like oxygen were initially published already in 2001 \cite{14-allendeprieto:2001}.  
In addition  to   using  3D   RHD  atmosphere  models,   non-local  thermodynamic
equilibrium has been taken into account when computing line formation for some key
elements such as C, N, and O.  Also, and this is of particular importance for
oxygen, blends in  the solar spectrum that had  been previously unnoticed were
identified and  taken into  account in the  determination of  abundances.  The
most relevant result in the context of solar models and neutrinos is that 
abundances of CNO elements (also Ne, but this is mostly because its abundance ratio 
to oxygen is assumed fixed) have been revised down by  30 to 40~\%.  Combining the 
abundance of all
metals, the present-day metal-to-hydrogen ratio that has been obtained is
$\zxsun=0.0178$  \cite{14-asplund:2009}.  This represents  a large  decrease in
comparison with 
previously   accepted  values,   0.0245  \cite{14-grevesse:1993}   and  0.0229
\cite{14-grevesse:1998},  that have  been widely  used in  solar  modeling. We
note, however, that  results by Asplund have not  been unchallenged.  In fact,
also based on 3D RHD model  solar atmospheres, larger CNO abundances have been
derived \cite{14-caffau:2011} to yield  $\zxsun= 0.0209$, much closer to older
determinations. Discrepancies between authors seem to have their origin at the
preferred set of spectral lines each group uses and on using either a spectral
synthesis or equivalent width techniques to determine the final abundances.

In the  last decade there have been  two flavors in SSM  calculations.  In one
case a {\em high} solar metallicity from older determinations 
\cite{14-grevesse:1993,14-grevesse:1998} is adopted; we will generically refer
to these models 
as {\em high-Z} solar models.  In the other case a {\em low} $\zxsun$ 
\cite{14-asplund:2009,14-asplund:2005} is taken from  and we refer to these,
not surprisingly, as the {\em low-Z} solar models.  Differences in the 
structure of  {\em high-Z}  and {\em low-Z}  models are readily  noticeable in
quantities such as the internal sound speed and density profiles, the depth of
the  solar  convective  envelope,  and  the  surface  helium  abundance  among
others.  The deficit  that {\em  low-Z} models  have in  matching helioseismic
constraints has been named the solar abundance problem in the literature, in clear
analogy to the solar neutrino problem. We discuss it in some detail in Section~\ref
{14-sec:helios}.

\subsubsection{Radiative Opacities}

The most  widely used calculations of atomic  radiative opacities, appropriate
for solar interiors, are those from OPAL \cite{14-iglesias:1996}. However, the
Opacity  Project (OP)  released in  2005 a  completely independent  set of
atomic radiative  opacities for stellar  interiors \cite{14-badnell:2005}.  In
the  case of the  solar radiative  interior, differences  between OPAL  and OP
Rosseland mean  opacities are  of the order  of a  few percent, with  OP being
larger by about 3\% at the base of the convective zone and 1 to 2\% smaller in
the central regions (see Fig.  7 in \cite{14-badnell:2005}).
At low  temperatures, at which  molecules can form,  neither OP or  OPAL atomic
opacities  are  adequate and  have  to  be  complemented by  low-temperature
opacities   \cite{14-ferguson:2005}.  Due   to  the   relatively   high  solar
temperature,  their influence  in the  properties  of solar  models is  rather
limited.

\subsubsection{Nuclear reactions cross sections}

Experimental  and  theoretical work  on  the  determination  of nuclear  cross
sections  have  been  very  active  fields  with  a  strong  impact  on solar 
model predictions of solar neutrino fluxes. Recently, a set of recommended rates 
and uncertainties, expressed  through the $S$-factor\footnote{A
non-resonant charged-particle induced reaction cross section can be written as
$\sigma (E)=\frac{S(E)}{E} 
{\rm exp \left[  -2 \pi \eta(E)\right]}$ where $\eta(E)=  Z_1 Z_2 \alpha/v$ is
the Sommerfeld  parameters, $v=\sqrt(2E/\mu)$, $\alpha$  the fine structure
constant  in natural  units, and  $\mu$ the  reduced mass  of  the interacting
nuclei.  The nuclear physics is isolated in $S(E)$, the 
astrophysical or S-factor, a slowly
varying  function of  energy that  can  be more  accurately extrapolated  from
experimental data down to the energy of the Gamow peak.},  for
all the reactions  both in the pp-chains and CNO-bicycle  that are relevant to
solar  modeling and neutrino production,  has  been
published (Solar  Fusion II, \cite{14-adelberger:2011},  hereafter SFII).  The
results  presented  in  SFII  reflect  the progress  made  in  laboratory  and
theoretical nuclear  astrophysics over the last decade,  since the publication
of  the  seminal Solar  Fusion  I  (SFI)  article \cite{14-adelberger:1998}  .
Unfortunately, for  reasons of  space, here we  cannot review in  detail every
reaction.   Instead, we  provide in  Table~\ref{14-tab:nucrates}  the standard
$S$-factors at zero energy, $S(0)$, and the  uncertainties recommended in SFII 
for the most relevant reactions. For comparison, with results  from SFI are also 
shown. The impact of  changes in key  reactions on the  production of neutrino  
fluxes is discussed in  Section~\ref{14-sec:fluxes}. The reader is referred  to the 
SFII paper and references  therein for details on the  experimental and theoretical
developments  in nuclear  astrophysics  related  to the  Sun  during the  last
decade.

\begin{table}[h]
\begin{center}
%\begin{minipage}{120mm}
\begin{tabular}{llcc}
\hline \hline
\multicolumn{2}{c}{Reaction} & SFII & SFI \\ 
 & & $S(0) \ \left[\hbox{keV b}\right]$ & $S(0) \ \left[\hbox{keV b}\right]$ \\
\hline
${\rm S_{11}}$ & $p(p,{\rm e}^+\nu_e)d$ & $4.01\times 10^{-22} \left(1 \pm
  0.010\right)$ & $4.00\times 10^{-22} \left(1 \pm 0.005\right)$ \\
${\rm S_{33}}$ & ${\rm ^3He(^3He},2p){\rm ^4He}$ & $5.21\times 10^{3} \left(1
  \pm 0.052\right)$ & $5.4\times 10^{3} \left(1 \pm 0.074\right)$ \\
${\rm S_{34}}$ & ${\rm ^3He(^4He},\gamma){\rm ^7Be}$ & $5.6\times 10^{-1}
  \left(1 \pm 0.054\right)$ & $5.3\times 10^{-1} \left(1 \pm 0.094\right)$ \\
${\rm S_{hep}}$ & ${\rm ^3He}(p,{\rm e}^+\nu_e){\rm ^4He}$ & $8.6\times 10^{-20}
  \left(1 \pm 0.30\right)$ & $2.3\times 10^{-20}$ \\
${\rm S_{17}}$ & ${\rm ^7Be}(p,\gamma){\rm ^8B}$ &  $2.08\times 10^{-2} \left(1
  \pm 0.077\right)$ & $1.9\times 10^{-2} \left(1 ^{+ 0.20}_{-0.10}\right)$ \\
${\rm S_{1,14}}$ & ${\rm ^{14}N}(p,\gamma){\rm ^{15}O}$ & $1.66 \left(1 \pm
  0.072\right)$ & $3.5 \left(1 ^{+0.11}_{-0.46} \right)$ \\
\hline
\end{tabular}
\caption{Standard  astrophysical  factors and  uncertainties  for key  nuclear
  reactions   in  the   pp-chains   and  CNO-bicycle.   SFII  represents   the
  state-of-the-art  \cite{14-adelberger:2011};  SFI \cite{14-adelberger:1998}  shows,  
  for  comparison, the  situation
  around 1998. \label{14-tab:nucrates}}
%\end{minipage}
\end{center}
\end{table}

\subsection{Solar Models: Helioseismology}\label{14-sec:helios}

Helioseismology, the study of the  natural oscillations of the Sun, provides a
unique  tool to  determine  the structure  of  the solar  interior.  The  '90s
witnessed a  rapid development of  helioseismic observations and  analysis 
techniques, which led,  in very few  years, to an  accurate characterization of  the solar
interior  \cite{14-jcd:2002}.   The agreement  between  SSMs and  helioseismic
inferences of the  solar structure \cite{14-jcd:1996,14-bahcall:2001} provided
a  strong support  to the  accuracy with  which SSMs  could predict  the $^8$B
neutrino  flux and,  therefore, a  strong indication,  before Kamland  and SNO
results found evidence  of neutrino flavor oscillations, that  the solution to
the solar neutrino problem had to be found in the realm of particle physics.

In  the  context  of the  present  article,  the  most relevant  results  from
helioseismology   are   the   following.    The  depth   of   the   convective
envelope\footnote{The Sun is characterized by  an outer region where energy is
  transported  by convection.  The  boundary between  this region,  located at
  $\rcz$,   and  the  radiative   interior  can   be  accurately   located  by
  helioseismology because  the discontinuity in  the slope of  the temperature
  gradient accross this  boundary leaves its imprint in  the solar sound speed
  profile.   The depth  of the  envelope  can be  located by  helisoeismology
because properties  of solar oscillations  are sensitive to the  derivative of
 the sound  speed as a function  of depth.}  is $\rcz=0.713  \pm 0.001\, \rsun$
  \cite{14-basu:1997a}  and  the  surface  helium  abundance  $\ys=0.2485  \pm
  0.0034$ \cite{14-basu:2004}.   The sound  speed differences between  the Sun
  and a reference solar model can be obtained by inversion from the oscillation
  frequencies with a formal error of a few parts per $10^{-4}$ for most of the
  solar     interior     $0.07     \lesssim     R/\rsun     \lesssim     0.95$
  \cite{14-basu:1997b,14-kosovichev:1997}.  Most recently, using a time series
  4752  days-long  from the  Birmingham  Solar  Oscillation Network,  improved
  results  on  the   sound  speed  in  the  solar   core  have  been  obtained
  \cite{14-basu:2009}.   The  density  profile  can also  be  determined  from
  inversion of frequencies, but with worse precision than for the sound speed,
  and we  therefore assign to  it a secondary  role in constraining  the solar
  structure.

As  mentioned previously,  metals determine  to a  large extent  the radiative
opacity  in  the solar  interior  and, in  this  way,  define the  temperature
stratification from below of convective envelope inwards, to the solar center.
At the  base of the  convective zone, for  example, metals are  responsible of
about 70\%  of the total radiative  opacity with O,  Fe and Ne being  the main
contributors. In  the solar core,  where light metals are  completely ionized,
the contribution from Fe and, to a  lesser extent Ni, Si and S, is still above
30\%.   In view  of  this,  it is  not  surprising that  the  low  CNO and  Ne
abundances determined  from 3D model atmospheres  have a strong  impact in the
structure of the solar interior. 

It has  been clear since initial  works where {\em low-Z}  SSMs were presented
that low  $\zxsun$ values posed a  problem, later named the solar  abundance problem, 
for solar  modeling  \cite{14-montalban:2004,  14-turckchieze:2004,  14-basu:2004,
  14-bahcall:2005a}. In  short, all  helioseismic predictions of  these models
are in  disagreement with observations. On  the other hand,  {\em high-Z} SSMs
have consistently  reproduced earlier success  \cite{14-jcd:1996}.  The solar
abundance  problem  represents  the  incompatibility between  the  best  solar
atmosphere  and  interior models  available  \cite{14-delahaye:2006}. In  this
review, we will base the  presentation and discussion of results on the
most up-to-date
standard   solar   models that  we   identify   as   SFII-GS98   and   SFII-AGSS09
\cite{14-serenelli:2011}, representative  of {\em high-Z} and  {\em low-Z} SSM
families  defined in  Section~\ref{14-sec:solarcompo} respectively.   With the
exception made on small quantitative variations, results based on these models
are  extensible  to   results  for  all  SSMs  available   in  the  literature
corresponding to each of the two families.

The most important characteristics of the SFII-GS98 and SFII-AGSS09 models are
summarized in  Table~\ref{14-tab:heliossm}. Helioseismic constraints  are also
included for comparison when appropriate. The disagreement between SFII-AGSS09
and  helioseismic  data is  evident  in  the  surface metallicity  and  helium
abundances,  $\zs$ and  $\ys$, and  in the  depth of  the  convective envelope
$\rcz$.  When  model  uncertainties  are  included,  the  discrepancy  between
SFII-AGSS09  and seismic  results are,  for each  of the  quantities mentioned
above,  of  the  order  3  to  $4-\sigma$  \cite{14-serenelli:2009}.  On  the
contrary, the SFII-GS98 model performs very well, within $1-\sigma$ when model
uncertainties are accounted for.  

\begin{table}[h]
%\begin{minipage}{105mm}
\begin{center}
\begin{tabular}{lccc}
\hline \hline
 & SFII-GS98 & SFII-AGSS09 & Helioseismology\\ 
\hline
$\zxsun$ & 0.0229 & 0.0178 & --- \\
$\zs$ & 0.0170 & 0.0134 & $0.0172 \pm 0.002$ \cite{14-antia:2006}\\
$\ys$ & 0.2429 & 0.2319 & $0.2485 \pm 0.0034$ \cite{14-basu:2004} \\
$\rcz/\rsun$ & 0.7124 & 0.7231 & $0.713 \pm 0.001$ \cite{14-basu:1997a} \\
$\left< \delta c / c \right> $ & 0.0009 & 0.0037 & --- \\
$\left< \delta \rho / \rho \right> $ & 0.011 & 0.040 & --- \\
$\zc$ & 0.0200 & 0.0159 & --- \\
$\yc$ & 0.6333 & 0.6222 & --- \\
$\left< \mu_C \right> $ & 0.7200 & 0.7136 & $0.7225 \pm 0.0014$ 
\cite{14-chaplin:2007} \\ 
$\zini$ & 0.0187 & 0.0149 & --- \\
$\yini$ & 0.2724 & 0.2620 & --- \\
\hline
\end{tabular}
\caption{Main characteristics  of SSMs  representative of {\em  high-Z} (GS98)
  and  {\em low-Z}  (AGSS09)  solar compositions.  Models  have been  computed
  including        the       most        up-to-date        input       physics
  \cite{14-serenelli:2011}.   Helioseismic    constraints   are   given   when
  available. See text for details. 
 \label{14-tab:heliossm}} 
%\end{minipage}
\end{center}
\end{table}

\begin{figure}[h!]
%\vspace{-15truecm}
\begin{center}
\hspace{-2truecm}
\includegraphics[width=16.5cm]{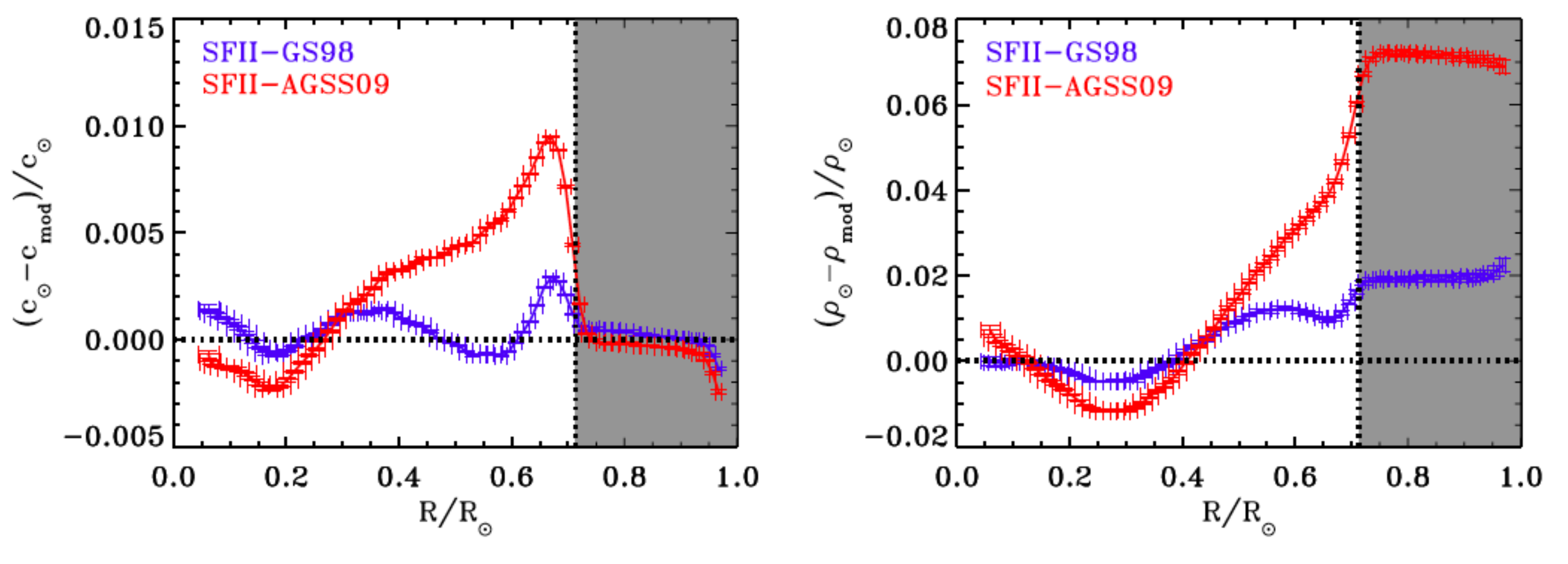}
\caption{Sound  speed and  density relative  differences between  solar models
  and    the    Sun     as    determined    from    helioseismic    inversions
  \cite{14-serenelli:2009}. The  convective envelope  is depicted by  the grey
  area. \label{14-fig:inversions}}
\end{center}
\end{figure}

Very explicit manifestations of the solar abundance problem are shown in the
plots in Figure~\ref{14-fig:inversions}, where  degradation in the sound speed
and   density    profiles   found   in   {\em   low-Z}    SSMs   are   clearly
evident. Particularly  the peak in  the sound speed profile  differences found
right  below  the  convective zone  is  4  times  larger  in the  {\em  low-Z}
SFII-AGSS09 than in the {\em high-Z}  SFII-GS98 model. The reason is the wrong
location of $\rcz$  in the model, caused by the lower  opacity which, in turn,
is due  to the low abundance of  metals.  The density profile  also shows very
large discrepancies, but they are  less telling. Density inversions include as
a constraint the known value of the solar mass and for this reason
small differences in
the core, where density is large,  translate into the large difference seen in
the  outer  envelope.  The  average   rms  in  the  sound  speed  and  density
differences,  $\left< \delta  c /  c\right>$ and  $\left< \delta  \rho  / \rho
\right>$, also show that {\em low-Z}  models are about 4 times worse than {\em
  high-Z} models.

Low-degree  helioseismology  provides   useful  information  about  the  solar
innermost  regions.  Specific combinations  of  mode  frequencies enhance  the
signal  that the  structure  of the  solar  core imprints  on the  oscillation
pattern  \cite{14-roxburgh:2003}. This  has been  used to  determine  the mean
molecular   weight    averaged   over   the   innermost    20\%   solar   core
\cite{14-chaplin:2007}, $\left< \mu_C \right>$ in Table~\ref{14-tab:heliossm}.
Comparison with SSMs  results shows that $\left< \mu_C \right>$  is too low in
{\em low-Z} models as a result of the lower helium abundance ($\yc$).  
This is due to the lower temperature in the solar core and the 
constraint  imposed by  the  solar luminosity.  The  decreased nuclear  energy
production originated by a smaller core temperature has  to be compensated by
an increased  hydrogen mass fraction,  therefore leading to a  lower molecular
weight.  It  is interesting to  note this puts  a stringent constraint  in the
amount of rotational mixing that can take  place in the solar core if the {\em
  low-Z} abundances  are correct, since  any mixing would lower  the molecular
weight even more, by bringing fresh hydrogen from outer regions, and make the 
agreement with helioseismic data worse.

The   current  situation   regarding  SSMs   and  their   performance  against
helioseismic inferences on  the solar structure can be  summarized as follows.
SSMs   that  use   solar  abundances   derived  from   1D   model  atmospheres
\cite{14-grevesse:1993, 14-grevesse:1998}, i.e. {\em high-Z} models, reproduce
overall  the most  important seismic  constraints. Improvements  in  the input
physics,  e.g.  radiative  opacities  and nuclear  reaction  rates, that  have
occurred  over the  last 10  years introduce  only small  changes to  the solar
structure as seen by helioseismology.  On the other hand, the solar abundance
problem arises  if the solar surface  composition used to  construct SSMs are
derived  from the  most sophisticated  3D  RHD solar  model atmospheres.  The
family  of {\em  low-Z} SSMs does not  match any helioseismic constraint. 

Have we reached the limit where the  paradigm of the SSM is not good enough as
a  model of  the  solar interior?  Are  the 3D-based  determinations of  solar
abundances  systematically underestimating  the metallicity  of the solar surface?
Does the microscopic input physics  in solar models, e.g. radiative opacities,
need to be thoroughly revised? It  is not possible to advance answers to these
questions,  but solar  neutrino  experiments  can play  an  important role  in
guiding research towards the solution of the solar abundance problem. In the
next section we  discuss the current status on  the theoretical predictions of
solar neutrino fluxes and the prospects of using solar neutrinos to constraint
the properties of the solar core. 
%{\bf Carlos, somewhere it should go the part
% on the determination of solar fluxes from the available experimental data. I
%will sketch that part only to give structure to the section.} 

\subsection{Solar Models: Neutrino Fluxes}\label{14-sec:fluxes}

\subsubsection{Production}

Based  on  theoretical arguments  and  indirect  evidence,  it has  long  been
believed that  the source of  energy of the  Sun is the conversion  of protons
into  helium, $4{\rm p  \longrightarrow ^4He  + 2e^+  + 2\nu_e  +\gamma}$. The
original quest for solar neutrinos  was indeed the search for the experimental
confirmation of this  hypothesis. In more detail, hydrogen  burning in the Sun
(and  in all  other hydrogen-burning  stars)  takes place  either through  the
pp-chains  or the  CNO-bicycle\footnote{Under peculiar  conditions  reached in
  advanced phases of stellar evolution,  hydrogen can be converted into helium
  by other cycles like the  NaMg-cycle. While important for nucleosynthesis or
  intermediate mass elements, these  processes are not energetically relevant}
\cite{14-clayton:1984, 14-bahcall:1989}.  Proton  fusion through the pp-chains
is a primary process because only protons need be present in the star.  On the
contrary, the CNO-bicyle is secondary  because proton fusion relies on, and is
regulated  by, the abundance  of C,  N, and  O which  act as  catalyzers. This
qualitative  difference is very  important, since  it renders  neutrino fluxes
from the  CNO-bicyle a very  good diagnostic tool  to study properties  of the
solar core,  particularly its  composition, as it  will be discussed  below. A
general discussion on the production of solar neutrinos is out of the scope of
the present review, but can be found elsewhere \cite{14-bahcall:1989}.

SSM calculations of neutrino fluxes  have been affected by developments in the
input  physics discussed in  previous sections.  The two  areas that  have the
strongest impact  on the neutrino fluxes  predicted by models  are: changes in
nuclear    cross   sections,    and    the   new    solar   composition.    In
Table~\ref{14-tab:nufluxes} we  list the results  for neutrino fluxes  for the
up-to-date SSMs SFII-GS98  and SFII-AGSS09. For comparison we  include, in the
last  column, results  from the  BP04 SSM  \cite{14-bahcall:2004}.

\begin{table}[ht]
\begin{center}
%\begin{minipage}{132mm}
\begin{tabular}{lcccc}
\hline \hline
Flux & SFII-GS98 & SFII-AGSS09 & Solar & BP04 \\ 
\hline
pp & $5.98(1 \pm 0.006)$ & $6.03(1 \pm 0.006)$ & $6.05(1^{+0.003}_{-0.011})$ &
$5.94(1\pm 0.01 )$\\
pep & $1.44(1 \pm 0.012)$  & $1.47(1 \pm 0.012)$ & $1.46(1^{+0.010}_{-0.014})$
& $1.40 (1 \pm 0.02)$\\
hep & $8.04(1 \pm 0.30)$ & $8.31(1 \pm 0.30)$ & $18(1^{+0.4}_{-0.5})$ & 
$7.8 (1 \pm 0.16) $ \\
$^7$Be & $5.00(1 \pm 0.07)$ & $4.56(1 \pm 0.07)$ & $4.82(1^{+0.05}_{-0.04})$ &
$4.86 (1 \pm 0.12) $ \\
$^8$B & $5.58(1 \pm 0.13)$ & $4.59(1 \pm 0.13)$ & $5.00(1\pm 0.03)$ & 
$5.79 (1 \pm 0.23 )$\\ 
$^{13}$N & $2.96(1 \pm 0.15)$ & $2.17(1 \pm 0.13)$ &$\leq 6.7$ & 
$5.71 (1 \pm 0.36)$ \\ 
$^{15}$O & $2.23(1 \pm 0.16)$ & $1.56(1 \pm 0.15) $ &$\leq 3.2$ & 
$5.03 (1 \pm 0.41)$ \\ 
$^{17}$F & $5.52(1 \pm 0.18)$ & $3.40(1 \pm 0.16)$ & $\leq 59.$ & 
$5.91 (1 \pm 0.44) $\\
\hline
$\chi^2/P^{\rm agr}$& 3.5/90\% & 3.4/90\%& --- & --- \\
\hline
\end{tabular}
\caption{SSM predictions for solar  neutrino fluxes (second and third columns)
  and  solar  neutrino fluxes  (fourth  column)  inferred  from all  available
  neutrino data.  Units are, in ${\rm cm^{-2} s^{-1}}$, as usual: $10^{10}$ 
 (pp),  $10^9$  ($^7$Be),  $10^8$  (pep, $^{13}$N,  $^{15}$O)  $10^6$  ($^8$B,
  $^{17}$F), and $10^3$ (hep).  
\label{14-tab:nufluxes}}
\end{center}
%\end{minipage}
\end{table}

The  most striking  difference  is the  large  reduction in  the $^{13}$N  and
$^{15}$O  fluxes between the  SFII-GS98 and  BP04 models,  which use  the same
solar composition.   This reduction comes as  a result of the new determination of
${\rm  S_{1,14}}$,  mostly  by  the LUNA  experiment  \cite{14-formicola:2004,
  14-marta:2008}, that has halved its  value with respect to previous results
(Table~\ref{14-tab:nucrates}).  If  correct, the new expectation  value of the
combined  $^{13}$N+$^{15}$O fluxes  poses an  even more  challenging  task for
neutrino experiments to detect CNO fluxes. 
By comparing  fluxes in  Table~\ref{14-tab:nufluxes} for models  computed with
the same solar composition (SFII-GS98 and  BP04), it can be seen that in terms
of  flux values, those  associated with  the pp-chains  have not  changed much
since 2004, despite improvements in the input physics entering solar model
calculations. Few percent changes are present and are the result of changes in
the nuclear cross  sections discussed before and also of  the new OP radiative
opacities.  This is an encouraging  situation; it implies that neutrino fluxes
are  robust predictions of  solar models  and, as  experimental data  on solar
neutrinos accumulate, it will be possible to start fulfilling the initial goal
posed by  Davis and Bahcall: to use  solar neutrinos to learn  about the solar
interior.

\begin{figure}
%\vspace{-2truecm}
\includegraphics[width=15cm]{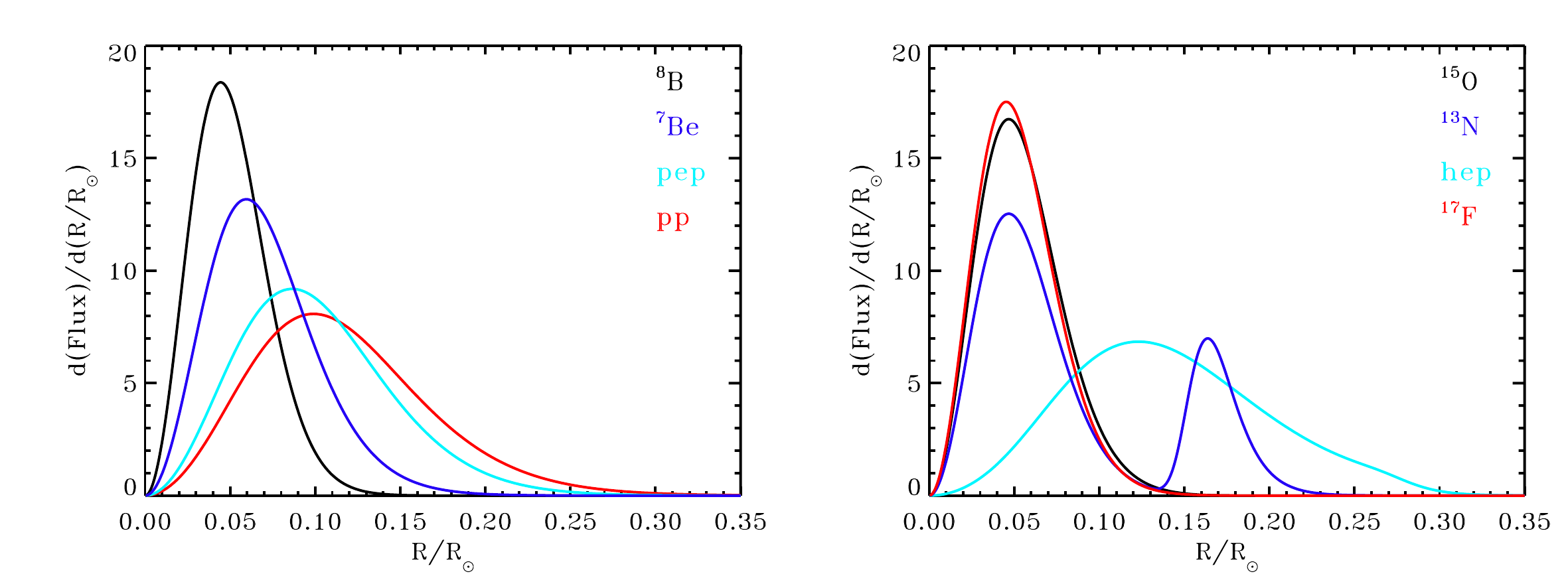}
\caption{Normalized production  profiles of solar  neutrinos as a  function of
  solar radius. \label{14-fig:nuprofiles}}
\end{figure}

In Figure~\ref{14-fig:nuprofiles} we show the
distribution of the solar neutrino fluxes as a function of solar radius. Together 
with the electron density profile, provided also by solar models (and neutron density 
profiles for sterile neutrino studies), these quantities are of fundamental 
importance for neutrino oscillation studies. It is
worth noting  that the $^{13}$N  flux has two  components. The larger  one is
associated with  the operation  in quasi steady-state  of the CN-cycle  in the
innermost solar core  ($R < 0.1\, \rsun$), and for  this reason coincides with
the   production    region   of   the   $^{15}$O   flux    (right   panel   in
Figure~\ref{14-fig:nuprofiles},  blue  and  black curves  respectively).  This
component  of the  $^{13}$N  flux, as  well  as the  total  $^{15}$O flux,  is
linearly  dependent  on ${\rm  S_{1,14}}$.  The  additional  component of  the
$^{13}$N flux comes from the residual burning of $^{12}$C by
the  reactions ${\rm  ^{12}C(p,\gamma)^{13}N(\beta^+)^{13}C}$  at temperatures
not high enough to close the CN-cycle with a proton capture on $^{14}$N. 
This component is
completely independent of ${\rm S_{1,14}}$. The careful
reader will notice that the ratio of $^{13}$N and $^{15}$O fluxes is different
in the SFII-GS98 and BP04 models, despite having the same solar composition.
Whereas the added $^{13}$N+$^{15}$O neutrino flux is linearly proportional to the C+N 
abundance in the solar core and also linearly proportional to ${\rm S_{1,14}}$, this
degeneracy can be broken, at least theoretically, if the two fluxes can be 
experimentally isolated from one another.

The impact  of the {\em  low-Z} solar composition  on the production  of solar
neutrinos  can  be  grasped  by   comparing  results  of models  SFII-GS98  and
SFII-AGSS09  shown  in Table~\ref{14-tab:nufluxes}.  As  stated before,
metals shape the solar  structure  through  the radiative  opacity.  The  lower
abundance of metals  in the AGSS09 composition is  responsible for a reduction
of the temperature in the solar core of about 1\%. Because of the
extreme temperature sensitivity of some  of the neutrino fluxes this is enough
to produce  large changes in  the total fluxes.  The most extreme case  is, of
course,  $^8$B, with  the SFII-AGSS09  value  being $\sim  20\%$ smaller.  For
$^7$Be the  reduction is of $\sim  9\%$. 

Given the small  uncertainties in the
experimental determination  of these  fluxes, it would be tempting to think these 
neutrino fluxes have
the potential to discriminate between the two flavors of solar composition and
contribute, in  this way, to the  solution of the solar  abundance problem. As
can  be seen  in  Table~\ref{14-tab:nufluxes}, unfortunately,  the $^7$Be  and
$^8$B fluxes determined from experiments  lie almost right in between the 
{\em high-Z} and {\em low-Z} models. 

In any case, since it is known  that {\em low-Z} solar models do not reproduce
well the solar structure as discussed in the previous section, it is dangerous
to extract conclusions from comparing neutrino  fluxes of this model to  
experimental results. Regardless
of what the  solution to the solar abundance problem is,  since it will modify
the solar interior structure, it will also change the expected values for the 
neutrino fluxes.  In  this regard, CNO  fluxes are particularly  interesting.  
Although
they are of  course affected by temperature variations  to a comparable degree
as  the $^8$B flux  is, they  carry an  extra linear  dependence on  the solar
composition  that is  not  related to  temperature  variations. Of  particular
interest is the  linear dependence of the $^{13}$N and  $^{15}$O fluxes on the
combined C+N abundance\footnote{The $^{17}$F  flux is linearly dependent on O,
  but  unfortunately  the  flux is  too  low  to  be detectable  with  current
  experimental  capabilities.}.  It  is  this dependence  that enhances  their
capability  as a diagnostic tool.  In fact,  differences between  SFII-GS98 and
SFII-AGSS09  models for  these two  fluxes are  of the  order of  30\% (taking
SFII-GS98 as reference)  and, what is more important,  a large contribution to
these differences  does not have  an origin on temperature  differences between
the models.

The last  row in Table~\ref{14-tab:nufluxes}  shows the results of  a $\chi^2$
test for the two  models against the solar fluxes also shown  in the table. It
is  clear that  both SSMs  give  very good  agreement with  current data.   We
emphasize  again,  however, that  the  four  fluxes  that are  currently  well
determined  from data  and  the  luminosity constraint,  depend  on the  solar
composition  only in  an indirect  manner. Experimental  determination  of the
combined  $^{13}$N+$^{15}$O  flux  will  therefore provide  qualitatively  new
information  on  the solar  structure  and  composition.   In fact,  one  can 
take advantage of  the similar response  to temperature variations that  CNO 
fluxes and   the   $^8$B   flux    has. This  has   been   exploited
\cite{14-haxton:2008} to develop  a very simple method to  determine the solar
core C+N abundance that  minimizes environmental uncertainties in solar models
(that is, sources of uncertainty  that affect the solar core temperature). The
idea  is simple: the  temperature dependences  are cancelled  out by  using an
appropriate ratio between the  $^8$B and the combined $^{13}$N+$^{15}$O fluxes
where SSM  fluxes only act  as normalization values  and the overall  scale is
determined  by   an  actual  $^8$B  flux  measurement.   The  only  additional
requirement  is that  a  measurement of  the  combined $^{13}$N+$^{15}$O  flux
becomes  available.   The current  upper  limit  on  this combined  flux  from
Borexino \cite{14-pepBX} places an upper limit on  the C+N central mass fraction of 
$X_{\rm C+N} < 0.072$. Results  for the SFII-GS98 and SFII-AGSS09  models are 
$X_{\rm C+N}=0.048$ and 0.039 respectively.

%{\bf Somewhere  in this section we should fit the  description by Carlos  of 
% the derivation of solar fluxes.}

\subsubsection{Uncertainties}

Uncertainties in the  model predictions of solar neutrino  fluxes are given in
Table~\ref{14-tab:nufluxes}. For deriving  the total uncertainty basically two
approaches can be  used. On one hand, all contributions  of uncertainty can be
treated    simultaneously    by     doing    a    Monte    Carlo    simulation
\cite{14-bahcall:2006}.  The  advantage is that  intrinsic non-linearities are
captured   in  the  total   error.   The   disadvantage  is   that  individual
contributions to the  total uncertainty are hardwired in  the final result and
can not be  disentangled. Fortunately, for the current  level of uncertainties
entering SSM calculations, non-linearities seem to be negligible and the total
uncertainty  in neutrino fluxes  can be  obtained (adding  quadratically) from
individual contributions.  To compute  the latter, the expansion 
of   fluxes   as  a   product   of   power-laws   in  the   input   parameters
\cite{14-bahcall:1989} around central values is  a widely  used, practical, 
insightful  and accurate approach. Uncertainties in the model fluxes listed in
Table~\ref{14-tab:nufluxes} have been obtained in this way.

The most  important change  introduced in the  estimation of  uncertainties is
related to  the treatment of the  solar composition.  Up until  the BP04 model
\cite{14-bahcall:2004},  the uncertainty  in the  solar composition  was taken
into account by  considering variations of the total  solar metallicity (to be
more  precise, changes in  the $\zxsun$  value used  to construct  SSMs). This
leads to an overestimation of  the neutrino uncertainties.  The reason is that
metals dominating  the error  budget in $\zxsun$  (C, N,  O, and Ne)  have, at
most,  a  moderate  impact on  the  neutrino  fluxes  because of  their  small
contribution to the radiative opacity,  and therefore a rather small impact on
temperature, in  the region  where most neutrinos  are produced. On  the other
hand, elements such  as Fe, S and Si are second  order in determining $\zxsun$
but play a  fundamental role as sources  of opacity in the solar  core.  It is
important,    therefore,   to    treat   metal    uncertainties   individually
\cite{14-bahcall:2005c}.   Of  course, in  the  case  of  the CNO  fluxes  the
situation is different because CNO  elements catalyze the CNO-bicycle and this
overimposes an  almost linear dependence of  the $^{13}$N and  $^{15}$O on the
C+N  content of  the  solar and  a similar  dependence  of $^{17}$F  on the  O
abundance. The  uncertainties in the  neutrino fluxes given for  the SFII-GS98
and  SFII-AGSS09 SSMs  have been  computed  using the  uncertainties for  each
relevant  element given in  the original  publications \cite{14-grevesse:1998,
  14-asplund:2009}. As a result, for  either family of solar models, i.e. {\em
  high-Z} or  {\em low-Z}  models, the solar  composition is not  the dominant
source of uncertainty for  any of the fluxes of the pp-chains.  In the case of
the CNO fluxes,  the linear dependence mentioned above  is the dominant source
of uncertainty: the  combined C+N abundance contributes to  a 12\% uncertainty
for both the $^{13}$N and the $^{15}$O  fluxes, and the O abundance to 15\% in
the $^{17}$F flux.

In the case  of the non-composition uncertainties, the  situation has improved
in  some  cases  thanks  to  more precise  measurements  of  nuclear  reaction
rates. This is  the case, in particular, for  the ${\rm ^3He(^4He},\gamma){\rm
  ^7Be}$ reaction,  which now  contributes only 4.7\%  and 4.5\% of  the total
uncertainty in the $^7$Be and  $^8$B fluxes respectively.  For comparison, the
analogous   contributions   in  the   BP04   model   were   8.0\%  and   7.5\%
\cite{14-bahcall:2004}. Significant progress  has also been achieved regarding
${\rm  ^{14}N}(p,\gamma){\rm ^{15}O}$, which  now introduces  uncertainties of
only 5\% and 7\%  in the $^{13}$N and $^{15}$O fluxes, half  the amount it did
in 2004.  An important contribution to  the uncertainty in the  $^8$B flux now
comes  from ${\rm ^7Be}(p,\gamma){\rm  ^8B}$ because  the uncertainty  of this
reaction  has  been revised  upwards  \cite{14-adelberger:2011}.  Even if  the
uncertainty   in   this    rate   is   now   smaller   than    in   SFI   (see
Table~\ref{14-tab:nucrates}), it is larger than  that used for the BP04 model,
which was taken considering only one experimental result for this reaction.

While  progress  has been  done  in  some cases,  others  have  not seen  much
development,  particularly  diffusion  and  the delicate  issue  of  radiative
opacities.    In   Table~\ref{14-tab:fluxuncert}   we  give   the   individual
contributions to flux uncertainties for  the most relevant sources. The reader
can compare directly to the situation in 2004 \cite{14-bahcall:2004}.

\begin{table}
%\begin{minipage}{85mm}
\begin{center}
\begin{tabular}{lccccccc}
\hline \hline
 & ${\rm S_{11}}$& ${\rm S_{33}}$& ${\rm S_{34}}$& ${\rm S_{17}}$ 
 & ${\rm S_{1,14}}$ & Opac & Diff \\ 
\hline 
pp & 0.1 & 0.1 & 0.3 & 0.0 & 0.0 & 0.2 & 0.2 \\
pep & 0.2 & 0.2 & 0.5 & 0.0 & 0.0 & 0.7 & 0.2 \\
hep & 0.1 & 2.3 & 0.4 & 0.0 & 0.0 & 1.0 & 0.5 \\
$^7$Be & 1.1 & 2.2 & 4.7 & 0.0 & 0.0 & 3.2 & 1.9 \\
$^8$B & 2.7 & 2.1 & 4.5 & 7.7 & 0.0 & 6.9 & 4.0 \\
$^{13}$N & 2.1 & 0.1 & 0.3 & 0.0 & 5.1 & 3.6 & 4.9 \\
$^{15}$O & 2.9 & 0.1 & 0.2 & 0.0 & 7.2 & 5.2 & 5.7 \\
$^{17}$F & 3.1 & 0.1 & 0.2 & 0.0 & 0.0 & 5.8 & 6.0 \\
\hline
\end{tabular}
\caption{Percentage contribution of selected individual sources of uncertainty
  to the neutrino fluxes. \label{14-tab:fluxuncert}}
%\end{minipage}
\end{center}
\end{table}

\section{Neutrino flavor conversion in vacuum and matter}

Neutrino flavor conversion has been reviewed  by A. Yu. Smirnov in this volume
and we  refer the reader for  a detailed physics discussion  and references to
his article.  Here we just summarize the basic features and formulae of flavor
conversion relevant to solar neutrinos. 

We consider mixing of the three flavor
neutrinos. The  description of flavor conversion of  solar neutrinos traveling
through a  medium is  simplified because a)  the hierarchy in  mass splittings
determined by  solar and atmospheric  data leads to  a reduction of  the three
neutrino flavor conversion to an effective two flavor problem, b) the neutrino
parameters, the  mixings and  solar mass splitting,  lead to  adiabatic flavor
conversion in  solar matter and to  cancel the interference  term by averaging
out. Therefore,  the physics  of the flavor  conversion of solar  neutrinos is
described  by simple  expressions with  very good  accuracy. In  practice, the
survival probability  is computed numerically to correctly  include the number
density of  scatterers along  the trajectory of  neutrinos from  production to
detection  and to  average  over  the neutrino  production  region.  In  solar
neutrino flavor conversion, $\nu_\mu$ and $\nu_\tau$ are indistinguishable and
therefore the survival probability of  electron neutrinos is the only function
needed to describe the flavor composition of the solar neutrino flux.

Solar  neutrino   survival  or   appearance  probabilities  depend   on  three
oscillation  parameters:  the  solar  oscillation  parameters  ($\theta_{12}$,
$\Delta m^2_{21}$), and $\theta_{13}$. The survival probability in the absence
of Earth--matter effects, i.e., during the day, is well described by
\begin{eqnarray}
P_{ee}^{D} =  \cos^4\theta_{13} \left(  \frac{1}{2} + \frac{1}{2} \cdot
\cos2\theta_S \cdot \cos2\theta_{12} \right) +  \sin^4\theta_{13}.
\label{Pday}  
\end{eqnarray}
Here $\theta_S$ is the mixing angle at the production point inside the Sun:  
\begin{eqnarray}
\cos 2\theta_S \equiv \cos2\theta_m(\rho_S) 
\label{eq}
\end{eqnarray}
where $\theta_m(\rho)$ is the mixing angle in matter of density $\rho_{S}$, 
\begin{eqnarray}
\cos2\theta_S  = { \cos2\theta_{12} - \xi_{S} \over( 1 -2\xi_{S} \cos 2\theta_{12} +
\xi_{S}^2 )^{1/2}}.
\label{cos2t12}
\end{eqnarray}
In  (\ref{cos2t12}),  $\xi_{S}$  is  defined  as the  ratio  of  the  neutrino
oscillation length  in vacuum,  $l_\nu$, to the  refraction length  in matter,
$l_0$:
\begin{eqnarray}
\xi_{S} \equiv \frac{l_\nu}{l_0} &=& 
\frac{2 \sqrt{2} G_F \rho_{S} Y_e  \cos^2\theta_{13}}{m_N} 
\frac{E}{\Delta m^2} 
\nonumber \\ 
&=& 
0.203 \times 
\cos^{2} \theta_{13} 
\left(  \frac{E}{ \rm{ 1 MeV} } \right) 
\left(  \frac{ \rho_{S} Y_e }{ \mbox{\rm 100 g\ cm}^{-3}  } \right),
\label{xi-def}
\end{eqnarray}
where 
\begin{eqnarray}
l_\nu\equiv\frac{4\pi E}{\Delta m^2}, 
\hspace{10mm}
l_0\equiv \frac{2\pi m_N}{\sqrt2 G_F \rho_{S} Y_e \cos^2\theta_{13}}.
\label{length}
\end{eqnarray}
In (\ref{xi-def}) and (\ref{length}),  $\rho_{S}$ is the solar matter density,
$Y_{e S}$  is the number  of electrons per  nucleon, and $m_N$ is  the nucleon
mass. The electron  solar density and neutrino production  distribution of the
neutrino  fluxes  are derived  from  solar  models  as discussed  in  previous
section.  In the last line in  (\ref{xi-def}) we have used the best fit values
of the global  analysis $\Delta m^2= 7.5 \times 10^{-5}$  eV$^2$. The ratio of
the parameter  $\rho_{S}$ to $\cos  2\theta_{12}$, separates the  region where
the flavor conversion corresponds to vacuum averaged oscillations from the one
of matter dominated conversion.

The $\nu_{e}$ survival probability at night during which solar 
neutrinos pass through the Earth can be written as 
\begin{eqnarray}
P_{ee}^{N} = 
P_{ee}^{D} - \cos 2\theta_{S} \cos^2 \theta_{13} \langle f_{reg}
\rangle_{\rm{zenith}}
\label{Pnight}
\end{eqnarray}
where $P_{ee}^{D} $  is the one given in  (\ref{Pday}).  $f_{reg}$ denotes the
regeneration  effect  in the  Earth,  and  is given  as  $f_{reg}  = P_{2e}  -
\sin^2\theta_{12}  \cos^2  \theta_{13} $,  where  $P_{2e}$  is the  transition
probability of second mass eigenstate to $\nu_{e}$.
Under the constant density approximation in the Earth, $f_{reg}$ is given by 
\begin{eqnarray} 
f_{reg} = \xi_{E}  \cos^2 \theta_{13} \sin^2 2\theta_{E} 
\sin^2 \left[ a_{E} \cos^2 \theta_{13}  (1-2 \xi_{E}^{-1}\cos^2
\theta_{12}+\xi_{E}^{-2})^{\frac{1} {2} } 
\left( \frac{L}{2} \right) \right] 
\label{freg}
\end{eqnarray}
for passage of distance $L$, where we have introduced 
$a_{E} \equiv  \sqrt{2} G_{F} n_{e}^{Earth} =  \frac{ \sqrt{2} G_{F} \rho_{E} Y_{e
E} } {m_N} $.

In ($\ref{freg}$), $\theta_{E}$  and $\xi_{E}$ stand for the  mixing angle and
the $\xi$ parameter [see (\ref{xi-def})] with matter density $\rho_{E}$ in the
Earth.  Within the range of  neutrino parameters allowed by the solar neutrino
data, the oscillatory  term averages to $\frac{1}{2}$ in  a good approximation
when integrated over zenith angle. Then, the equation simplifies to
\begin{eqnarray}
\langle f_{reg} \rangle_{\rm{zenith}} = 
\frac{1} {2} \cos^2 \theta_{13}  \xi_{E}  \sin^2 2\theta_{E}.  
\label{freg-ave}
\end{eqnarray}
At $E=7$  MeV, which  is a typical  energy for $^8$B  neutrinos, $\xi_{E}=3.98
\times  10^{-2}$  and $\sin  2\theta_{E}  =  0.940$  for the  average  density
$\bar{\rho}_{E} = 5.6  \rm{g/cm}^3$ and the electron fraction  $Y_{e E} = 0.5$
in the  Earth.  Then,  $\langle f_{reg} \rangle_{\rm{zenith}}$  is given  as $
\langle f_{reg} \rangle_{\rm{zenith}} = 1.76  \times 10^{-2}$ for the best fit
neutrino parameters.  This result is in reasonable agreement with the computed
Earth-matter factor using the best estimates on the Earth-matter density.

\section{Recent solar neutrino measurements}

\subsection{The SNO and SK legacy}\

After  the  results  and  analyses  from  2002, it  was  clear  that  the  LMA
oscillation was the right solution  of the long standing solar neutrino puzzle
\cite{14-afterSNO2002},  but  the  activity  of  the SNO  and  SK  experiments
continued in  the following  years. The data  obtained from  these experiments
were very  important in making the  LMA solution more robust  and in improving
the accuracy and precision of the mixing parameters determination.

The so-called SNO II experiment  began in June  of 2001 with the  addition of
2000~kg of NaCl  to the 1000 tons of $D_2 O$ and  ended in October 2003 when
the NaCl was removed. The addition of salt  significantly increased  SNO's
efficiency (by a factor
$\sim$\,3 with respect to the pure $D_2 O$ phase) in the  detection of neutrons 
produced in  the neutral current (NC)
disintegration of  deuterons by solar  neutrinos and, by enhancing the  energy of
the  $\gamma$-ray  coming  from  neutron  capture,  allowed  a  more  precise
measurement of this interaction channel, well above the low-energy radioactive
background. Moreover,  the isotropy of  the multiple $\gamma$-ray  emission by
neutron capture  on $^{35}{\rm Cl}$ is  different from the one  of the \v{C}erenkov
light  emitted by  the single  electron  of the  charged current  interaction;
therefore, by studying the  event isotropy, it has been  possible to separate the
neutral from the  charged current events without any  additional assumption on
the neutrino  energy spectrum.  The salt  phase results have  been reported in
two main  publications. In  \cite{14-SNOII}, referring to  the first  254 live
days, a global  analysis including all the solar  and reactor neutrino results
rejected the maximal mixing hypothesis at a $5.4 \, \sigma$ level and gave a
value of the  $^8$B neutrino flux in agreement  with previous measurements and
with SSMs. These  results were essentially confirmed (even if
with a  small shift towards  larger values of  the mixing angle) by  the second
publication  \cite{14-SNOII_2005}, which included  the full  data of  the salt
phase (391 live days), analyzed  in terms of the  CC spectra (starting from  
$5.5 \, \rm{MeV}$
kinetic  energy)  and NC  and  ES integrated  fluxes  separately  for day  and
night. The day-night asymmetry in the  neutral current rate, which would be an
indication  of oscillation to  sterile neutrinos  or non  standard interaction
with matter in the earth, came out to be consistent with zero.

This  result confirmed also  the outcome  of the  study performed  for elastic
scattering  (ES) interaction  above $5  \, \rm{MeV}$  by  the Super-Kamiokande
collaboration   \cite{14-SK2003SMY}.    The  full   SK-I   low  energy   data,
corresponding to  1496 live days  until July 2001, were  investigated analyzing
the  time variations  of  the ES  rates  and fitting  them  to the  variations
expected from active two neutrino oscillations. The day-night asymmetry turned
out to be $A_{DN} = \frac{2 (D-N)}{D  + N} = -0.021 \pm 0.020 \, {\rm (stat.)}
\, ^{+0.013}_{-0.012} \, {\rm (syst.)}$,  which is consistent with zero within
$0.9  \,  \sigma$.   This value  was  in  good  agreement  also with  the  LMA
oscillation   solution,  which   (for  the   best  fit   parameter)  predicted
\cite{14-SK2003SMY}  $A_{D   N}=  -0.018  \pm  0.016  \,   {\rm  (stat.)}   \,
^{+0.013}_{-0.012}     \,     {\rm      (syst.)}$.     The     SK     analysis
\cite{14-SK2003SMY,14-SKI-2005} also  showed that  the energy spectrum  of the
recoiling electron  was consistent  with an undistorted  solar $^8$B neutrino
spectrum and did not  find any anomalous periodic time  variation of the rates,
apart  from  the  expected  seasonal   variation  due  to  the  Earth's  orbit
eccentricity.  The  SK best fit point was  in quite a good  agreement with the
SNO results, even if SK would  favor slightly larger values of ${\rm tan}^2\theta$.
A      SNO-only     analysis     gave      the     following      best     fit
parameters\cite{14-SNOII_2005}:  $\Delta  m_{12}^2  =  5.0 \times  10^{-5}  \,
\rm{eV}^2$,  $\rm{tan}^2 \theta_{12}  =0.45$.  Including  all the  other solar
neutrino  and  the KamLAND  results  the best  fit  was  obtained for  $\Delta
m_{12}^2  =  8.0^{+  0.6}_{-0.4}  \times 10^{-5}  \,  \rm{eV}^2$,  $\rm{tan}^2
\theta_{12} = 0.452^{+0.088}_{-0.070}$.  The effect of KamLAND data was mainly
to increase the value of $\Delta m^2$  and to restrict the allowed region in the
mixing parameter plane.  The main difference of the  global analysis done with
the SNO salt  phase data with respect to previous  studies was the possibility
to exclude at $95 \% \rm{C.L.}$  the secondary region at even larger values of
the mass differences  (the so called LMA II solution,  with $\Delta m_{12}^2 >
10^{-4} \, \rm{eV}^2$).

In  the third  SNO phase  (November  2004-November 2006)  the neutral  current
signal  neutrons  were  mainly detected  by  means  of  an  array of  $^3  He$
proportional  counters  deployed  in  the  $D_2  O$ and  looking  at  the  gas
ionization  induced by  neutron capture  on $^3$He. In  this way  the fluxes
correlation was reduced and the accuracy in the mixing angle determination was
improved.  The total active $^8$B neutrino flux was found \cite{14-SNOIII} to
be $5.54^{+0.33}_{-0.31} (\rm{stat}) ^{+0.36}_{-0.34} (\rm{syst}) \times 10^6
\, {\rm  cm^{-2}} {\rm s}^{-1}$,  in agreement with previous  measurements and
SSMs. The ratio  of the $^8$B neutrino flux measured with
CC and  NC reaction was  ${\rm \Phi^{SNO}_{CC}/\Phi_{NC}^{SNO}} = 0.301  \pm 0.033$.
The global solar neutrino experiment analysis included, in this case, also the
first results coming from the Borexino experiment \cite{14-Arpesella2008}, that we
discuss in subsection $\ref{Borexino}$.  The best fit point moved
to $\Delta m_{12}^2 =  4.90 \times 10^{-5} \rm{eV}^2$, $\rm{tan}^2 \theta_{12}
= 0.437$  and the uncertainty  in the mixing  parameter plane was  still quite
large.  Adding the KamLAND  data,  the allowed  region was  significantly
restricted  (mainly for  $\Delta  m^2$)  and the  marginalized  $1 \,  \sigma$
regions   were  $\Delta   m_{12}^2  =   7.59^{+0.19}_{-0.21}   \times  10^{-5}
\rm{eV}^2$, $\rm{tan}^2 \theta_{12} = 0.469_{-0.041}^{+0.047}$.

A subsequent  joint reanalysis of SNO  I and SNO  II data, known as  LETA (Low
Energy   Threshold   Analysis)   \cite{14-LETA},  succeeded,   with   improved
calibration  and analysis  techniques, in  lowering the energy threshold, with  
respect  to previous analyses (\cite{14-SNOphaseI_analisi2007,14-SNOII_2005}),  
down  to an  effective  electron kinetic  energy  of $T_{\rm eff}  =  3.5 \,  {\rm
  MeV}$. The main effect was to increase the statistics of CC and ES and, above
all, of NC events, and to increase significantly the precision on both the total
$^8$B  neutrino flux and  the neutrino mixing  parameters. The value  for the
total $^8$B neutrino  flux extracted from  neutral current was  $\Phi_{NC} =
5.14^{+0.21}_{-0.20}  \times 10^6 \,  {\rm cm^{-2}}  {\rm s}^{-1}$,  where the
error,   obtained  by summing   in  quadrature   the  statistic   and  systematic
contributions,  was reduced  by more  than  a factor  of two  with respect  to
previous publications.  For SNO data  alone (LETA plus  SNO III) the  best fit
point moved to  the LOW region of parameter space,  but the significance level
was very similar to the one of the usual LMA solution. A global fit, including
all  the solar  and  the  KamLAND data,  essentially  confirmed, instead,  the
previous results \cite{14-SNOIII} for $\Delta m_{12}^2$ and it made possible a
further  improvement  in  the  angle  determination, giving,  in  a  2  flavor
analysis, $\rm{tan}^2 \theta_{12} = 0.457_{-0.028}^{+0.041}$.
 
In the last  five years also the Super-Kamiokande  collaboration presented new
analyses,  including the  data  of the  different working phases of  this
experiment: Super-Kamiokande II \cite{14-SKII}  (from December 2002 to October
2005)   and   Super-Kamiokande  III   (from   July   2006   to  August   2008)
\cite{14-SKIII}.  Due  to  the  2001  accident,  which  damaged  some  of  the
photomultiplier tubes,  the detector sensitivity  was reduced with  respect to
SK-I and  therefore it was important  to improve the methods  adopted for data
collection (particularly  for vertex event  reconstruction, angular resolution
and background  reduction) and analysis. In  this way, during the  548 days of
SK-III  a $2.1  \%$ systematic  uncertainty on  the total  flux (corresponding
roughly to  two thirds of  the SK-I value)  was reached. The second  and third
Super-Kamiokande  phases  essentially confirmed  the  SK-I  results, for  what
concerns  the absence  of significant  spectral distortion,  the total  $^8$B
measured flux and the day-night asymmetry.

Since  September  2008,  Super-Kamiokande  is  running  with  modernized  data
acquisition system (DAQ) and electronics, which allow a wider dynamic range in
the measured charge and is read out via Ethernet. This phase of the experiment
is denoted as  Super-Kamiokande-IV \cite{14-SuperKamiokandeIV}.  Thanks to the
fast DAQ every  hit can be recorded and the resulting  data stream analyzed by
an online  computer system that finds  timing coincidences which  are saved as
triggers. As  a consequence, Super-Kamiokande's  energy threshold is  now only
limited by  computing speed  and the event  reconstruction. The  present event
reconstruction is able  to reconstruct electrons with a total  energy of 3 MeV
or more. The computing speed limits  the energy threshold to ~4.2 MeV which is
just  below the threshold of Super-Kamiokande-I and  III (4.5 MeV).  The
same water flow techniques  developed during Super-Kamiokande-III result in an
observed solar  neutrino elastic scattering peak  between 4 and  4.5 MeV total
recoil electron  energy. Special techniques are developed  to discriminate the
signal from the  background, taking advantage from the  fact that the background is
mainly due to  $\beta$ emission from $^{214}$Bi and it  is characterized by a
larger Coulomb multiple scattering.  This  makes possible a reduction of about
$10-15 \%$ of the statistical uncertainty  and this method can also be applied
to previous phases of the experiment. The additional systematic uncertainty of
this method is under investigation.

\subsection{The impact of KamLAND results on solar neutrino physics}\

Even if it  is based on the analysis of a reactor  antineutrino beam, the KamLAND
experiment  played a fundamental  role in  the solution  of the  long standing
solar neutrino puzzle.  In fact, the first KamLAND  data \cite{14-firstKL} were
determinant, in conjunction with  the previous solar neutrino experiments (and
mainly with  SNO) and assuming  CPT invariance, to  prove the validity  of the
oscillation hypothesis and to select the LMA solution as the correct one.

\begin{figure}[h!!]
\begin{center}
%\vspace{}
\includegraphics[width=8cm,height=5.7cm,angle=0]{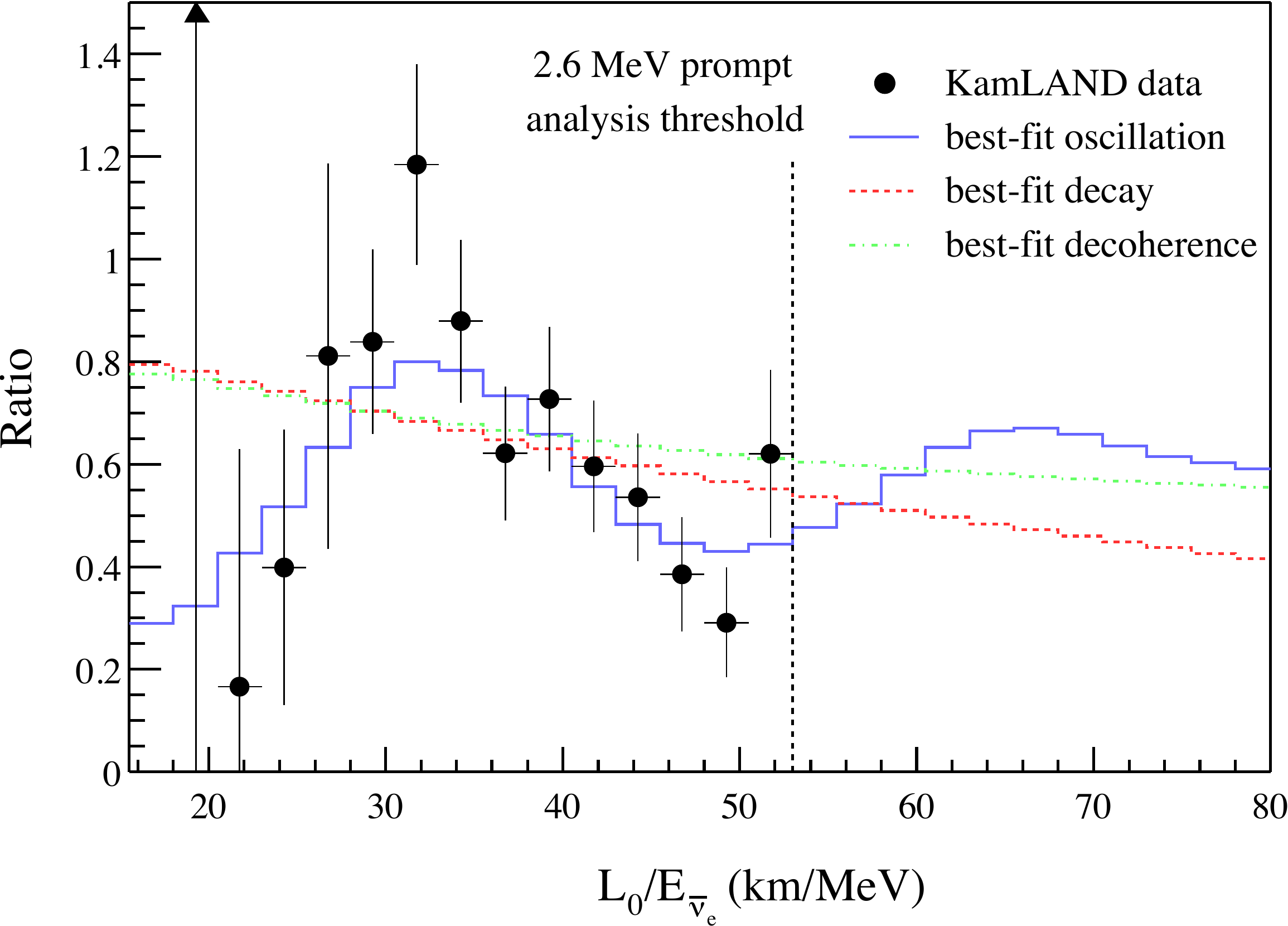}
\caption{Ratio of  the observed $\bar{\nu}_e$ spectrum to  the expectation for
  no-oscillation  versus L$_{0}$/E. The  curves show  the expectation  for the
  best-fit oscillation, best-fit decay and best-fit decoherence models, taking
  into account  the individual time-dependent flux variations  of all reactors
  and detector effects.
%The data points and models are plotted with L$_{0}$=180\,km, as if all anti-neutrinos
%detected in KamLAND were due to a single reactor at this distance.
Taken from \cite{14-KL2004}.
\label{fig-14:1}}
\end{center}
\end{figure}

Between  March 2002 and  January 2004  a new  set of  data were  collected and
the KamLAND collaboration  performed a study  including also a re-analysis  of the
previous  data. During the  2002-2004 campaign  important upgrades  were done
both  on  the  central  detector  (increasing  the  photocatode  coverage  and
improving the energy resolution) and  in the analysis techniques (reduction of
the background with better techniques in the event selection cuts based on the
time,  position  and  geometry  of  the  events).   The  number  of
antineutrino  events  above  $2.6  \,   {\rm  MeV}$  expected  in  absence  of
antineutrino  disappearance  was $365.2  \pm  23.7  {\rm(syst)}$  and the  258
observed events  corresponded to a $\bar{\nu_e}$  survival probability equal
to $0.658 \pm 0.044 ({\rm stat}) \pm 0.047 ({\rm syst})$.  The energy spectrum
analysis was  in disagreement  with the no  oscilation hypothesis  at 99.6$\%$
statistical  significance.   In  \cite{14-KL2004} the  KamLAND  collaboration,
looking at the $L_0/E$ spectrum dependence (where $L_0$ is the source-detector
distance and $E$  the $\bar{\nu}_e$ energy), performed also  an interesting study of
other  alternative  hypotheses  (like  decoherence  and  decay)  for  neutrino
disappearance.  The oscillation hypothesis offered by far the best explanation
of the spectrum shape, as one can see from Fig.(\ref{fig-14:1}).

\begin{figure}[h!t]
\begin{center}
\vspace{-10.5truecm}
\includegraphics[width=14cm,height=18cm,angle=0]{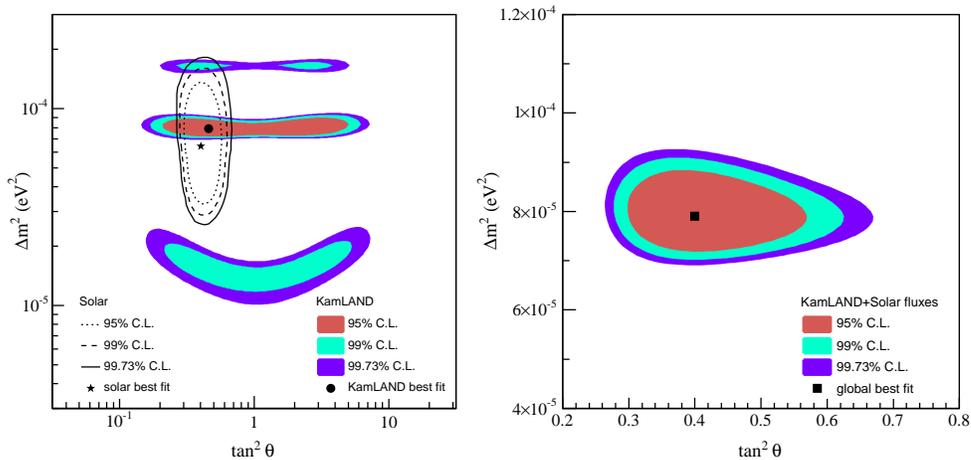}
\caption{(A)  Allowed region of the neutrino  oscillation  parameter from  KamLAND
  anti-neutrino data  (colored regions) and solar  neutrino experiments (lines)
  \cite{14-SNOII}.  (B) Result of a combined two-neutrino oscillation analysis
  of KamLAND  and the observed solar  neutrino fluxes under  the assumption of
  CPT invariance.  Taken from \cite{14-KL2004}.
\label{fig-14:2}
}
\end{center}
\end{figure}

As  shown  in Fig.(\ref{fig-14:2}A),  the  best  fit  obtained from  the  data
analysis was  in the so-called  LMAI region (with values  of $\Delta m_{12}^2$
around $8 \cdot 10^{-5} \, {\rm eV}^2$) and the alternative solution at higher
$\Delta  m_{12}^2$  (around  $2  \cdot  10^{-4} \,  \rm{eV}^2$)  was  strongly
disfavoured,  at $98 \%$  C.L., mainly  due to  the spectrum  distorsions. The
KamLAND data  alone were not sufficient  to solve completely  the ambiguity on
the mixing angle  values and to exclude maximal  mixing. However, including in
the  analysis also  the results  coming from  solar neutrino  experiments, the
allowed   values   of   the   angle   were   significantly   restricted   (see
Fig.($\ref{fig-14:2}$B)) and the two flavor combined  analysis gave 
$\Delta m_{12}^2 = 7.9^{+0.6}_{-0.5}
\cdot    10^{-5}   \,    \rm{eV}^2$   ,    \,   $\rm{tan}^2    \theta_{12}   =
0.40^{+0.10}_{-0.07}$ at a  $1 \, \sigma$ level.
%and 
%$\Delta m_{12}^2$ (around $2 \cdot 10^{-4} \, \rm{eV}^2$) was strongly %disfavoured, at $98 \%$ C.L. . 

The next KamLAND analysis \cite{14-KLfollowing2} included also, in addition to the
one  of \cite{14-firstKL,14-KL2004},  the  new data  collected up  to May
2007.  The increase in  data collection  was significant  (also thanks  to the
enlarging the radius of the fiducial volume from 5.5 to 6 m) and there was a reduction
of  systematic  uncertainties,  in  the  number  of  target  protons  and  the
background.  The total  uncertainty on  $\Delta m_{21}^2$  was around  $2 \%$,
mainly due  to the distortion of the  energy scale in the  detector. The total
uncertainty, $4.1 \%$, on the expected event rate was due to different sources
(above  all  the  definition  of  the  detector  fiducial  volume  and  energy
threshold, the  $\bar{\nu}_e$ spectra and  the reactor power) and  it affected
primarily the mixing angle determination.
%, as summarized in table \ref{KL2008-systematic} (taken from \cite{KLfollowing2}), and it affected 
%primarily the mixing angle determination.
%\begin{table}[h]
%\caption{\tiny Estimated systematic uncertainties relevant for the neutrino
%oscillation parameters $\Delta m_{12}^2$ and $\theta_{12}$.}
%\label{KL2008-systematic}
%\begin{tabular*}{12.0cm}{@{\extracolsep{\fill}}c|llll}
%\hline\hline
% & Detector-related (\%) &  & Reactor-related (\%) &  \\ \hline
% $\Delta m_{12}^2$  & Energy scale & 1.9 & $\bar{\nu}_e$-spectra~\cite{Achkar} & 0.6 \\ \hline
% \multirow{4}{*}{Event rate} & Fiducial volume & 1.8 & $\bar{\nu}_e$-spectra & 2.4 \\
%  & Energy threshold & 1.5 & Reactor power & 2.1 \\
%  & Efficiency & 0.6 & Fuel composition & 1.0 \\
%  & Cross section & 0.2 & Long-lived nuclei & 0.3 \\
% \hline\hline
% \end{tabular*}
% \end{table}
The different  background sources were  studied and reduced further. The most
important one  was the  ${\rm ^{13} C (\alpha,n)  ^{16}O}$ reaction, made possible 
by   the $\alpha$  decay  of  $^{210}$Po (a daughter  of  $^{222}$Rn)
introduced  in the liquid  scintillator during  the construction, which produces
neutrons with energies up to $7.3 \, {\rm MeV}$.

\begin{figure}[h!!]
\begin{center}
\vspace{-2.5cm}
\includegraphics[width=11cm,height=12cm,angle=0]{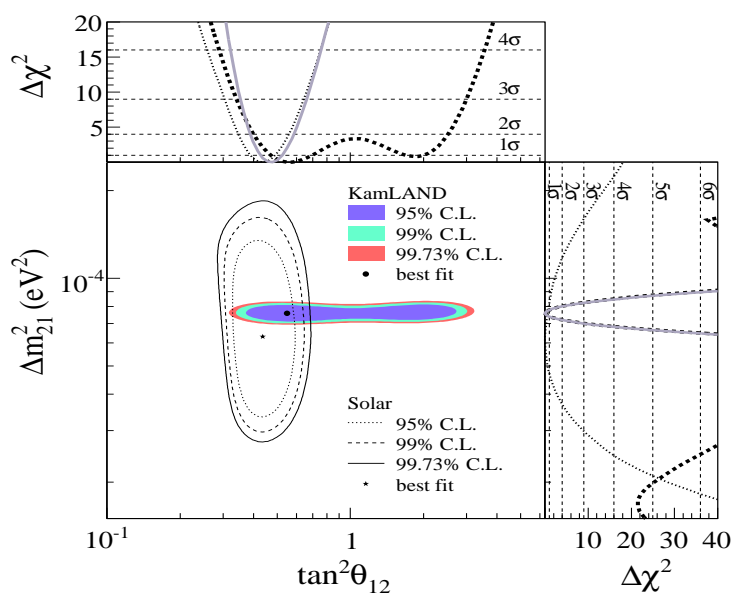}
\caption{ Allowed region for  neutrino oscillation parameters from KamLAND and
  solar   neutrino    experiments.   The   side-panels    show   the   $\Delta
  \chi^{2}$-profiles  for  KamLAND  (dashed)  and solar  experiments  (dotted)
  individually, as  well as  the combination of  the two (solid).   Taken from
  \cite{14-KLfollowing2}.
\label{fig-14:3}}
\end{center}
\end{figure}

The results of the  statistical analysis are reported in Fig.(\ref{fig-14:3}),
taken  from \cite{14-KLfollowing2}. The  allowed oscillation parameter  values
were  $\Delta  m_{21}^2  = 7.58^{+0.14}_{-0.13}  (\rm{stat})  ^{+0.15}_{-0.15}
(\rm{syst})  \cdot 10^{-5}  \, \rm{eV}^2$ for the mass eigenvalues and  
$ {\rm  tan}^2  \theta_{12} =0.56^{+0.10}_{-0.07}  ({\rm stat}) ^{+0.10}_{-0.06}  ({\rm syst})$,  
for ${\rm  tan}^2 \theta_{12} \, < \, 1$ and the no oscillation hypothesis was excluded
at $5 \, \sigma$. The extension to the three neutrino oscillation analysis had
the main effect  to enlarge the uncertainty on  $\theta_{12}$, leaving $\Delta
m_{12}^2$  substantially unchanged.  Figure (\ref{fig-14:3}),  taken from
\cite{14-KLfollowing2}, shows that the effect of the inclusion in the analysis
of  the  data  from  SNO  \cite{14-SNOII_2005}  and  previous  solar  neutrino
experiments was  essentially to reduce the interval  of allowed $\theta_{12}$
values and also to move the best fit point towards slightly lower values of the
mixing angle.

\begin{figure}[h!!]
\begin{center}
\vspace{-1cm}
\includegraphics[width=9.5cm,height=8.0cm,angle=0]{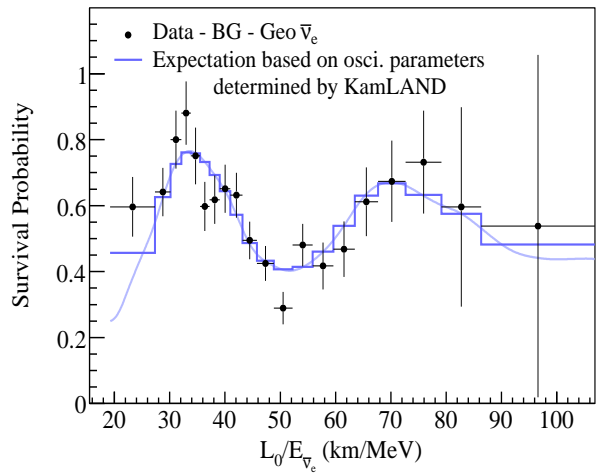}
\vspace{-0.5truecm}
\caption{  Ratio  of the  background  and geo-neutrino-subtracted  $\bar\nu_e$
  spectrum   to  the  expectation   for  no-oscillation   as  a   function  of
  $L_{0}/E$.  $L_{0}$  is the  effective  baseline  taken  as a  flux-weighted
  average ($L_{0}$\,=\,180\,km). The energy bins are equal probability bins of
  the best-fit  including all backgrounds.   The histogram and curve  show the
  expectation  accounting  for  the  distances  to  the  individual  reactors,
  time-dependent  flux  variations  and   efficiencies.  The  error  bars  are
  statistical  only and  do not  include, for  example,  correlated systematic
  uncertainties in the energy scale.  Taken from \cite{14-KLfollowing2}.
\label{fig-14:4}}
\end{center}
\end{figure}

Figure (\ref{fig-14:4})  (taken from \cite{14-KLfollowing2}) illustrates,
instead, the  $\bar{\nu}_e$ survival probability,  as a function of  the ratio
$L_0/E$ between  the average baseline  and the antineutrino energies.  One can
notice  that  the  observed  spectrum  (after subtraction  of  background  and
geo-neutrino signals), reproduces correctly  the general shape of the expected
oscillation  cycle, with  a slight  excess of  low energy  antineutrinos, that
could be interpeted as geo-neutrinos.

\subsection{Toward the sub-MeV analysis: the Borexino detector and its measurements}\label{Borexino}\

In the last  decade significant steps forward have been  done in the knowledge
of solar  neutrino properties,  thanks mainly to  the results obtained  by the
kiloton scale \v{C}erenkov detectors (SK and SNO) and by advent of the reactor
neutrino experiment KamLAND. However,  these experiments investigated only the
energy part of solar neutrino  spectrum above 5~MeV, which represents
a small  fraction of  the full spectrum.  The single components  of the neutrino
spectrum  cannot  be  determined  by  such techniques  at  low  energies  and,
therefore, up to  the last four years, low energy  neutrinos had been observed
only  via radiochemical  methods. A  significant  change took  place with  the
advent  of Borexino,  a real  time  experiment which  opened the  way to  the
investigation  of the  sub-MeV  region and  isolated  for the  first time  the
neutrinos corresponding to the monochromatic berillium line.

\subsubsection{ The Borexino detector}
\label{Borexinodetector}

\begin{figure}[h!!]
\begin{center}
\vspace{-4truecm}
\includegraphics[width=10.0cm,height=10.5cm,angle=0]{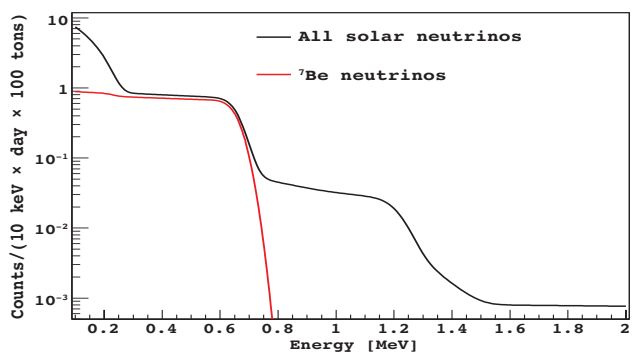}
\caption{Neutrino spectra expected in  Borexino (accounting for the detector's
  energy resolution).  The upper line  represents the neutrino signal  rate in
  Borexino  according to  the most  recent predictions  of the  Standard Solar
  Model~\cite{14-carlos}
% Update
% update the reference to SSM predictions or comment
including neutrino  oscillations with the  LMA-MSW parameters. The  lower line
illustrates the contribution due to ${^7}$Be~neutrinos. The {\it pp} neutrinos
contribute to  the spectrum below  0.3~MeV and the  edge at 1.2~MeV is  due to
{\it pep} neutrinos (from \cite{14-bxfirstresults}).
\label{fig-14:5}}
\end{center}
\end{figure}

Borexino is an ultra-high  radiopure large volume liquid scintillator detector
(using   pseudocumene  -PC-\footnote{1,2,4-trimethylbenzene}   as  aromatic
scintillation solvent,  and PPO\footnote{2,5-diphenyloxazole} as  solute at a
concentration  of 1.5~g/l)  located  underground at  the  italian Gran  Sasso
National  Laboratories  (LNGS),  under  about   1400  m  of  rock  (3800  mwe)
\cite{14-Bor09}.  The  employment of  a  liquid  scintillator  as target  mass
assures a  light production sufficient  to observe low energy  neutrino events
via  elastic scattering  by  electrons.   This reaction  is  sensitive to  all
neutrino  flavors, through  the  neutral current  interaction,  but the  cross
section for  $\nu_e$ is larger  than $\nu_\mu$ and  $\nu_\tau$ by a  factor of
5-6, due  the combination of charged  and neutral currents.  The  main goal of
Borexino is the measurement of the mono-energetic ($0.862 \, {\rm MeV}$) $^7$Be neutrinos, which have the basic  signature of the Compton-like edge of the
recoil electrons at 665~keV (see Fig.~\ref{fig-14:5}).

The high light yield typical of  a liquid scintillator makes it possible to reach
a low energy threshold, a good energy resolution of about $5\%$ at 1 MeV and a
pulse shape discrimination between $\alpha$ and $\beta$ decays. On the other hand, 
no directionality  is possible  and it  is also not  possible to  distinguish 
neutrino
scattered  electrons from  electrons due  to natural  radioactivity.  For this
reason, an extremely low level  of radioactive contamination is compulsory and
this  has been  one of  the main  tasks and  technological achievements  of the
experiment.  The background  due to the presence of  $\beta$~decay of $^{14}$C
($\beta_{end-point}$ 156~keV), intrinsic  to the scintillator, limits neutrino
observation  to  energies  above  200~keV.  Techniques  for  the  scintillator
purification  are based  mainly on  methods  developed and  tested in  earlier
studies with the  Counting Test Facility (CTF), a  4-ton prototype of Borexino
which demonstrated  for the  first time the  feasibility of achieving  the low
backgrounds   needed   to   detect   solar   neutrinos  in   a   large   scale
scintillator~\cite{14-ctf1,14-ctf2,14-ctf3}.     For    Borexino,   a    larger
purification plant was  developed similar to the CTF  system, but with several
improved  features including  the use  of high  vacuum and  precision cleaning
techniques.

\begin{figure}[h]
\begin{center}
\begin{minipage}{0.8\textwidth}
\centering{\includegraphics[width=0.6\textwidth]{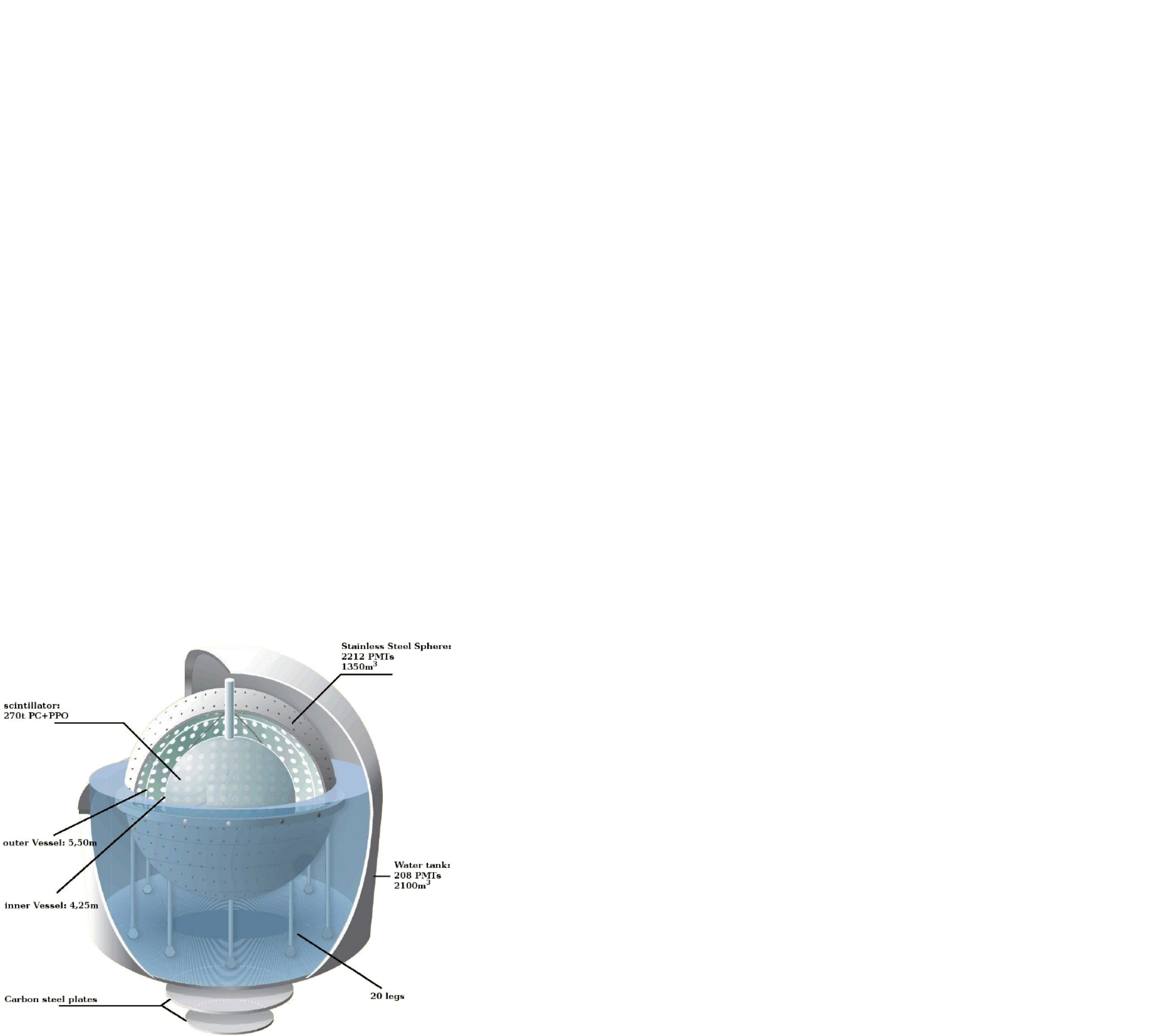}}
\caption{Schematic view of the Borexino detector.}\label{fig-14:Borexino_detector}
\end{minipage}
\end{center}
\end{figure}

The  design  of  Borexino  is  based  on the  principle  of  graded  shielding
(onion-like structure - see Fig.~\ref{fig-14:Borexino_detector}).  

% Update: I think we can cut this figure

The  scintillator ($\approx$~300~tons)  is  contained in  a  thin nylon  Inner
Vessel (IV), of radius 4.25~m, at the  center of a set of concentric shells of
increasing radiopurity and it is surrounded by an outer vessel (OV),
%with radius 5.50~m, which is 
filled with PC and 5.0~g/l DMP \footnote{dimethylphthalate}, a material 
% Update: check this sentence
which is able to quench the residual scintillation of PC and acts as a passive
shield against radon and  other background contaminations originating from the
external parts.  A third more external vessel is composed of a stainless steel
sphere (SSS),
%with radius 6.85~m, 
enclosing the passive shield (PC-DMP), and the entire detector is contained in
a  dome-shape  structure  16.9~m  high  with  a radius  of  9~m,  filled  with
ultra-pure water, denominated Water Tank (WT).
% Update
% This part about the light collection and the veto systems should be reduced
% In all the part above, we may probably cut the information about the 
%dimensions
The  scintillation  light  is   recorded  by  2212  8-inches  photomultipliers
distributed  on the inner  part of  the SSS~\cite{14-pmts1,14-pmts2};  1828 of
them are equipped  with aluminum light concentrators designed  to increase the
light collection efficiency~\cite{14-cones}.   \v{C}erenkov light and residual
background  scintillation in  the  buffer  are thus  reduced.  The others  384
photomultipliers without  concentrators are used to study  this background and
to identify muons  that cross the buffer and not the  Inner Vessel.  The Water
Tank is equipped with~208 8-inches photomultipliers and acts as a \v{C}erenkov
muon detector.
% Update: Probably the reference to the figure is not needed here
% (see figure.~\ref{fig:detector})
Although   the  muon  flux   is  reduced   by  six   order  of   magnitude  by
the~3800~m.w.e.  depth of  the  Gran Sasso  Laboratory,  is still  significant
(1.1~$\mu$~m$^{-2}$~h$^{-1}$).  An  additional  reduction,  of  the  order  of
about~$10^{4}$,    has    been     necessary;    for    more    details    see
Ref.~\cite{14-bxdetector}.

In order to remove contaminants from dust (U, Th, K), air ($^{39}$Ar,
$^{85}$Kr)  and cosmogenically  produced isotopes  (${^7}$Be), different
purification techniques were applied,  such as distillation, water extraction,
nitrogen stripping  and ultra-fine filtration. The  pseudocumene was distilled
in-line during the  detector filling at 80~mbar and at  a temperature of about
90--95\,$^{\circ}$C. Distilled pseudocumene was  stripped in a 8~m-high (15~cm
in diameter)  packed column with  specially prepared ultra-low  Ar/Kr nitrogen
(0.005~ppm   Ar   and  0.06~ppt   Kr,   see  Ref.~\cite{14-lakn}).    Position
reconstruction  of the events,  as obtained  from the  photomultipliers timing
data via  a time-of-flight algorithm,  allowed to define a  fiducial spherical
volume,  corresponding approximately  to  1/3  (i.e. about  100  tons) of  the
scintillator  volume  in  order  to reject  external~$\gamma$~background.  The
others 2/3 of the scintillator act as an active shield.

\subsubsection{ The measurement of the ${^7}$Be line}\

The Borexino collaboration  started taking data in May~2007  and after only 3
months (47.4~live  days) it was able  to extract the ${^7}$Be  signal from the
background.   The best value  estimate for  the rate  was $47  \pm 7  \, ({\rm
  stat})  \pm  12  \,  ({\rm syst})$~counts/(day~$\cdot$~100~ton),  where  the
systematic   error  is  mainly   due  to   the  fiducial   mass  determination
\cite{14-bxfirstresults}.  An update of the ${^7}$Be signal was reported after
9 months from an  analysis of 192 live days (from May  $16^{th}$ 2007 to April
$12^{th}$ 2008), corresponding to 41.3~ton$\cdot$yr fiducial exposure to solar
neutrinos.

The severe cuts  that had to be passed  by the events in order  to be selected
and  enter the  analysis were  mainly designed  to avoid  pile up  of multiple
events, reject the events originated by muons and their daughters and the ones
due  to radon  daughters preceding  the  $\alpha-\beta$ Bi-Po  delayed
coincidences.  Moreover, severe cuts  (radial and  based on  the z-coordinates)
were  finalized to  reduce the  external $\gamma$  background.   The remaining
fiducial mass  was of 78.5 tons.   Important background sources  were the fast
coincidence decays from the $^{238}$U chain (contamination level of $(1.6 \pm
0.1) \, 10^{-17}$ g/g) and the $^{232}$Th chain (contamination level of $(6.8
\pm 1.5) \,  10^{-18}$ g/g) and the $^{85}$Kr  contained in the scintillator that produces the rare decay sequence ${\rm ^{85}Kr} \to \, {\rm ^{85 m}Rb}  \, + \,
e^+ \, + \nu_e \, , \, {\rm ^{85 m}Rb} \to \, {\rm ^{85}Rb} + \gamma$.
% Update : Forse meglio togliere la tabella su errori sistematici 
%
%\begin{table}[b!]
%\caption{Estimated Systematic Uncertainties.}
%\begin{tabular}{ll}
%\hline
%Total Scintillator Mass      & 0.2 \%   \\
%Live Time		   	        & 0.1 \%   \\
%Efficiency of Cuts		& 0.3 \%   \\
%Fiducial Mass Ratio		& 6.0 \%  \\
%Detector Resp. Function	& 6.0 \%	\\
%\hline
%Total Systematic Error	& 8.5 \% \\
%\hline\hline
%\end{tabular}
%\label{tab:systerr}
%\end{table}
The  total estimated  systematic error  was 8.5\% \cite{14-Arpesella2008},
mainly determined by two sources, introducing an uncertainty of 6\% each:
the  total uncertainty  on  the fiducial  mass  and the  one  on the  response
function.  The best value for the  interaction rate of the 0.862~MeV $^7$Be
solar   neutrinos   was   $49  \,   \pm   3   ({\rm   stat})  \pm   4   ({\rm
  syst})$~counts/(day$\cdot$100~ton).   This  result excludes  at  the $4  \,
\sigma$ C.L. the  no oscillation hypothesis for $^7$Be solar neutrinos, which
in the  high metallicity SSM \cite{14-SSM06,14-carlos} would
imply  $74  \, \pm  4$  counts/(day$\cdot$100~ton).  The  Borexino result  is,
instead, in very  good agreement with the predictions  of the LMA oscillation
solution: $48 \, \pm 4$~counts/(day$\cdot$100~ton).\\
%Update
%{\bf  Insert results  on the  neutrino  magnetic moment  and eventually  other
%  results on the global situation.  }

In order to reduce the systematic uncertainties and to tune the reconstruction
algorithm and Monte Carlo simulations, a calibration campaign was performed in
2009  introducing  inside  the   Borexino  detector  several  internal  radiosources
$\alpha$'s, $\beta$'s, $\gamma$'s, and  neutrons, at different energies and in
hundreds of different  positions, which were determined with  a precision better
than 2~cm.   The previous systematic error on $^7$Be  solar neutrino flux was
estimated to  be \cite{14-Arpesella2008} at the  level of $6 \%$  for both the
fiducial  volume and  the  energy  scale.  In  the  calibration campaign,  the
detector  energy response  was studied  with  eight $\gamma$  sources and  Am-Be
neutron source\footnote{When thermal neutrons are captured by protons a 2.2~MeV
  $\gamma$-ray is  generated.} and  comparing the  calibration data  and Monte
Carlo  simulations at different  energies within  the solar  neutrinos energy
region.
%n figure (\ref{fig-14:6}) the comparison between calibration data 
%nd Monte Carlo at different energies within the energy region in solar 
%eutrino analysis are shown. 
The energy scale  uncertainty, obtained with these studies,  was determined to
be less than $1.5\%$.
%\vspace{-0.25truecm}
%
%begin{figure}[h!!]
%begin{center}
%vspace{-8.0truecm}
%includegraphics[width=14.0cm,height=16cm,angle=0]{campaign_sources.pdf}
%caption{Total photoelectron distribution of various gamma sources: comparison between calibration data 
%nd MonteCarlo.
%label{fig-14:6}}
%end{center}
%end{figure}

The inaccuracy of the position (reduced  by means of studies with $\alpha$ and
$\beta$ events) was  less than 3 cm, equivalent to a  systematic error of $1.3
\%$ for the overall fiducial volume in the $^7$Be solar neutrino energy region.  
The
analyzed data  set run from May 2007  to May 2010, with  a fiducial exposure
equivalent  to 153.6 ton$\cdot$year.  In order  to extract  the $^7$Be solar
neutrino  signal, the  spectral fit  was  applied assuming  all the  intrinsic
background components such as  $^{85}$Kr, $^{210}$Bi, $^{14}$C, $^{11}$C.  The
$^{7}$Be solar  neutrino rate was  evaluated to be  ${\rm 46.0 \pm 1.5  (stat) \pm
1.3(syst)}$~counts/day$\cdot$100~ton   \cite{14-Bellini2011rx}.    Thanks  to   the
calibration  campaign, the  systematic error  was reduced  to $2.7\%$  and the
total uncertainty to $4.3\%$.

\begin{figure}[h!]
%\begin{center}
\vspace{-0.5truecm}
\hspace{-3.0cm}
\includegraphics[width=11.0cm,height=11.5cm,angle=0]{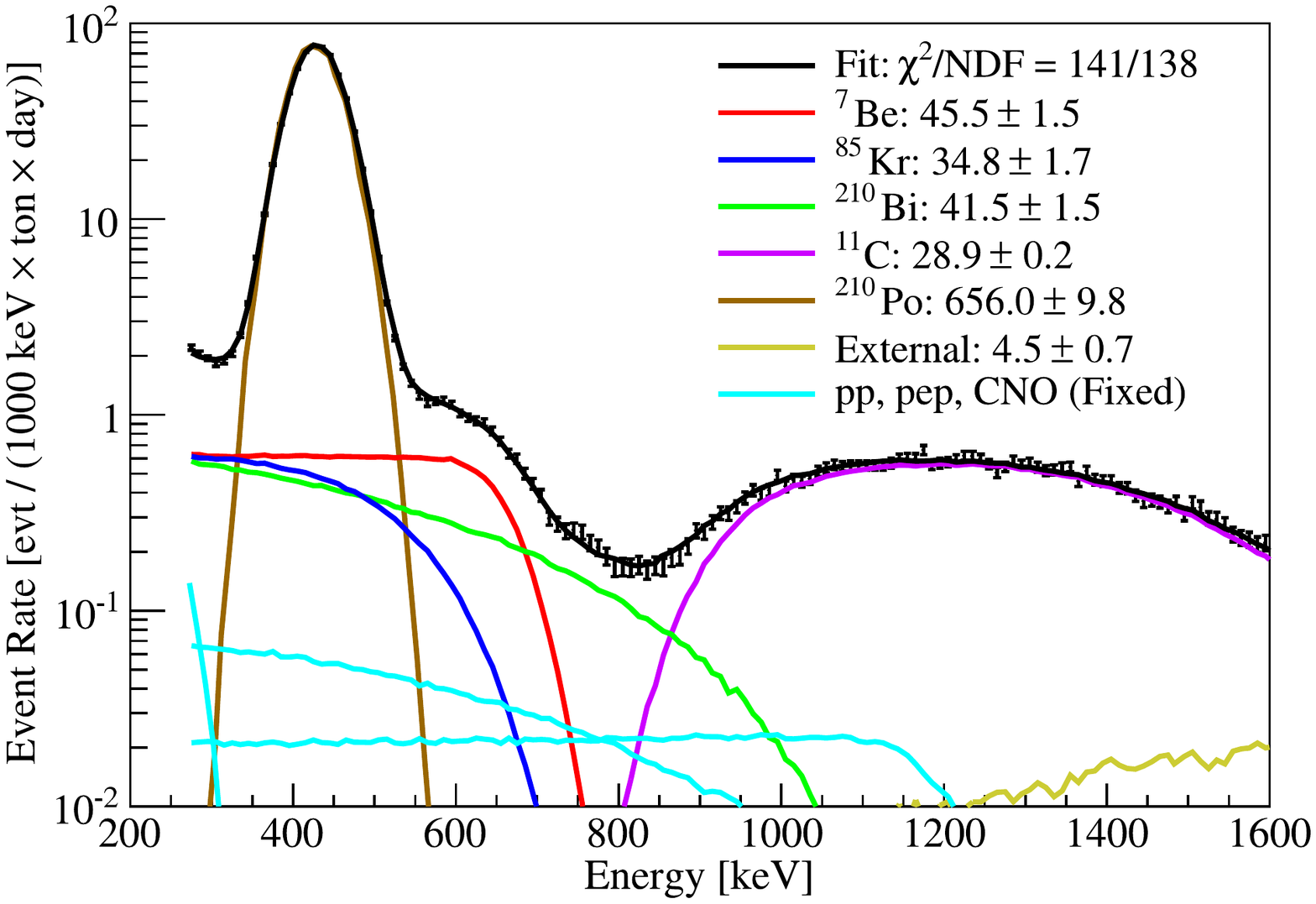}
\hspace{-2.0cm}
\includegraphics[width=11.0cm,height=11.5cm,angle=0]{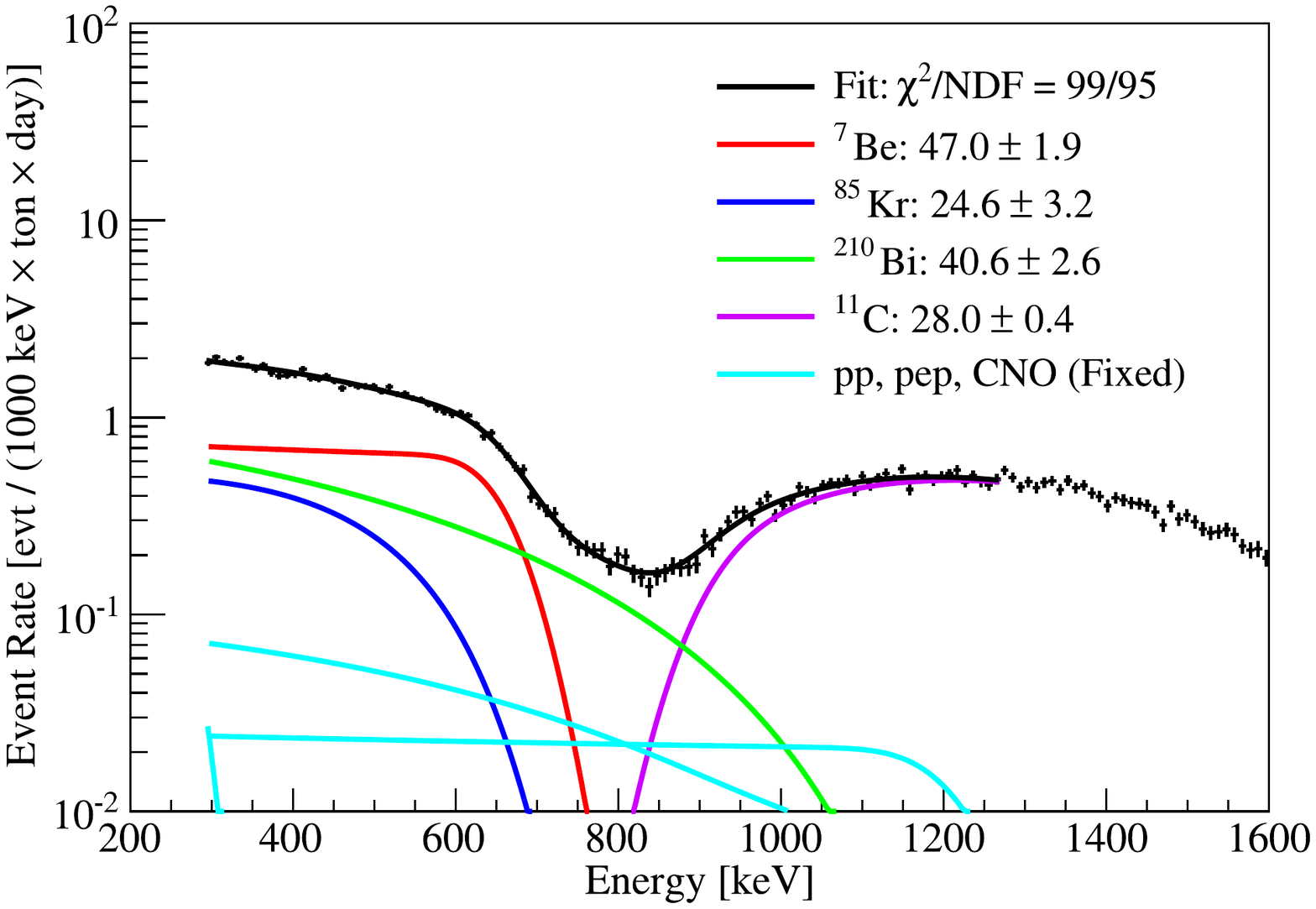}
\vspace{-3.5truecm}
\caption{  Examples of fitted  spectra; the  fit results  in the  legends have
  units  [counts/(day$\cdot$100\,ton)]. Left  panel: A  Monte Carlo  based fit
  over the energy region 270--1600~keV to  a spectrum from which some, but not
  all, of the $\alpha$ events have been  removed using a PSA cut, and in which
  the event  energies were estimated using  the number of  photons detected by
  the PMT array.   Right panel: An analytic fit  over the 290--1270~keV energy
  region to a  spectrum obtained with statistical $\alpha$  subtraction and in
  which the event energies were  estimated using the total charge collected by
  the PMT array.  In all cases the  fitted event rates refer to the total rate
  of  each   species,  independently  from   the  fit  energy   window  (from
  \cite{14-Bellini2011rx}).
\label{fig-14:6}}
%\end{center}
\end{figure}

%\vspace{2truecm}

\subsection{The {\mbox{\it pep}} and  {\mbox{CNO}} neutrinos measurement in 
Borexino}\ 

In the SSM, due to the solar luminosity constraint and their intimate link to 
the {\it pp} neutrinos \cite{14-bahcall:1989,14-adelberger:1998}, the
mono-energetic  1.44\,MeV \mbox{\it pep}  neutrinos have  one of  the smallest
uncertainties (1.2\%) \cite{14-serenelli:2011}.  For this reason, after 
the \mbox{\it  pp} neutrinos, they  constitute the ideal probe to  test SSM
 hypotheses.  On
the  other hand, the  detection of  neutrinos within  the CNO-bicycle is
central to probe    the  solar core  metallicity and contribute in this way to the 
solution of the solar metallicity  problem~\cite{14-metallicity, 14-serenelli:2011}. Also, they are believed  to fuel massive  stars with mass
greater  than  $\sim 1.2\, {\rm M_{\odot}}$ during main sequence evolution and also 
stars with lower masses in more advanced stages of evolution. The  energy  spectrum  
of  neutrinos from  the CNO-bicycle is the  result of  three continuous  spectra with  
end point
energies   of  1.19\,MeV  ($^{13}$N),   1.73\,MeV  ($^{15}$O)   and  1.74\,MeV
($^{17}$F).   Despite  their relevance,  until  2011,  no  \mbox{\it pep}  and
\mbox{CNO} neutrinos had been detected directly.
%Update: check eventual superposition with the part written by Carlos and Aldo

\begin{figure}[h!!]
\begin{center}
\hspace{-1truecm}
%\vspace{1cm}
\includegraphics[width=12.0cm,height=10.5cm,angle=0]{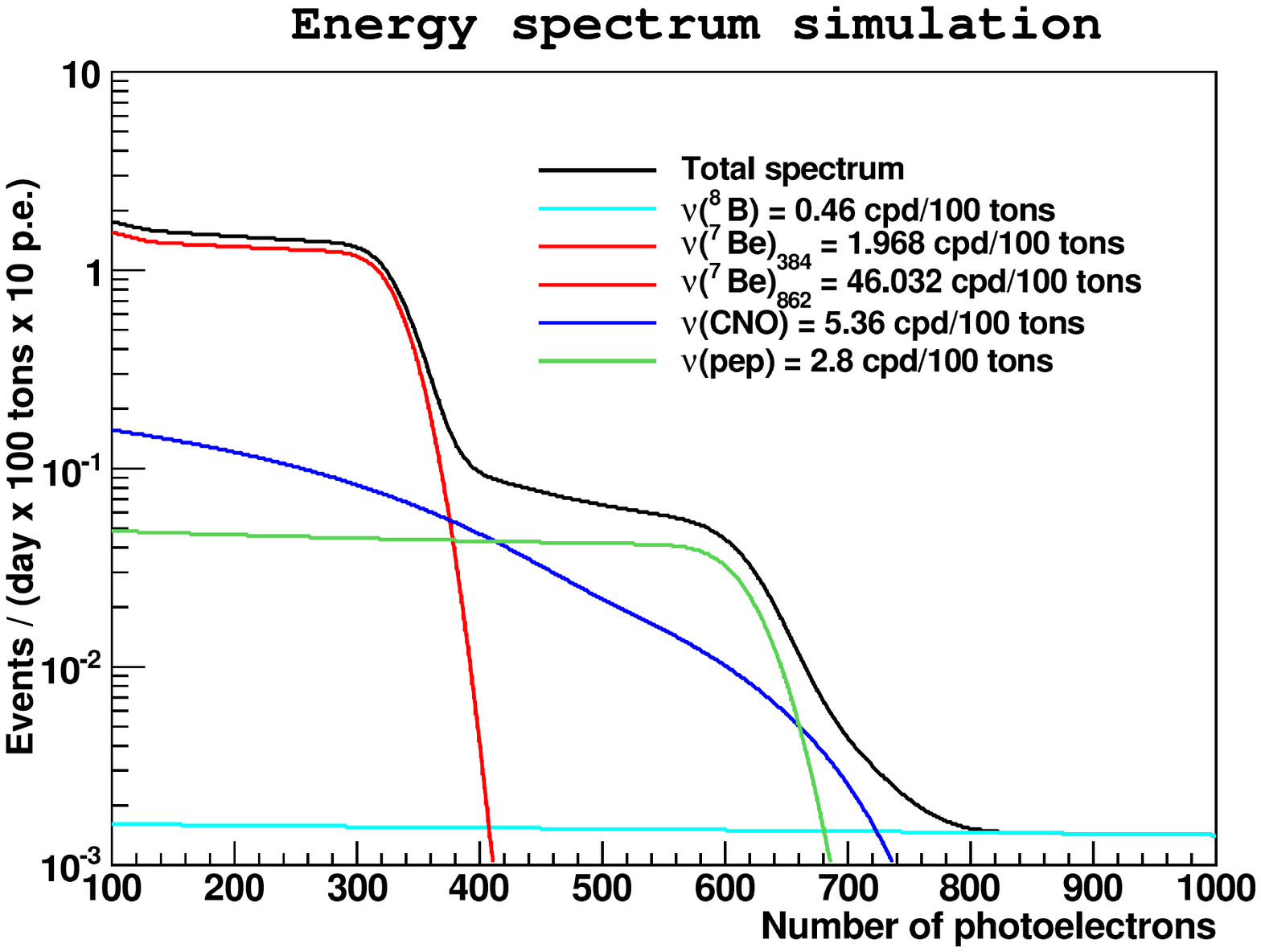}
\vspace{-2.5truecm}
\caption{   The   neutrino-induced  electron   recoil   spectra  expected   in
  Borexino. The total rates are those predicted by the latest {\em high-Z}
  solar model \cite{14-serenelli:2011}. The {\it  pep} and CNO neutrinos recoil
  spectra with end points in the region 1.2-1.5~MeV are shown. Also the $^7$Be
  neutrinos (measured  in \cite{14-Bellini2011rx}), with a count  rate about
  10 times larger,  are shown for comparison. Note that the  variable on the x
  axis is not directly the energy value. Taken from \cite{14-Davini}.
\label{fig-14:7}}
\end{center}
\end{figure}

The electron recoil energy  spectrum from \mbox{\it pep} neutrino interactions
in Borexino is a Compton-like shoulder with end point of 1.22\,MeV, as one can
see  from Fig.(\ref{fig-14:7}), showing  the \mbox{\it  pep}  and \mbox{CNO}
contribution in Borexino .

As         already        mentioned,        very         low        background
levels~\cite{14-bxfirstresults,14-Arpesella2008}   are   required  to   detect
$^{7}$Be neutrinos;  the detection of \mbox{\it pep}  and \mbox{CNO} neutrinos
is   even  more  challenging,   as  their   expected  interaction   rates  are
$\sim$10~times lower.  The expected rate is on the order of a few counts per day
in a 100\,ton  target.  To detect \mbox{\it pep}  and \mbox{CNO} neutrinos the
Borexino  Collaboration adopted  a novel  analysis procedure  to  suppress the
dominant  background  in the  1--2  MeV  energy range,  due to  the
cosmogenic $\beta^+$-emitter \mbox{$^{11}$C}
% having a lifetime of 29.4 min. 
produced within the scintillator 
%\mbox{$^{11}$C}%
by muon  interactions with {\mbox{$^{12}$C}}  nuclei.  The muon  flux crossing
the Borexino detector, $\sim$4300\,$\mu$/day,  yields a \mbox{$^{11}$C} production
rate of  $\sim$27 \mbox{counts/(day$\cdot$100\,ton)}.  This  background can be
reduced by performing a space and time veto following coincidences between signals
from  the muons  and  the cosmogenic  neutrons \cite{14-deutsch,  14-pep-ctf},
discarding exposure that is more  likely to contain \mbox{$^{11}$C} due to the
correlation  between the  parent muon,  the neutron\footnote{In  95\%~of the
  cases at least one free neutron is spalled in the \mbox{$^{11}$C} production
  process \cite{14-c11cris}, and then captured in the scintillator with a mean
  time  of 255\,$\mu$s \cite{14-bxmuon}.}  and the  subsequent \mbox{$^{11}$C}
decay (the  Three-Fold Coincidence, TFC).  The  TFC technique is  based on the
reconstructed  track  of  the  muon  and the  reconstructed  position  of  the
neutron-capture $\gamma$-ray \cite{14-bxmuon}.  The criteria of rejection were
applied to  obtain the best  compromise between \mbox{$^{11}$C}  rejection and
preservation  of fiducial  exposure, resulting  in a  \mbox{$^{11}$C}  rate of
(2.5$\pm$0.3)  count  per  day,  (9$\pm$1)$\%$  of the  original  rate,  while
preserving 48.5\% of the initial exposure.

\begin{figure}[h!!]
\begin{center}
%\vspace{1cm}
\includegraphics[width=9.0cm,height=8.0cm,angle=0]{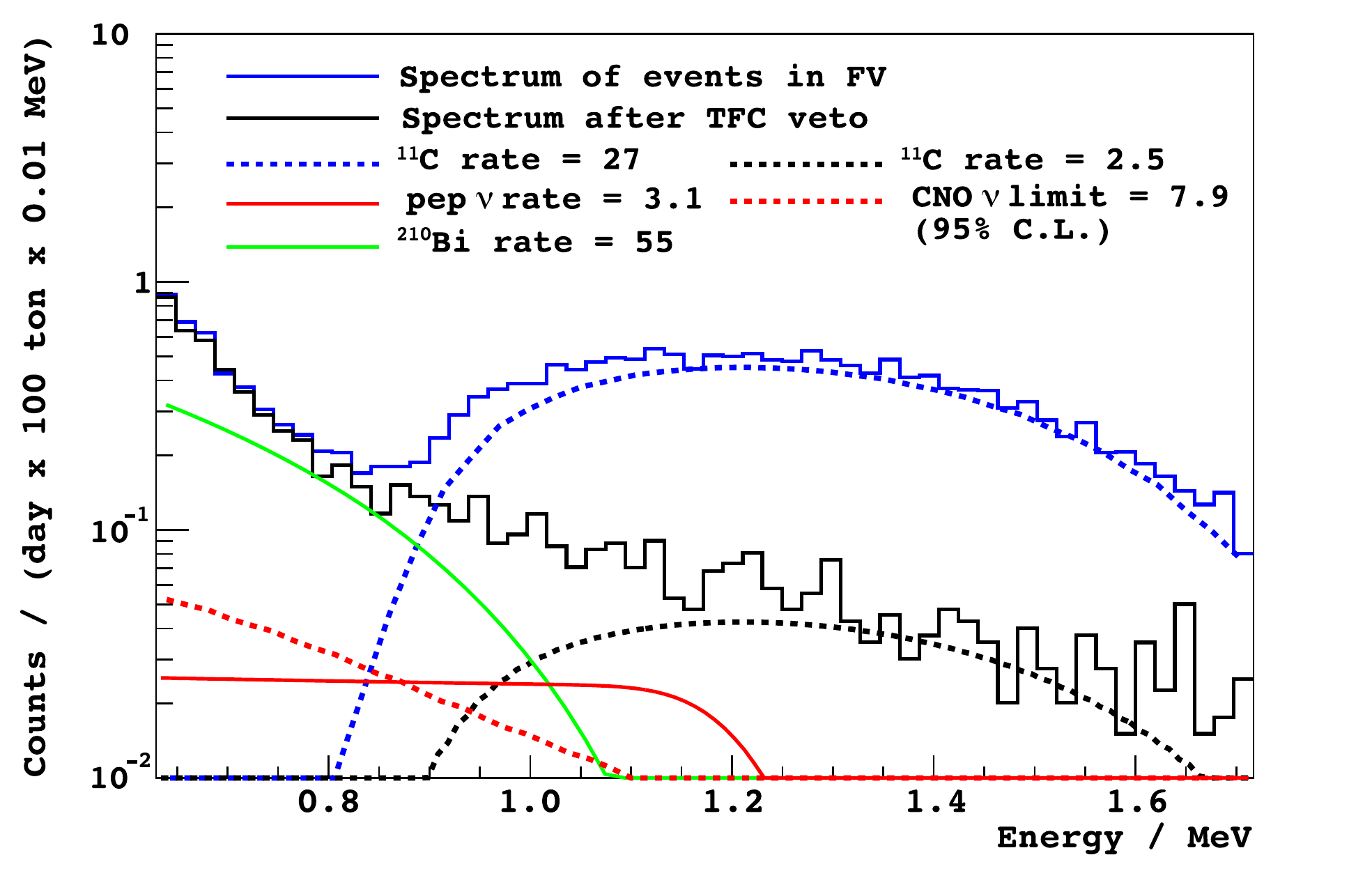}
\caption{Energy spectra of the events in  the FV before and after the TFC veto
  is applied.   The solid and  dashed blue lines  show the data  and estimated
  \mbox{$^{11}$C} rate before any veto is applied.  The solid black line shows
  the  data after  the procedure,  in which  the  \mbox{$^{11}$C} contribution
  (dashed)  has  been  greatly   suppressed.   The  next  largest  background,
  \mbox{$^{210}$Bi}, and the  electron recoil spectra of the  best estimate of
  the  \mbox{\it pep}  neutrino  rate and  of  the upper  limit of  \mbox{CNO}
  neutrino rate are shown for reference.  Rate values in the legend are quoted
  in counts/(day $\cdot$ 100 ton) from \cite{14-pepBX}.
\label{fig-14:8}}
\end{center}
\end{figure}

Figure~(\ref{fig-14:8})  shows the  resulting  spectrum  obtained with  data
collected  between January  2008 and  May  2010, corresponding  to a  fiducial
exposure of  20409 ton$\cdot$day \cite{14-pepBX}.   Despite the TFC  veto, the
number  of \mbox{$^{11}$C}  surviving events  still constituted  a significant
background.

To discriminate  \mbox{$^{11}$C} $\beta^+$ decays  from neutrino-induced $e^-$
recoils  and $\beta^-$ decays  the pulse  shape differences  between  $e^-$ and
$e^+$  interactions  in  organic liquid  scintillators  \cite{14-annihilation,
  14-positronium}  were exploited.   In fact  a small  difference in  the time
distribution of  the scintillation signal  arises from the finite  lifetime of
ortho-positronium as well as  from the presence of annihilation $\gamma$-rays,
which present  a distributed, multi-site  event topology and a  larger average
ionization  density  than electron  interactions.  The Borexino  Collaboration
employed  an optimized  pulse  shape parameter  using a  boosted-decision-tree
algorithm \cite{14-tmva},  trained with a TFC-selected  set of \mbox{$^{11}$C}
events  ($e^+$) and  \mbox{$^{214}$Bi}  events ($e^-$)  selected  by the  fast
\mbox{$^{214}$Bi-$^{214}$Po}  $\alpha$-$\beta$  decay  sequence.   In  a  work
published  in 2012  \cite{14-pepBX} the  Borexino Collaboration  presented the
results of an analysis based  on a binned likelihood multivariate fit performed
on   the  energy,  pulse   shape,  and   spatial  distributions   of  selected
scintillation  events  whose reconstructed  position  is  within the  fiducial
volume\footnote{less than  2.8\,m from the  detector center and  with a
  vertical  position  relative to  the  detector  center  between -1.8\,m  and
  2.2\,m.}.

The  energy spectra  and  spatial distribution  of  the external  $\gamma$-ray
backgrounds  have  been  obtained   from  a  full,  Geant4-based  Monte  Carlo
simulation,  and   validated  with  calibration  data   from  a  high-activity
$^{228}$Th source~\cite{14-maneschg} deployed  in the outermost buffer region,
outside  the active  volume.  $\alpha$  events  were removed  from the  energy
spectrum by  the statistical subtraction  method~\cite{14-bxfirstresults}.  In
the  energy region of  interest of  the fit  procedure all  background species
whose rates  were estimated  to be less  than 5\%  of the predicted  rate from
\mbox{\it pep}  neutrinos have  been excluded. All  rates were  constrained to
positive   values    and   thirteen   species   were   left    free   in   the
fit\footnote{electron recoils  from $^{7}$Be, \mbox{\it  pep}, and \mbox{CNO}
  solar  neutrinos,  internal  radioactive backgrounds  $^{210}$Bi,  $^{11}$C,
  $^{10}$C,  $^{6}$He,  $^{40}$K,  $^{85}$Kr,  and $^{234m}$Pa,  and  external
  $\gamma$-rays from  $^{208}$Tl, $^{214}$Bi, and $^{40}$K.}.  The  rate of the
radon   daughter   $^{214}$Pb  was   fixed   using   the   measured  rate   of
\mbox{$^{214}$Bi-$^{214}$Po} delayed coincidence events. The contribution from
\mbox{\it  pp} solar  neutrinos was  fixed to  the SSM  assuming  MSW-LMA with
$\tan^2\theta_{12}$=0.47$^{+0.05}_{-0.04}$,                             $\Delta
m^2_{12}$={(7.6$\pm$0.2)}$\cdot  10^{-5}$\,eV$^2$~\cite{14-pdg2010},  and  the
contribution   from  $^{8}$B  neutrinos   to  the   rate  from   the  measured
flux~\cite{14-LETA,14-SNOI+II+III}.

In  Table~\ref{tab:results-summary} the  results  for the  \mbox{\it pep}  and
\mbox{CNO} neutrino interaction  rates are shown.  The absence  of a \mbox{\it
  pep}  neutrino signal was  rejected at  98\%~C.L. Concerning  the \mbox{CNO}
neutrinos flux, its electron-recoil spectrum  is similar to the spectral shape
of $^{210}$Bi,  but the last one is  about 10 times greater;  therefore it has
only  been possible  to  provide an  upper  limit on  the \mbox{CNO}  neutrino
interaction      rate.      The      95\%~C.L.      limit     reported      in
Table~\ref{tab:results-summary}  has been  obtained from  a  likelihood ratio
test   with   the   \mbox{\it   pp}   neutrino   rate   fixed   to   the   SSM
prediction~\cite{14-serenelli:2011}  under    the    assumption  of    MSW-LMA,
(2.80$\pm$0.04)\,\mbox{counts/(day$\cdot$100\,ton)}.

\begin{table}[!ht]
\begin{center}
\begin{tabular}{lccc}
\hline     \hline     $\nu$     &Interaction    rate     &Solar-$\nu$     flux
&Data/SSM\\  &[\mbox{counts/(day$\cdot$100\,ton)}] &[$10^{8}  cm^{-2} s^{-1}$]
&ratio\\ \hline \mbox{\it pep} &$3.1  \pm 0.6_{\rm stat} \pm$ 0.3$_{\rm syst}$
&$1.6\pm0.3$  &$1.1\pm0.2$\\   \mbox{CNO}  &$<7.9$  ($<7.1_{\rm  stat\,only}$)
&$<7.7$ &$<1.5$\\ \hline \hline
\end{tabular}
\end{center}
\caption{  Best  estimates for  the  \mbox{\it pep}  and \mbox{CNO}  solar
  neutrino interaction  rates.  For the results  in the last  two columns both
  statistical and systematic uncertainties  are considered.  Total fluxes have
  been obtained assuming MSW-LMA  and using the scattering cross-sections from
  \cite{14-BahcallRadiativeCorrection,  14-pdg2010,   14-erlerRadCorr}  and  a
  scintillator $e^-$ density of (3.307$\pm$0.003)$\cdot 10^{29}$ \,ton$^{-1}$.
  The  last  column gives  the  ratio between  our  measurement  and the  
  {\em high-Z} (GS98) SSM~\cite{14-serenelli:2011}. Table taken from \cite{14-pepBX}.}
\label{tab:results-summary}
\end{table}

%Table~\ref{tab:results-summary} also shows the solar neutrino fluxes inferred from our best estimates of the 
%\mbox{\it pep}  and \mbox{CNO} neutrino interaction rates, assuming the MSW-LMA solution, and the ratio of these values 
%to the High Metallicity (GS98) SSM predictions \cite{bib:ssm2011}.  Both results are consistent with the predicted 
%High and Low Metallicity SSM fluxes assuming MSW-LMA.  Under the assumption of no neutrino flavor oscillations, we would 
%expect a \mbox{\it pep}  neutrino interaction rate in Borexino of (4.47$\pm$0.05)\,\mbox{counts/(day$\cdot$100\,ton)}; 
%the observed interaction rate disfavors this hypothesis at 97\%~C.L.  If this discrepancy is due to $\nu_e$ oscillation 
%to $\nu_\mu$ or $\nu_\tau$, we find \Pee=0.62$\pm$0.17 at 1.44\,MeV. This result is shown alongside other solar 
%neutrino \Pee\ measurements in Fig.~\ref{fig:pee}. The MSW-LMA prediction is shown for comparison.

%We have achieved the necessary sensitivity to provide, for the first time, evidence of the rare signal from \mbox{\it pep}  neutrinos and to place the strongest constraint on the \mbox{CNO} neutrino flux to date.  This has been made possible by the combination of the extremely low levels of intrinsic background in Borexino, and the implementation of novel background discrimination techniques.  This result raises the prospect for higher precision measurements of \mbox{\it pep}  and \mbox{CNO} neutrino interaction rates, if the next dominant background, \bite, is further reduced by scintillator re-purification.

\section{Phenomenological analysis}\
\label{14-sec:pheno-analysis}

\subsection{Status of the determination of the mixing parameters in a 3 flavor 
analysis}
\label{14-sec:status-mixing}

\begin{figure}[h!!]
\begin{center}
%\vspace{1cm}
\includegraphics[width=7.5cm,height=7.0cm,angle=0]{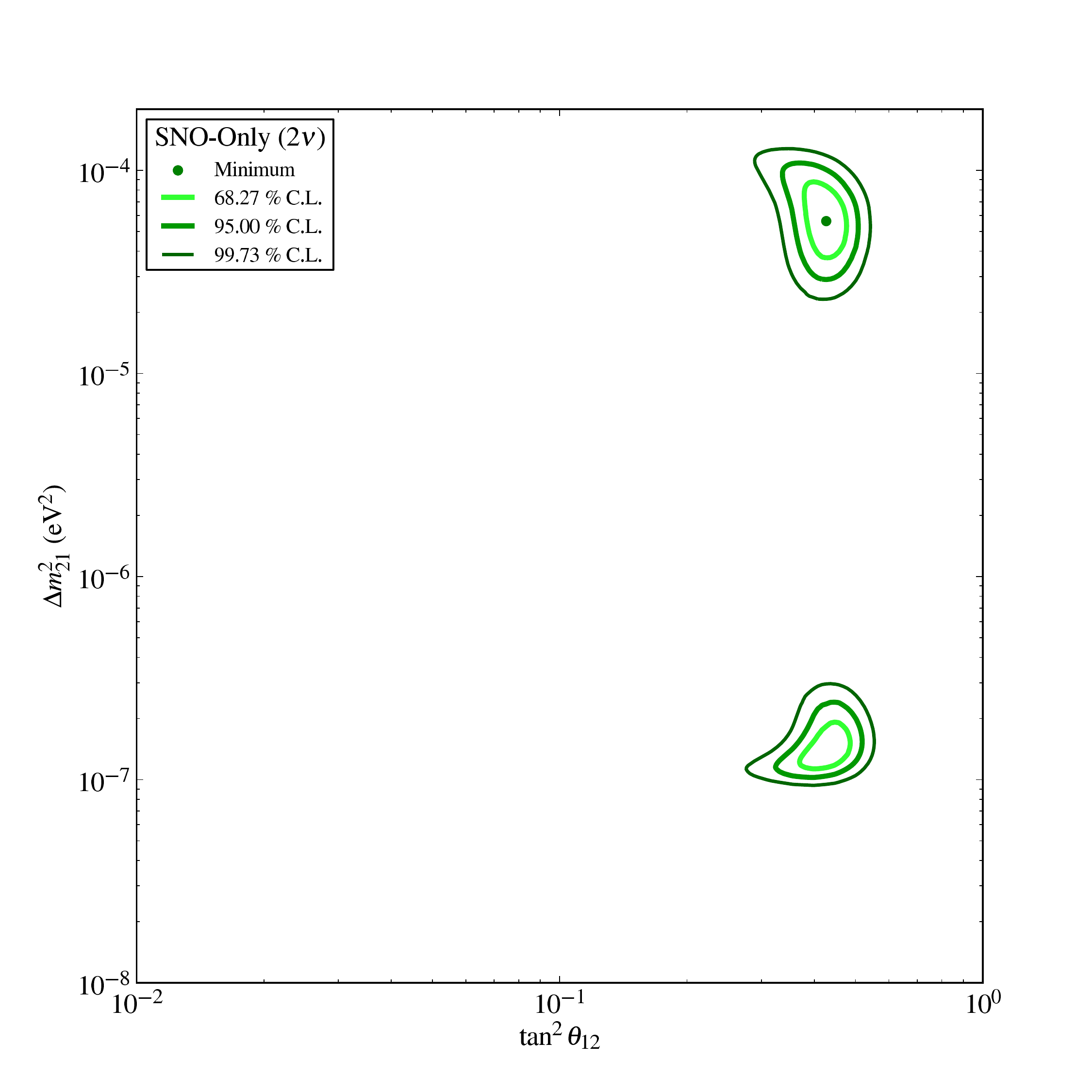}
\caption{Two-flavor neutrino oscillation analysis contour using only SNO data \, 
(taken from \cite{14-SNOI+II+III}).
\label{fig-14:9}}
\end{center}
\end{figure}

Recently, the SNO collaboration performed  a combined analysis of all the three
working phases of the  experiment \cite{14-SNOI+II+III} based on
a fit to Monte Carlo derived  probability density functions (PDFs) for each of
the  possible  signals  and backgrounds,  and also introduced  a new  way  to
parametrize  the  $^8$B   neutrino  signal.   Figure~(\ref{fig-14:9}),
reporting the results  of the two flavour (with  the assumption $\theta_{13} =
0$) SNO only analysis, shows  the further improvement in the mixing parameters
accuracy, but, at the same time, it confirms that the SNO results alone would
not be sufficient to completely exclude the LOW solution.

%\vspace{-0.2truecm}   

\begin{figure}[h!!]
\begin{center}
%\vspace{1cm}
\includegraphics[width=8.5cm,height=8.0cm,angle=0]{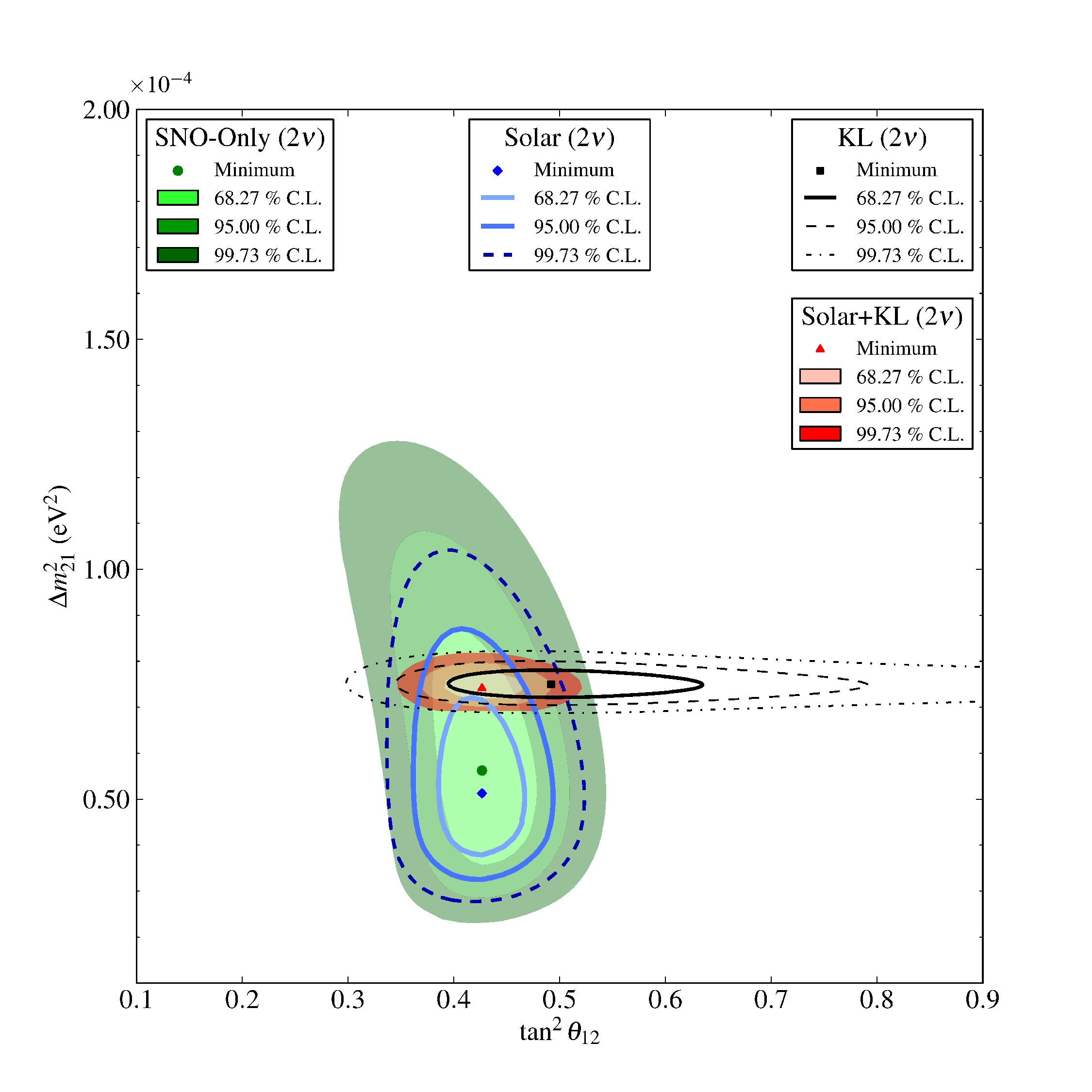}
\caption{Two-flavor neutrino oscillation analysis contour using both solar neutrino 
and KamLAND results \,(taken from \cite{14-SNOI+II+III}).
\label{fig-14:10}}
\end{center}
\end{figure}

This ambiguity  was definitely removed, as shown  in Figure~(\ref{fig-14:10}),
by including  in  the  analysis  the  results  of  all  previous  solar  neutrino
experiments \cite{14-Gallium09,14-Cleveland1998,14-SKI-2005,14-SKII,14-SKIII},
the  $^7$Be  solar  neutrino  rate  measured   by  Borexino  
\cite{14-Bellini2011rx}, the    $^8$B    neutrino    spectra
\cite{14-Bellini2008mr} and the KamLAND data\footnote{The  KamLAND   data  were  
obtained  in   a  completely  independent
  experiment and,  therefore, the corresponding $\chi^2$  values, as functions
  of  the mixing  parameters,  were  directly summed  to  the $\chi^2$  values
  computed by direct solar neutrino analysis.} \cite{14-KLfollowing3}.

The  higher values  of  $\Delta m_{12}^2$  in  the LMA  region were  excluded,
together with the full LOW  solution, thanks mainly to the large discrimination
power of KamLAND. This experiment, however, did not contribute significantly to
improve  the mixing  angle determination  and the  accuracy on  this parameter
remained quite high.   The results of the two flavor  analysis are reported in
Table~(\ref{tab:global2nu}) (taken from \cite{14-SNOI+II+III}).  
\begingroup
\begin{table}
\centering
\begin{tabular}{lccc}
\hline\hline
Analysis & $\tan^{2} \theta_{12}$ & $\Delta m_{21}^2 [{\rm eV^{2}}]$ & $\chi^{2}/{\rm NDF}$\\
\hline
SNO only (LMA)&$0.427^{+0.033}_{-0.029}$& $5.62^{+1.92}_{-1.36}\times 10^{-5}$& $1.39/3$\\
SNO only (LOW) &$0.427^{+0.043}_{-0.035}$& $1.35^{+0.35}_{-0.14}\times 10^{-7}$& $1.41/3$ \\
Solar &$0.427^{+0.028}_{-0.028}$& $5.13^{+1.29}_{-0.96}\times 10^{-5}$& $108.07/129$\\
Solar+KamLAND&$0.427^{+0.027}_{-0.024}$& $7.46^{+0.20}_{-0.19}\times 10^{-5}$& \\
\hline\hline
\end{tabular}
\caption{Best-fit neutrino  oscillation parameters from  a two-flavor neutrino
  oscillation analysis.  Uncertainties listed are  $1 \, \sigma$ after the
  $\chi^2$  was minimized  with respect  to all  other parameters  (taken from
  \cite{14-SNOI+II+III}).}
\label{tab:global2nu}
\end{table}
\endgroup

\begin{figure}[h!!]
\begin{center}
\hspace{-0.50cm}
\includegraphics[width=7.6cm,height=7.6cm,angle=0]{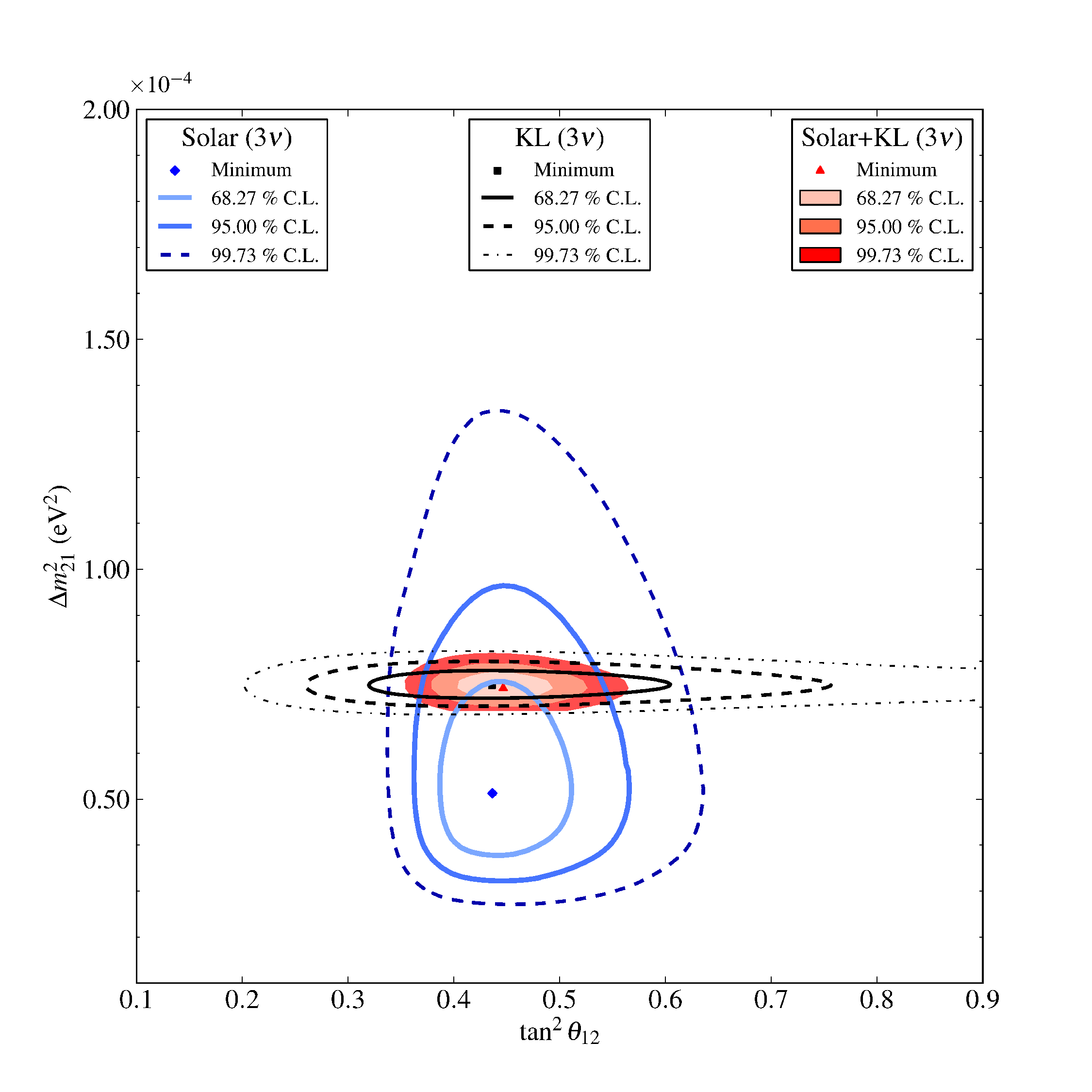}
\hspace{0.1cm}
\includegraphics[width=7.6cm,height=7.6cm,angle=0]{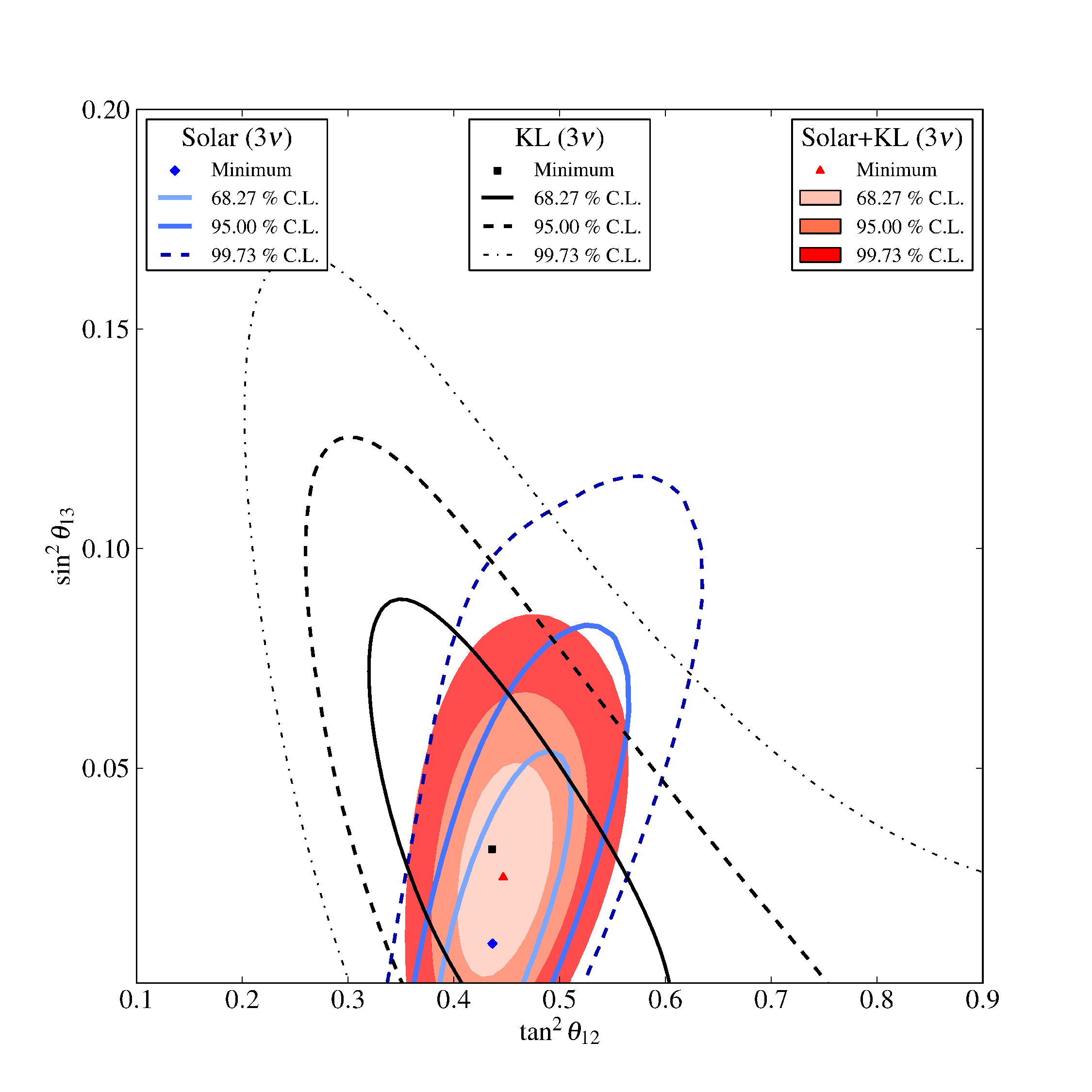}
%\vspace{-2truecm}
\caption{Three-flavor neutrino  oscillation analysis contour  using both solar
  neutrino and KamLAND results. Taken from \cite{14-SNOI+II+III}.
\label{fig-14:11}}
\end{center}
\end{figure}

The  slight tension  between the  solar neutrino  experiments and  KamLAND was
significantly reduced by extending the analysis to the 3 flavor oscillation case
as shown in Figure~(\ref{fig-14:11}), from which  it is clear that the best global
fit is obtained for values of $\theta_{13}$ different from zero.

\begin{figure}[h!!]
\begin{center}
%\vspace{1cm}
\includegraphics[width=7.5cm,height=7.5cm,angle=0]{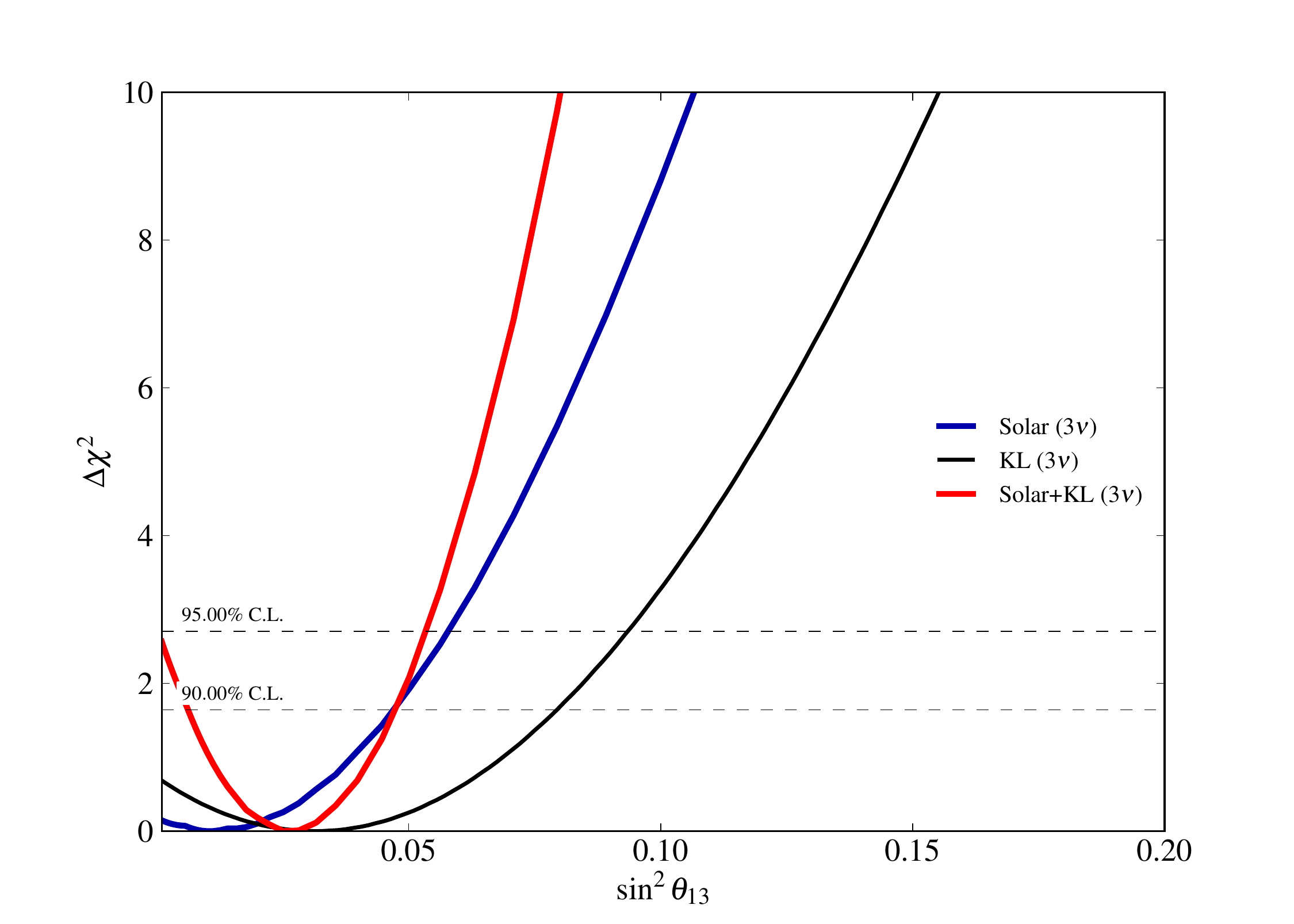}
\vspace{-0.5cm}
\caption{Projections of the three-flavor neutrino oscillation parameters. 
%determined from figure (\ref{fig14:12}). 
The horizontal lines represent  the $\Delta\chi^2$ for a particular confidence
level. Taken from \cite{14-SNOI+II+III}.
\label{fig-14:12}}
\end{center}
\end{figure}

A detailed analysis of the  $\chi^2$ behavior proved also that the combination
of  solar  experiments  and  KamLAND  enables  to  improve  significantly  the
discriminating  power  on  the  $\theta_{13}$  mixing  parameter  (see  Figure~
\ref{fig-14:12} and Table~\ref{tab:chi3nu}).

\begin{figure}[h!!]
\begin{center}
%\vspace{1cm}
\includegraphics[width=7.5cm,height=6.8cm,angle=0]{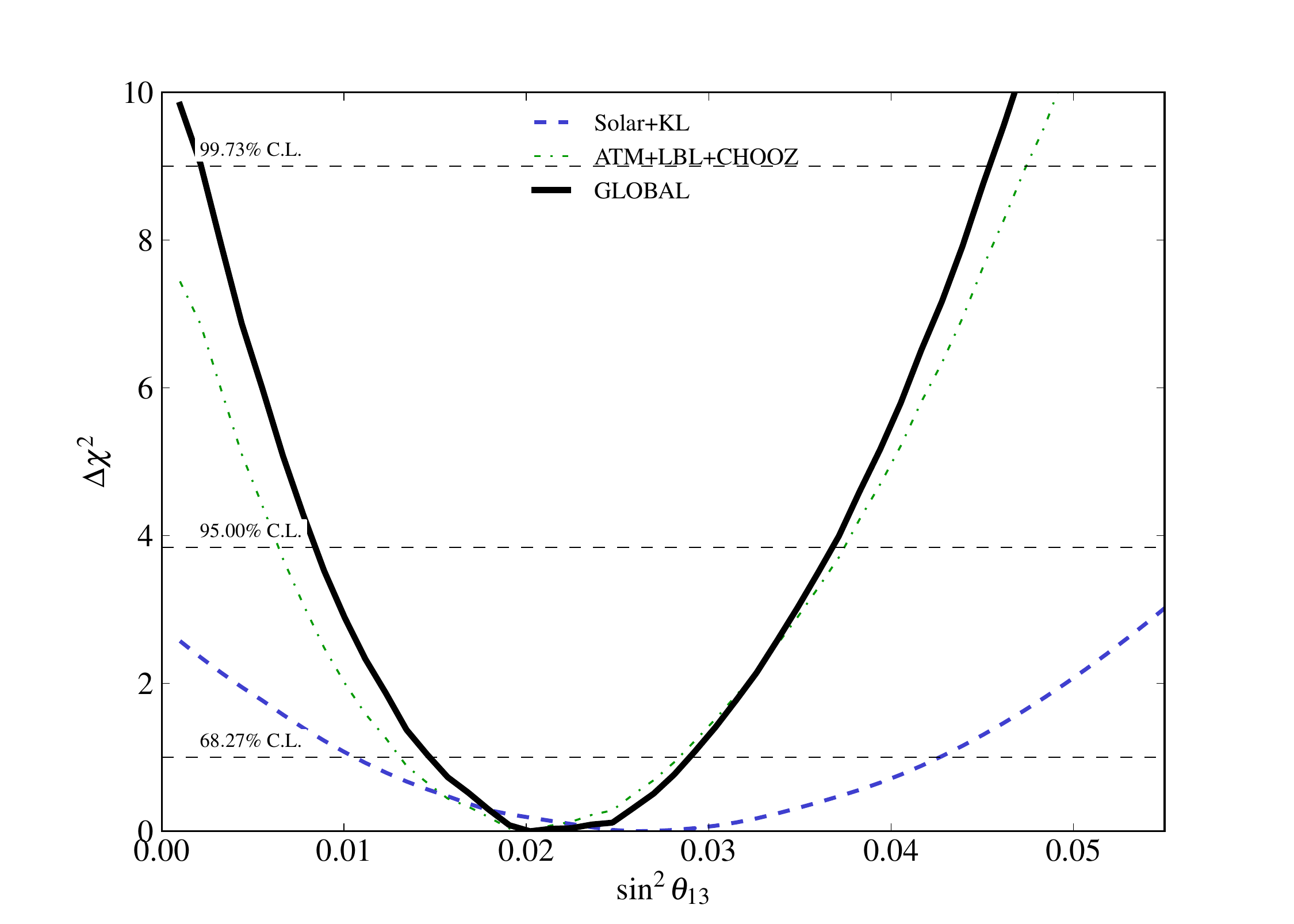}
\caption{Projection  over   $\sin^{2}\theta_{13}$  combining  the  projections
  obtained  by  analyzing  data  from  all neutrino  sources.  The  data  from
  atmospheric,  short-baseline   experiments  and  long-baseline  experiments
  (ATM+LBL+CHOOZ)     was     determined     from     Figure~2 (left panel) in
 \cite{14-FogliLisi2011}   which   already   includes  the   latest
  T2K~\cite{14-T2K2011} and MINOS~\cite{14-MINOS2011} results.
\label{fig-14:13}}
\end{center}
\end{figure}

\begingroup
\begin{table}
%\centering
\begin{tabular}{lccc}
\hline\hline
Analysis & $\tan^2 \theta_{12}^2$ & $\Delta m_{12}^2 [{\rm eV^{2}}]$ & $\sin^2
\theta_{13} \times 10^{-2}$\\
\hline
Solar & $0.436^{+0.048}_{-0.036}$ & $5.13^{+1.49}_{-0.98}\times 10^{-5}$ & $ <5.8$ (95\% C.L.)\\
Solar+KL& $0.446^{+0.030}_{-0.029}$ &$7.41^{+0.21}_{-0.19}\times 10^{-5}$& $2.5^{+1.8}_{-1.5}$\\
& & & $<5.3$ (95\% C.L.)\\
Global & & & $2.02^{+0.88}_{-0.55}$\\
\hline\hline
\end{tabular}
\caption{Best-fit neutrino oscillation parameters from a three-flavor neutrino
  oscillation analysis.   Uncertainties listed are $\pm 1\,  \sigma$ after the
  $\chi^2$  was minimized  with respect  to all  other parameters.  The global
  analysis includes Solar+KL+ATM+LBL+CHOOZ.}
\label{tab:chi3nu}
\end{table}
\endgroup 
The indication in favor  of $\theta_{13}$ being different from zero was in
agreement  with the  recent results  from the  long-baseline  experiments T2K
\cite{14-T2K2011} and MINOS \cite{14-MINOS2011}, and with the combined analysis
performed in \cite{14-FogliLisi2011},  including also the atmospheric neutrino
and the  CHOOZ \cite{14-CHOOZ03} data. Moreover the validity of  this hint has
been corroborated by the data obtained this year by the short baseline neutrino
reactor   experiments   \cite{14-DoubleChooz,   14-DayaBay,  14-RENO},   which
established that $\theta_{13}  > 0$ at about $5 \sigma$ (and  even more in the
Daya  Bay  case  \cite{14-Daya_Bay_neu2012}). These experiments  
found  values  of  $\sin^2
\theta_{13}$  centered between $0.020$  and $0.030$; very promising results for 
future  experiments looking for  leptonic CP  violation \cite{14-Fogli2012ua}.
The impact  and the  possible consequences of  these recent results  have been
discussed,  among the  others, in  the following  papers \cite{14-Fogli2012ua,
  14-Tortola2012te,14-Schwetz2012}.   The  different   accuracy  that  can  be
reached in the  determination of the mixing angle between  the first and third
generation, according  to the different kind of  neutrino experiments included
in the analysis, is represented in Figure~(\ref{fig-14:13}).

The combined  analysis of  the different  SNO phases was  also very  useful to
obtain a precise determination of  the $^8$B solar neutrino flux, $\Phi_{^8
  B}  = 5.25  \pm  0.16  (\rm stat)^{+0.11}_{-0.13}  (\rm  syst) \times  10^6
\,{\rm cm^{-2} \, s^{-1}}$,  with an important  reduction of the  systematic 
uncertainty.
This  result was consistent  with, but  more precise  than, both  the {\em high-Z}
BPS09(GS), $\Phi =  (5.88 \pm  0.65) \times 10^6  {\rm \, cm^{-2} \, s^{-1}}$,  and 
{\em low-Z} BPS09(AGSS09),
$\Phi  =  (4.85  \pm  0.58)  \times  10^6  {\rm \, cm^{-2}  \,  s^{-1}}$,  solar  
model predictions \cite{14-Serenelli2009yc}.

\begin{figure}[h!!]
\begin{center}
%\vspace{1cm}
\includegraphics[width=8.5cm,height=7.0cm,angle=0]{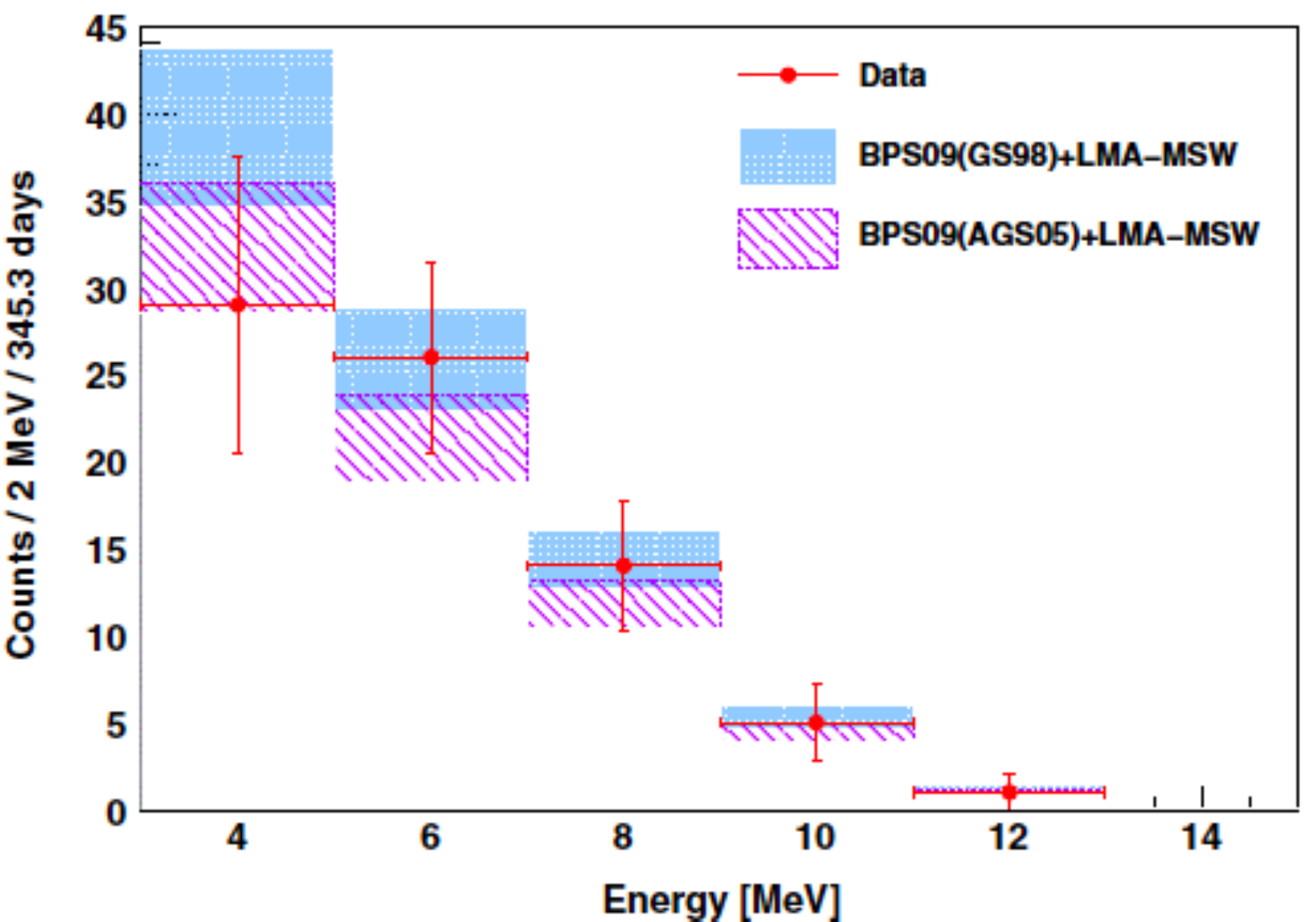}
\caption{
%Caption still to write. 
Taken from (\cite{14-Bellini2008mr}). 
\label{fig-14:14}}
\end{center}
\end{figure}

The combination of  the LETA analysis by the SNO  collaboration \cite{14-LETA} and
of the Borexino measurements  \cite{14-Bellini2008mr} made possible a detailed
study of  the low energy  part of the  $^8$B solar neutrino spectrum.  Even if
characterized  by a larger uncertainty (mainly due  to a more limited
statistics), Borexino data confirm  the LETA indication of low energy data
points  lower  than  the  theoretical  expections  based  on  matter  enhanced
oscillation and solar models as shown in Figure~\ref{fig-14:14} (taken from
\cite{14-Bellini2008mr}).     These   results   agreed    also   with   the
Super-Kamiokande observation  \cite{14-SKI-2005} of flat  spectrum, consistent
with the undistorted spectrum hypothesis.  The emergence of this slight tension
between theory and experiments seems to indicate  the presence of new
subdominant  effects  and  also suggests   the possibility  of  non-standard
neutrino interactions (like those studied in \cite{14-Friedland:2004pp}) or
the  mixing  with a  very  light  sterile neutrino  \cite{14-deHolanda2010am}.
Future solar  neutrino experiments, like SNO+,  could shed more  light on this
subject, by performing precision measurements of lower energies solar neutrinos 
(like the \mbox{\it pep} neutrinos).

%\vspace{4 truecm}

\subsection{Free flux analyses}
\label{14-subsection:freeflux}

The increasing data of solar neutrinos allow to independently test the astrophysics of the solar interior 
and the physics of neutrino propagation. The analysis discussed in previous sections can be modified by also varying the solar neutrino fluxes 
in order to accommodate all neutrino data, while all the functional dependences are maintained as predicted 
by the standard model dependences. A key step in this kind of analysis is the 
imposition of the luminosity constraint~\cite{14-spirovignaud,14-luminosity}, which implements in a global way for the Sun the constraint of conservation of energy for nuclear fusion among light elements. Each neutrino flux is associated with a 
specific amount of energy released to the
star and therefore a particular linear combination of the solar neutrino fluxes is equal to the solar
luminosity (in appropriate units). One can write the luminosity constraint as
\begin{equation}
{L_\odot\over 4\pi (A.U.)^2} = \sum\limits_i \alpha_i \Phi_i~, \label{eq:genconstraint}
\end{equation}
where $\L_\odot$ is the solar luminosity measured at the earth's surface, 1 $A.U.$ is the average
earth-sun distance,  and the coefficient $\alpha_i$ is the amount of energy provided to the star by
nuclear fusion reactions associated with each of the important solar neutrino fluxes, $\Phi_i$. The
coefficients $\alpha_i$ are calculated accurately in ref.~\cite{14-luminosity}.

The  model independent determination of the solar neutrino fluxes \cite{14-concha:2010, 14-roadmap} shows that 
present solar neutrino data leads to accurate results  for four  fluxes and  also the  correlations between them.  
This  information allows for a consistent global comparison of SSM fluxes with the inferred fluxes by neutrino 
data. Present data leads to the values for the inferred solar neutrino fluxes
reported in the fourth column (labelled as ``Solar'') of 
Table~\ref{14-tab:nufluxes} in Section~\ref{14-section:SSM}.
% :\\ $6.05(1^{+0.003}_{-0.011})$ (pp), 
% $1.46(1^{+0.01}_{-0.014})$ (pep), $18(1^{+0.4}_{-0.5})$ (hep),  
% $4.82(1^{+0.05}_{-0.04})$ ($^7 Be $), $5.0(1 \pm 0.03)$ ($^8$B), 
% $\leq 6.7$ ($^{13}$N), $\leq 3.2$ ($^{15}$O) and $\leq 59$ ($^{17}$F) 
% where the  neutrino   fluxes   are   given  in   units   of $10^{10}$(pp), 
% $10^9$($^7$Be), $10^8$(pep, $^{13}$N, $^{15}$O), 
% $10^6$($^8$B, $^{17}$F) and $10^3$(hep)~\mbox{cm$^{-2}$ s$^{-1}$}. 
 The precision of the $^7$Be and $^8$B neutrino fluxes is driven by the Borexino and SNO (SK) neutrino experiments, 
 while the precision of the pp and pep neutrino fluxes mainly comes by the imposition of the luminosity constraint. 
 The neutrino data directly demonstrates that the Sun shines by the pp chain. The CNO cycle only contributes to the 
 total luminosity at the percent level.
 
The reader may wonder how much these inferences are affected by the luminosity constraint. The idea that the Sun 
shines because of nuclear fusion reactions can be tested accurately by comparing the observed photon luminosity of 
the Sun with the luminosity inferred from measurements of solar neutrino fluxes. Moreover, this same comparison will 
test a basic result of the standard solar model, namely, that the Sun is in a quasi-steady state in which the current 
energy generation in the interior equals the current luminosity at the solar surface.  The free flux analysis, without 
imposing luminosity constraint, permits an estimation of the solar luminosity  inferred by neutrino data, 
which agrees with the directly measured one within 15 \% (1 $\sigma$). 
 
\section{Future solar neutrino experiments}\label{14-sec:future}

\subsection{The near future: improvement of {\it pep} measurements and CNO 
detection} 

\begin{figure}[h!!]
\begin{center}
%\vspace{1cm}
\hspace{-4truecm}
\includegraphics[width=14.0cm,height=12.0cm,angle=0]{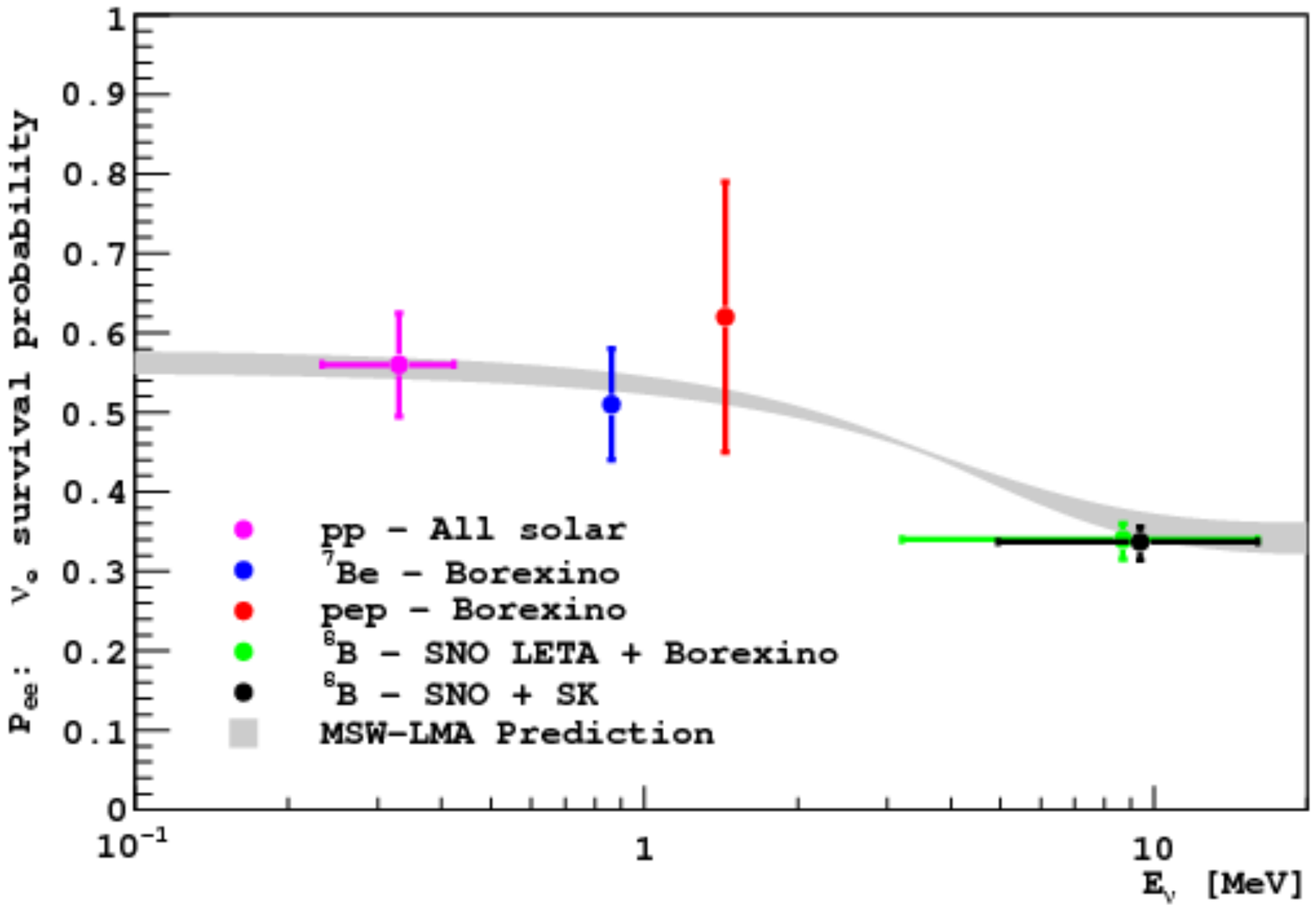}
\vspace{-3.2truecm}
\caption{The  $\nu_e$ survival  probability is  represented as  a  function of
  neutrino energy. The gray band represent the MSW-LMA prediction.  The higher
  survival  probability  region  at  low energies  is  where  vacuum-dominated
  oscillations occur. As the  neutrino energy increases, matter effects become
  important  and the lower  survival probability  at high  energies is  due to
  matter-enhanced oscillations. The reported data correspond to solar nuetrino
  flux   measurements   performed  by   different   experiments.  Taken   from
  (\cite{14-Chavarria2012sd}).}
\label{fig-14:15}
\end{center}
\end{figure}

In the  last decades  the  intensive study  of $^8$B and,  more
recently, $^7$Be  solar neutrinos made possible fundamental  steps forward in
the solution of the solar neutrino  puzzle and the determination of the neutrino 
mixing parameters.  Nevertheless,  many key features of the  oscillation models (like
the transition between the vacuum dominated sub-MeV region and the spectral region  
between  1 and  3  MeV, where  matter  effects become
relevant) still have to be tested or verified with better accuracy and precision 
(see Figure~\ref{fig-14:15}, taken from \cite{14-Chavarria2012sd}).

The apparent partial  deficit of events in  the low energy part of  the $^8$B
spectrum suggested the introduction of new theoretical models (as discussed in
section \ref{14-sec:pheno-analysis}). Also for these reasons, the experimental
efforts  in the  last years  focused  on the  detection of  neutrinos of  ever
decreasing energies, to fully confirm the validity of the MSW-LMA solution and
verify the  fluxes predicted by SSMs, discriminating between
different version  of these models.   The fluxes of  the medium and  high energy
neutrinos of  the {\it pp}  chains ($^7$Be, $^8$B  and hep) are  predicted with
quite large  uncertainties, mainly due  to the uncertainties in  nuclear cross
sections and solar opacity (Table~\ref{14-tab:fluxuncert}). The {\it pp} and {\it 
pep} fluxes,
instead,  are strongly  correlated  between themselves  and  their values  are
predicted with the highest precision because SSMs predict that {\it pp} chain 
reactions are responsible for more than 99\% of the energy powering the Sun 
\cite{14-bahcall:2006}. 
Therefore,  the  measurements  of  these  components  
%(which represent also 99\% of the  full neutrino spectrum) 
would be the most stringent test of the SSM.
The tight correlation between {\it pep} and {\it pp} neutrinos is theoretically
well established and, therefore, even in the pessimistic hypothesis that 
{\it pp} neutrinos could not be measured with the desired accuracy, a 
significant improvement in the {\it pep} neutrinos measurement with respect to 
data presently available would make possible to reduce significantly the 
$ 15 \%$ indetermination on the solar luminosity (see subsection \ref{14-subsection:freeflux}) and to test 
indirectly the SSM's predictions that almost $100 \%$ of solar energy is
produced by nuclear burning. 

As already mentioned, water \v{C}erenkov detectors, which played  a fundamental 
role in the solution
of  the  solar neutrino  problem,  are characterized  by  a  low photon  yield
\cite{14-Boger1999bb, 14-Fukuda2003s}  and therefore can  detect only the
higher part of the spectrum (hep  and $^8$B neutrinos with a threshold around
$3.5  \,  \rm{MeV}$).  The radiochemical  experiments  \cite{14-Cleveland1998,
  14-Abdurashitov1999zd}  are limited,  instead,  by their ability to  measure  
 only the
integrated neutrino rate above  the charge-current interaction threshold (down
to $0.23 \, \rm{MeV}$ for the Gallium experiments), without the possibility to
discriminate  between   the  different  spectrum   components.   Therefore, an
important contribution should come from  the present and future organic liquid
scintillator  detectors,   planned  to  perform  low   energy  solar  neutrino
spectroscopy.   To reach  this goal, they will  take advantage  from  the high
values of light  yield (about $10^4$ photons per MeV  of deposited energy) and
from the  possibility to assemble very  large masses of  high purity material.
The excellent levels of radiopurity, reached for instance at Borexino, and the
typical geometry of  these detectors (which are unsegmented  and can be easily
adapted to the definition of a  fiducial volume) are fundamental to reduce the
impact of the background, that is so critical due to the feebleness of the
low energy signal.

In the  near future significant  contributions are expected from  Borexino and
SNO$+$  \cite{14-Kraus2010zzb}   experiments.  Borexino  has already   proved  its
importance in this kind of  analysis performing the first measurements of {\it
  pep} and  CNO  neutrinos (even if  the level  accuracy is not  yet the
desired  one) and  further reducing,  with the  purification  campaign started
since  July 2010,  the level  of  contamination from  almost all  of the  main
radioactive background sources\footnote{The main problem still surviving seems
  to be  the reduction  of $^{210}$Pb.}.  The purification efforts  are still
ongoing  and  should  make  possible a  further  improvement  on the accuracy 
of the signal extraction.  The SNO+ experiment,  that should start taking data
soon in the SNOLAB, should take advantage from the location (about two times deeper
underground than  the Gran Sasso  laboratory), with the consequent lower  muon flux
and a strongly reduced $^{11}$C rate. Moreover, thanks  to the detector mass
(about three times  larger than in Borexino), it should  be able to reach
a higher  counting rate. This  could determine a fundamental  improvement at
least  in  the  case  of the {\it  pep} neutrino  measurement,  where  a  $5  \%$
uncertainty is expected, to a level  that should make possible a significant test
of the MSW transition region.

In the  more optimistic scenarios it may  be also possible to  attach the main
problem of measuring lowest energy  parts of the solar neutrino spectrum, that
is the {\it pp}  neutrinos and the $0.38 \, {\rm MeV}$  Berillium line. In any
case  the  presence  in  organic  scintillators  of  an  intrinsic  $^{14}$C
background will make this very  low energy measurements an extremely hard task
and they may require the introduction  of new techniques, like the ones we are
going to describe in the next subsections.

\subsection{The far future: experimental challenges}

The challenge for all future  experiments aimed at measuring the low energy
part of  solar neutrino  spectrum is that  of assembling  experimental devices
with low  energy thresholds suitable to detect  a low rate signal  in a region
characterized by different potential  sources of radioactive background.  This
difficult  experimental task  is common  also to  the experiments  looking for
neutrinoless  double $\beta$  decay or  for  dark matter  signals (search  for
signatures  of WIMPs, a stable  or long-lived weakly  interacting
elementary  particle,  produced in  the  early  Universe,  whose existence  is
predicted in extensions of the Standard  Model).  In fact, some  of the solar 
neutrino
experiments planned for the  future are multipurpose experiments designed also
for the other above-quoted topics.

They are  all characterized by  a very large  detector target mass and  by the
need to reach  very high levels of radiopurity. The common  feature is that of
using  scintillator   detectors,  but  they  differ  for   the  chosen  active
scintillator material,  which can vary from  traditional organic scintillators
(developed with the use of innovative technological devices) to new materials,
like the noble gases.
 
\subsubsection{Noble liquid detectors: CLEAN and XMASS}\

One  of  the  possible future  frontiers  is  the  idea to  use  scintillation
detectors  with  liquid  noble  gases,  like xenon,  argon  and  neon.   These
materials have the  advantage of being relatively inexpensive,  easy to obtain
and dense and  it is not too difficult to build  large homogeneous detectors of
this kind;  moreover, they can be  quite easily purified, offer  very high
scintillation  yields (about  $30-40$ photons/keV) and do not absorb their own 
scintillation light.

\begin{figure}[h!]
\begin{center}
%\vspace{1cm}
\includegraphics[width=8.5cm,height=7.3cm,angle=0]{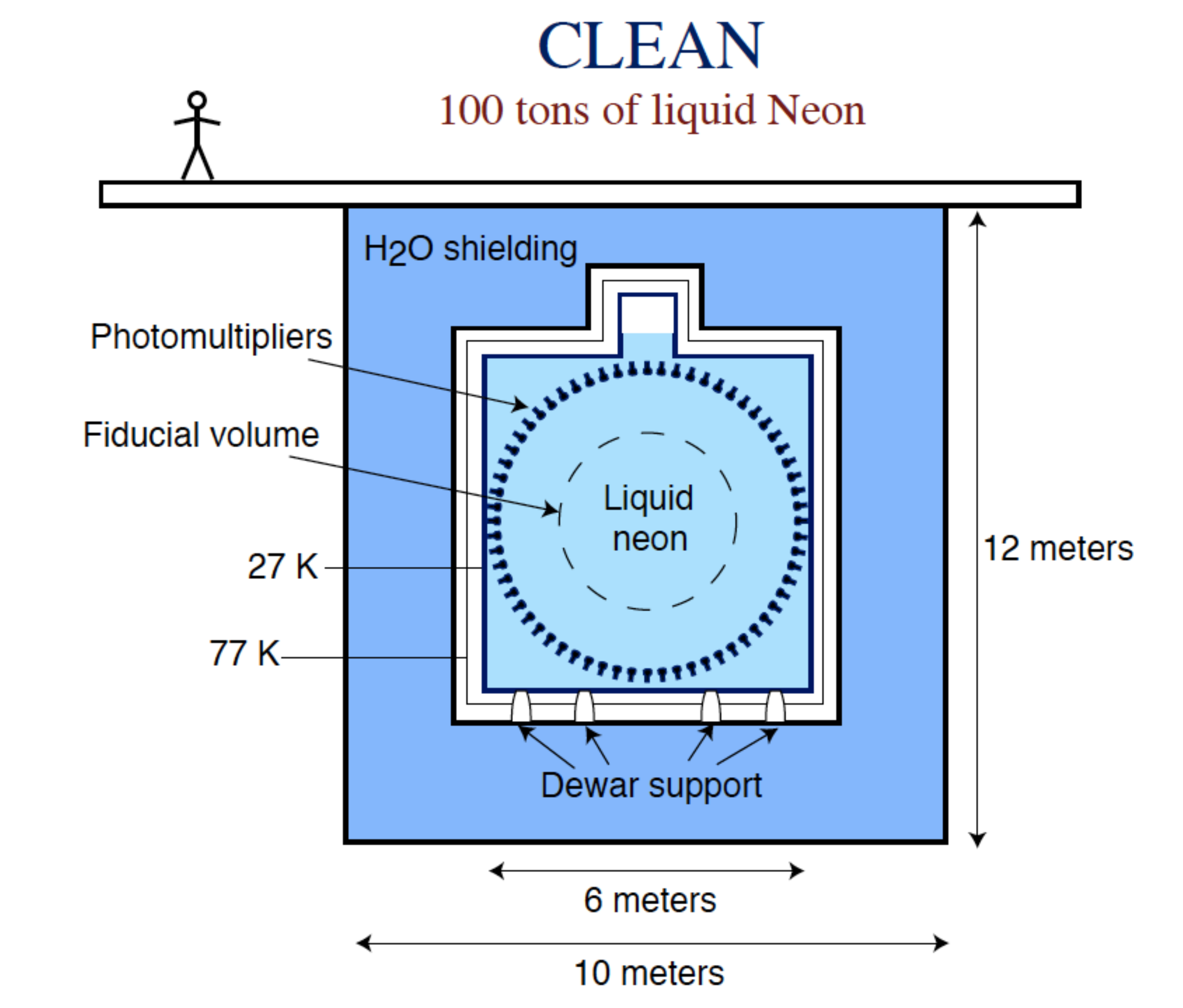}
%\vspace{-5truecm}
\caption{Scheme of the CLEAN detector. Taken from \cite{14-CLEAN2004}.
\label{fig-14:CLEAN}}
\end{center}
\end{figure}

A  first example is  offered by  the CLEAN/DEAP  family, a  series of
detectors  based entirely  on scintillation  in liquid  neon (LNe)  and liquid
argon (LAr). They have been realized  using a scaleable technology in order to
reach  increasing  sensitivities  in  the  different  prototypes  realized  and
installed  in  the SNOLAB  (Pico-CLEAN,  Micro-CLEAN,  DEAP-I, Mini-CLEAN  and
CLEAN/DEAP) with the aim to search for dark matter and to perform (through the
analysis of elastic neutrino-electron  and neutrino-nucleus scattering) a real
time measurement  of the  ${\it pp}$ solar  neutrino flux. The  final detector
CLEAN (Cryogenic Low Energy Astrophysics with Noble gases) \cite{14-CLEAN2004}
(see figure  \ref{fig-14:CLEAN}) will  be made by  a stainless steel  tank, of
about 6  meters of diameter,  filled with 100~tons of cryogenic  liquid neon;
only  the  central  part  of  it,  surrounded isotropically  by  a  series  of
photomultipliers, will  constitute the detector fiducial  volume.  An external
tank  of water,  10~metres  wide and  12~metres  high, will  act  as $\gamma$-ray
shielding,  neutron shielding  and muon  veto.  According  to  Monte Carlo
simulations, there should be a production of 15000 photons/MeV and it should be
possible  to reach  a  100\%  photon  wavelength shifter  efficiency and  a
statistical uncertainty on  the ${\it pp}$ measurements of the  order of $1 \,
\%$.

A precise  measurement of the  ${\it pp}$ component  and of the  ratio between
${\it pp}$  and $^7$Be fluxes would  be essential to test  the predictions of
SSMs. A high accuracy  on the ${\it pp}$ neutrino flux would
also make possible  a better determination of the  $\theta_{12}$ mixing angle,
which, complemented with the  results from previous solar neutrino experiments
and  from KamLAND  (essential for  the  $\Delta m_{12}^2$  measurement), would  
be fundamental to test the consistency of  the LMA solution also in the region of
transition between vacuum dominated and matter enhanced oscillations.  Finally,
CLEAN could  in principle try to  measure also the CNO  neutrino flux, through
the  analysis of  neutrino spectrum  from  0.7 to  1.0 MeV,  with an  estimated
accuracy between 10 and 15\%.

An  interesting alternative  to the  use of  neon is  offered by  liquid xenon
scintillator  detectors \cite{14-Aprile2009dv},  which take  advantage of the
fact that  among liquid rare  gases xenon has  the highest stopping  power for
penetrating radiation  (thanks to its high  atomic number, $A  \simeq 131$ and
density, $\rho = 3 {\rm g/cm^3}$)  and also the highest ionization and 
scintillation yield.  The  technological improvements of  the last twenty to 
thirty years made
possible significant  improvements in the cooling  and purification techniques
of  this kind  of detectors  and  in the  possibility of assembleing large mass
detectors, of  the order of some tons  (like in the case  of MEG \cite{14-MEG}
experiment, studying the $\mu \rightarrow e \gamma$ decay).\\

\begin{figure}
\begin{center}
\includegraphics[width=14.5cm,height=12.5cm,angle=0]{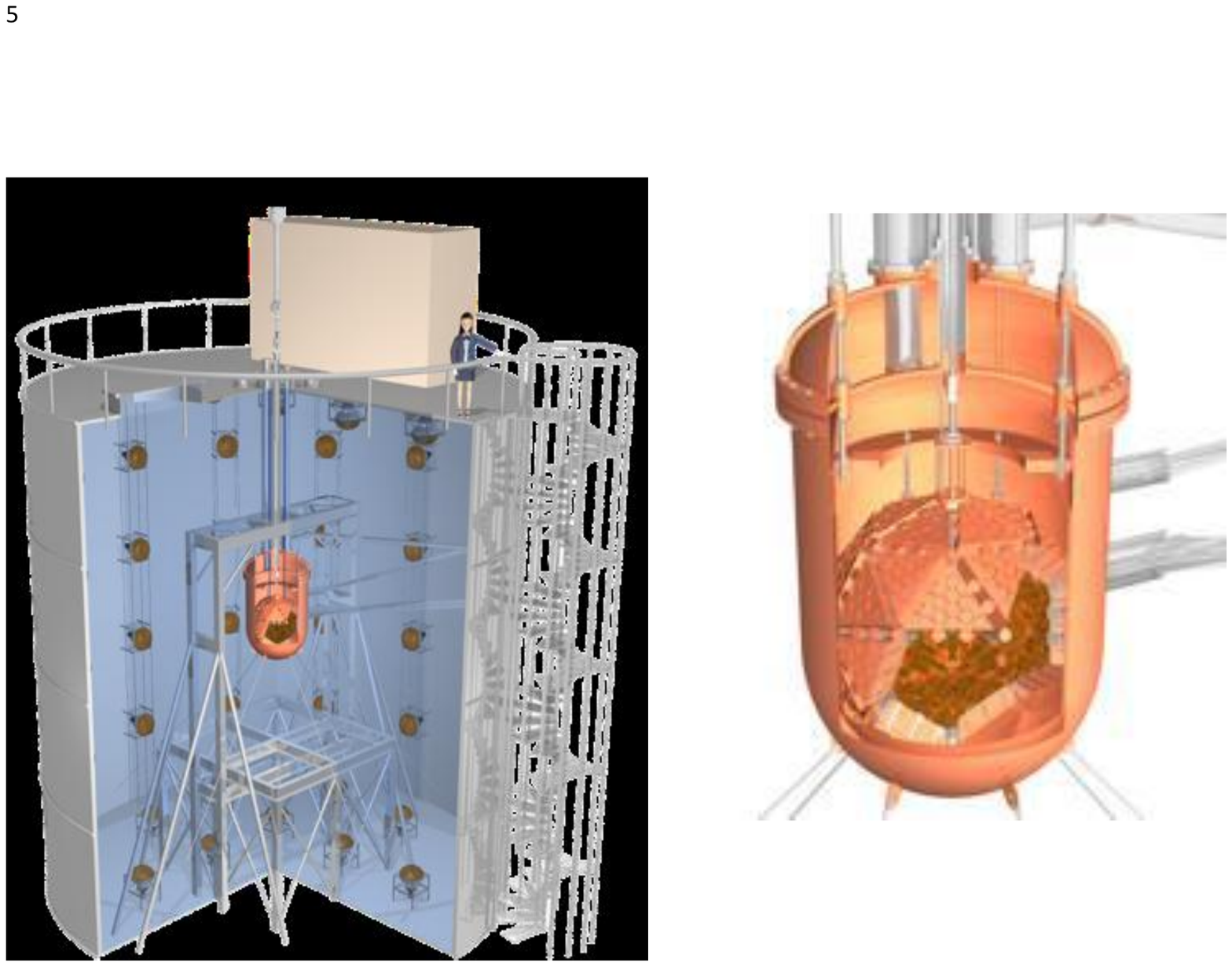}
\vspace{-4.2truecm}
\caption{Schematic view of the full XMASS facilty (left) and a detail of the inner
  detector (right panel), from which  one can see the particular configuration
  of the hexagonal photomultiplier tubes.  Taken from \cite{14-Moriyama2011zz}
  and \cite{14-XMASStalk}.
\label{fig-14:XMASS}}
\end{center}
\end{figure}

The  XMASS experiment (see  Figure \ref{fig-14:XMASS})  is a  multipurpose low
background and low  energy threshold experiment that will  use a large massive
liquid xenon  detector and  has been designed  to look for  WIMPs (dark  matter 
candidates),  search for  neutrinoless double
$\beta$ decay and study the ${\it  pp}$ and the $^7$Be solar neutrinos. After
two preliminary phases, during which smaller prototypes have been realized and
installed in the  Kamioka mine \cite{14-Moriyama2011zz}, and the  first data on
double beta  decay and dark  matter have been  taken, the full  XMASS detector
(that will  measure also solar neutrinos) will  have a total mass  of 20 tons,
with a  fiducial volume of  10 tons.  Special  efforts are required  mainly to
lower the background, by reducing  the radioactive contamination in the parts
used for detector construction (with special attention to the photomultipliers
and the copper material used for PMT holder), constructing a larger pure water
active shield  (for muons  and mainly neutrons  and $\gamma$ rays)  and, above
all,  developing  a distillation  system  for xenon  in  order  to reduce  the
contamination by $^{85}$Kr, the major source of radioactive background inside
the detector.

\begin{figure}[h!]
\begin{center}
\vspace{1cm}
\includegraphics[width=7.5cm,height=6.5cm,angle=0]{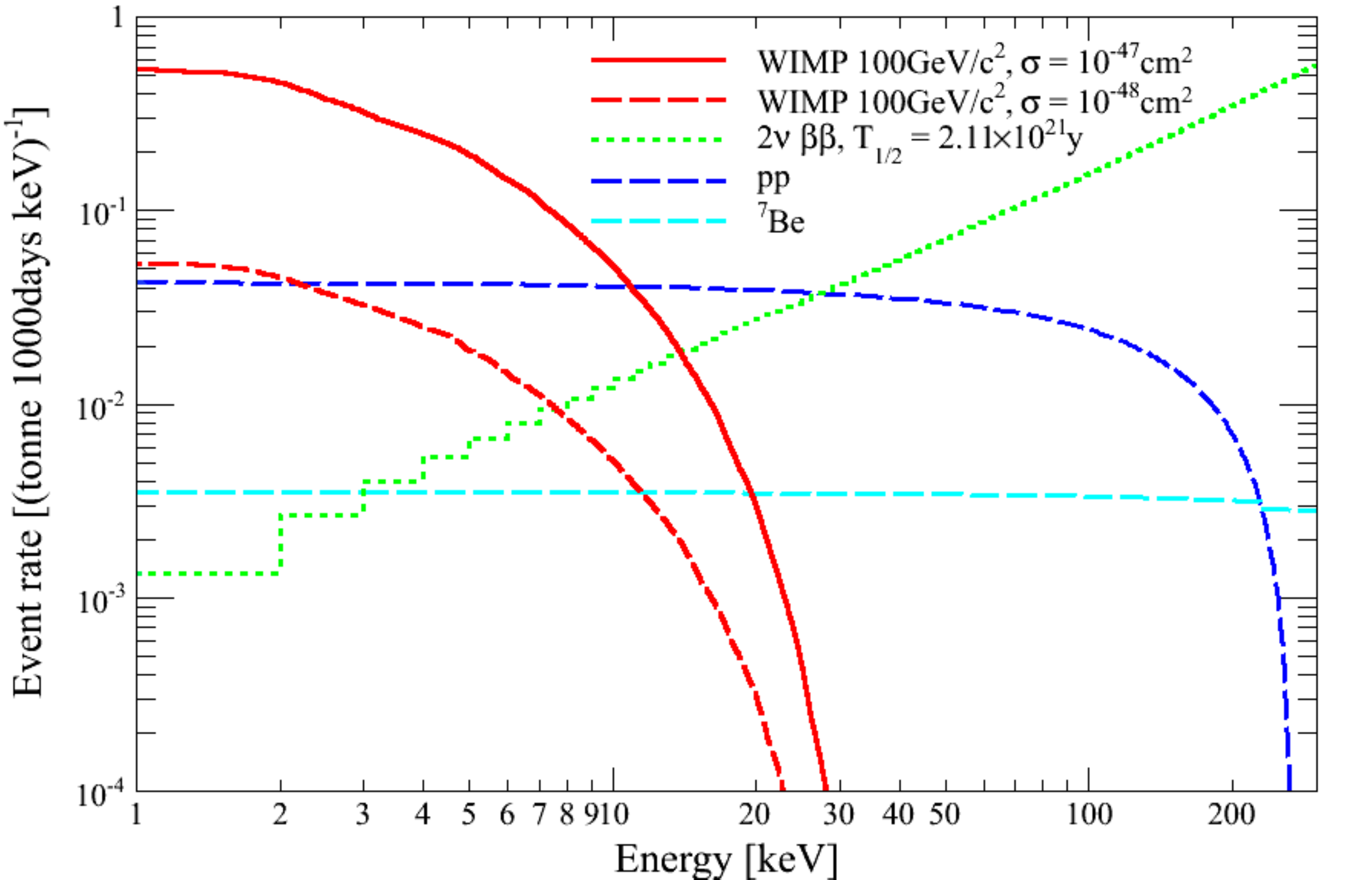}
\caption{Expected  nuclear recoil  spectrum from  WIMP scatters  in LXe  for a
  spin-independent  WIMP-nucleon  cross  section  of  10$^{-47}$\,cm$^2$  (red
  solid) and 10$^{-48}$\,cm$^2$ (red dashed) and a WIMP mass of 100\,GeV/c$^2$,
%%(using the standard halo model as in \cite{complementarity})
 along with  the differential  energy spectrum  for pp  (blue)  and $^{7}$Be
  (cyan)  neutrinos, and  the electron  recoil spectrum  from the  double beta
  decay of $^{136}$Xe (green).
%, assuming the natural abundance of 8.9\% and the recently measured half life 
%%of  2.1$\times$10$^{21}$\,yr \cite{exo_2011}.  
Assumptions are: 99.5\% discrimination  of electronic recoils, 50\% acceptance
of nuclear recoils, 80\% flat  analysis cuts acceptance. Taken from the second
paper of \cite{14-DARWIN}.
\label{fig-14:DARWIN}}
\end{center}
\end{figure}

Another  interesting experimental  project  based on  the  noble gases  liquid
scintillator technique is  that of DARWIN (DARk matter  WImp search with Noble
liquids)  \cite{14-DARWIN}, which  brings  together differen  European and  US
research groups working on existing experiments  and on the study for a future
multi-ton scale  LAr and LXe dark  matter search facility in  Europe. The main
goal of the experiment  is to look for a WIMP signal and to demonstrate its dark
matter nature,  taking advantage from  the fact of performing  the measurement
with multiple  different targets operating under similar  conditions.  In this
way, it  should be  possible to estimate  the dependence  of the rate  with the
target material  and, therefore, to  better determine the WIMP  candidate mass
and to distinguish between spin  independent and spin dependent couplings. The
energy region of the nuclear recoil  spectrum, below 200~keV, that should be
investigated by this future experiment  is of particular interest also for the
study of the ${\it  pp}$  solar neutrinos and, in  fact, the elastic scattering on
electrons by the  low energy component of the neutrino  spectrum would be one
of the main  background sources for WIMP searches in  liquid xenon detectors, as
shown in Figure~\ref{fig-14:DARWIN}.

DARWIN officially started in 2010; a technical design study should be ready in
Spring 2013 and the start of the first physics run is expected by mid 2017.

\subsubsection{Multi kiloton scale liquid scintillators: example LENA}\

\begin{figure}[h!!]
\begin{center}
%\vspace{1cm}
\includegraphics[width=8.8cm,height=8.0cm,angle=0]{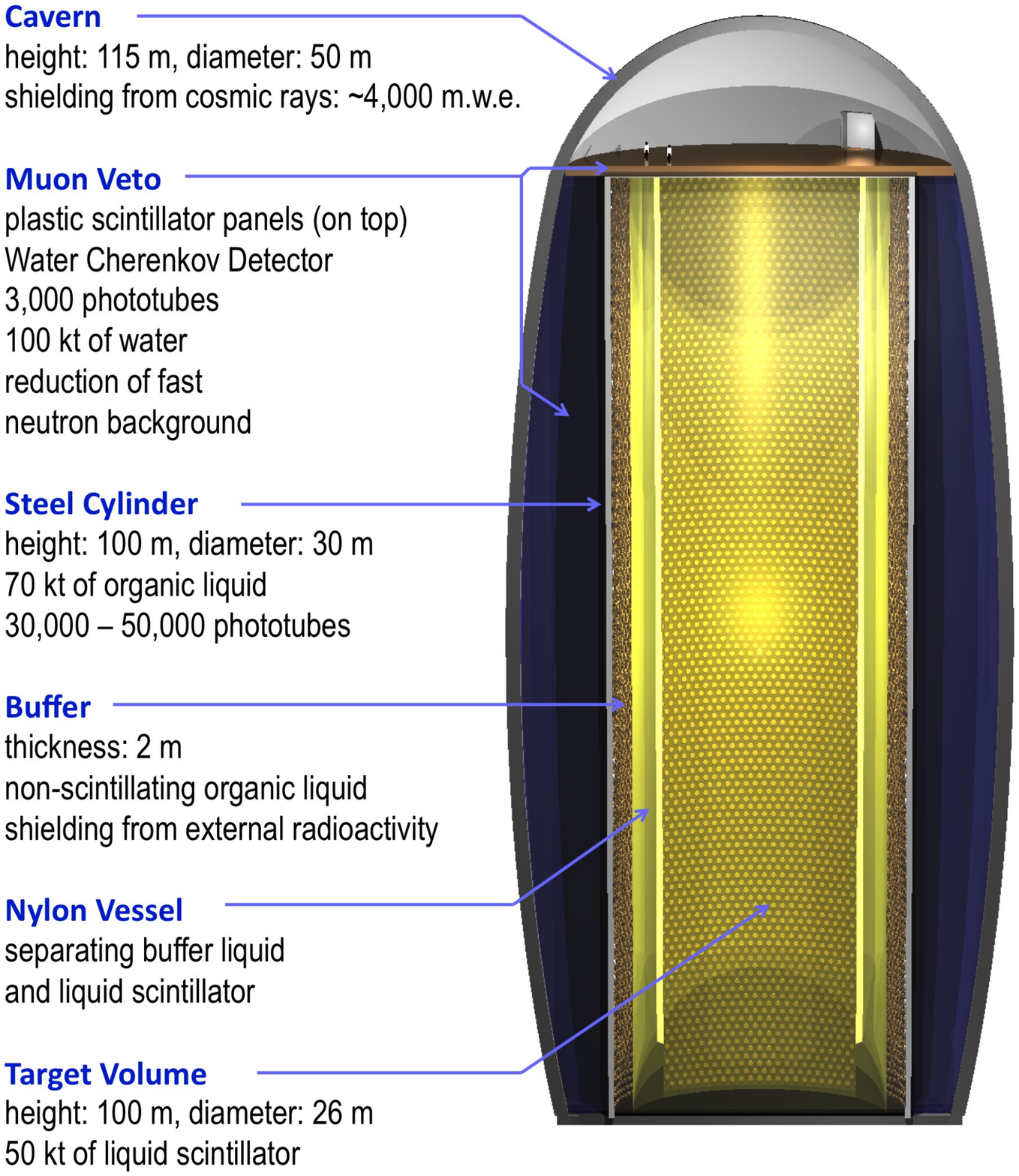}
\caption{Schematical view of the LENA detector. From  \cite{14-LENA}
\label{fig-14:16}}
\end{center}
\end{figure}

The Borexino    experiment    demonstrated   the    great    potential  of the
liquid-scintillator technique for the detection of low energy solar neutrinos.
Thanks to  this  experience,  a  next-generation neutrino  detector  has  been
proposed:  LENA (Low  Energy Neutrino  Astronomy) \cite{14-LENA}.   LENA  is a
multipurpose detector  aiming to study supernova  neutrinos, diffuse supernova
neutrino  background,  proton   decay,  atmospheric  neutrinos,  long-baseline
neutrino beams, geoneutrinos  and, last but not least,  solar neutrinos. The LENA
project foresees a cylindrical detector with a diameter of 30~m and a
length of about 100~m. Inside  the detector is foreseen an internal part (with
a diameter of about 26 m) containing about 50 kilotons of liquid scintillator,
separated from a non-scintillating buffer region by a nylon barrier.  Outside,
a tank (made in steel or  concrete) separates the inner detector from an outer
water tank;  it is used  both for  shielding and as  an active muon  veto.  To
collect  the  scintillation  light,  about  45,000  photomultipliers  (with  a
diameter  of 20 cm)  are mounted  to the  internal walls  of the  detector. To
increase  the optically active  area, the photomultipliers tubes  are equipped
with conic mirrors, the corresponding surface coverage is about 30\%.  Figure
(\ref{fig-14:16}) shows a schematic overview of the current LENA design.

Among the favored solvent for the liquid scintillator in LENA, the LAB (linear
alkylbenzene) is  currently the preferred one.  It has a high  light yield and
large attenuation length  and it has also the advantage  of being a non-hazardous
liquid. The attenuation lengths is on the order of  10 to 20~m (at a wavelength of 
430~nm) and the photoelectron yield  could be greater than 200 photoelectrons per
MeV (with  a scintillator mixture  containing 2g/l PPO  and 20 mg/l  bisMSB as
wavelength shifters).  Studies have been  carried out to test  the large-scale
light  transport and the  differences in  scintillator response  for $\alpha$,
$\beta$ and $\gamma$ particles. An  alternative solvent  option is  the well  studied 
PXE  \cite{14-PXE}  or a mixture of PXE and dodecane.

As already pointed out, Borexino  has splendidly demonstrated the potential of
the detection  technique  with  liquid  scintillator  based  detectors  for  solar
neutrino detection.   This technique offers the opportunity  for a spectrally
resolved measurement of the solar neutrino spectrum in the all energy range.

Because  the smaller  ratio  of surface  to  volume compared  to the  Borexino
detector\footnote{A smaller  ratio of surface to volume  decreases the chance
  that the scintillator is contaminated with radioimpurities}, in LENA it is
very likely to  reach the excellent background conditions  of Borexino.  Monte
Carlo  simulations of the  gamma background  due to  the uranium,  thorium and
potassium from the photomultipliers glass  shows that a fiducial volume of the
order of 30~ktons is achievable  for solar neutrino studies; LENA will be able
to address topics both in neutrino oscillations and in solar physics thanks to
its unprecedented statistics. A high statistics can be obtained in short times
and in both  Pyhsalmi and Frejus underground laboratories,  where the detector
could be hosted, where the cosmogenic background of $^{11}$C will be significantly
lower than in Borexino.

Monte  Carlo simulations show  that for  {\it pep},  CNO  and low-energy
$^{8}$B-$\nu$s detection  a fiducial mass  of $\sim$30\,kton is  
necessary, while the   fiducial   mass  for   $^{7}$Be-$\nu$s   and  high-energy   
($E>5$\,MeV) $^{8}$B-$\nu$s could be enlarged to 35\,kton or more.

In Table~\ref{tab:ratesinLENA}  are reported the expected rates  in 30\,kton for
the neutrinos emitted in the {\it pp} chain and the CNO-bicycle, using the
most recent  solar model  predictions. This evaluation  refers to  a detection
threshold set at about 250\,keV.
\begin{table}
\begin{tabular}{lcccr}
\hline Source & EW  [MeV] & $m_\textrm{fid}$ [kt] & Rate [cpd]  \\ \hline pp &
$>$0.25 & 30  & 40 \\ pep &  0.8$-$1.4 & 30 & 2.8$\times$10$^2$  \\ $^{7}$Be &
$>$0.25  & 35 &  1.0$\times$10$^4$ \\  $^{8}$B &  $>$2.8 &  35 &  79 \\  CNO &
0.8$-$1.4 & 30 & 1.9$\times$10$^2$ \\ \hline
\end{tabular}
\caption{Expected solar neutrino rates in LENA (channel $\nu e\to e\nu$).  The
  estimates    are    derived   from    the    existing   Borexino    analyses
  \cite{14-Arpesella2008mt,14-Bellini2008mr} as well as expectation values for
  the     respective      energy     windows     (EW) of     observation
  \cite{14-wur09phd,14-Dangelo2006,14-Ianni2005ki}. The quoted fiducial masses,
  $m_\textrm{fid}$,  in LAB  are based  on a  Monte Carlo  simulation  of the
  external  $\gamma$-ray background  in  LENA. Table  taken  and adapted  from
  \cite{14-LENA}. \label{tab:ratesinLENA}}
\end{table}

\subsubsection{New techniques with organic scintillators: LENS}\

The main goal  of the Low Energy Neutrino Spectroscopy  (LENS) detector is the
real  time measurement  of  solar neutrinos  as  a function  of their  energy,
focusing, in  particular, in  the analysis  of the  lowest  energy neutrinos
coming  from proton-proton  fusion  (i.e.  the  {$\it  pp$} neutrinos),  which
represent the main  contribution and the less known  component of the pp-chain
of fusion reactions inside the Sun.

In order to make an energy spectrum measurement on low energy neutrinos, it is
necessary to reach a low threshold for the charged current (CC) process and to
be  able to  discriminate  the  background from  radioactive  decays.  The  CC
process employed in  LENS is the neutrino induced  transition of $^{115}$In to
an excited state of $^{115}$Sn:

\begin{equation}
	\nu_{e} + ^{115}{\rm In} \rightarrow ^{115}{\rm Sn}^* + e^{-}  \ ({\rm E=E_\nu -114 keV})\, .
	\label{eq:lens1}
\end{equation}

\begin{equation}
	^{115}{\rm Sn}^*  (\tau = 4.76 \mu s) \rightarrow ^{115}{\rm Sn + \gamma(498 keV) + \gamma(116 keV)}\, .
	\label{eq:lens2}
\end{equation}

Thanks to that it is possible  to detect low energy neutrinos with a threshold
of  114  keV  and measure  their  energy,  following  an  idea that  has  been
investigated since the 1970's \cite{14-Raghavan1976yc}.

The primary interaction and the secondary cascade enable a triple coincidence,
correlated  in space and  time.  LENS  employs as  detection medium  a liquid
scintillator chemically  doped with natural indium ($^{115}$In  = 95.7\%).  In
order  to exploit  the  spatial correlation,  the  volume of  the detector  is
segmented into cubic cells (7.5 cm)  by clear foils (Teflon FEP) that have a
lower  index of  refraction than  the  liquid scintillator.  By  internal
reflection, the  scintillation light  produced in a  cell is channeled  in the
directions of the 6 cell faces.   The collected channeled light is read-out at
the edge of the detector by photomultiplier tubes.

LENS  should be able to  determine the  low energy  solar neutrino
fluxes with an accuracy $\leq 4 \%$, testing neutrino and solar physics with a
global  precision  better  than the  present  one  and  also looking  for  any
inconsistency in the LMA conversion mechanism \cite{14-Raghavan2004}.
   
\section{Open questions in solar neutrino physics}\ 

\subsection{The metallicity problem}\label{14-sec:metalproblem}

The solar abundance or solar metallicity problem has been around for some time now.
In  analogy  with the  solar  neutrino  problem,  there have  been  attempts
(although in most  cases, it is fair to say, of  somewhat less radical nature)
to  solve it  by introducing  modifications to  the input  physics of
SSMs.    To    mention   a   few of them, we could remember:\\   
- large   enhancement    of   Ne   abundance
\cite{14-antia:2005, 14-bahcall:2005},  important because of  its contribution
uncertainty  and a  weak bond  in solar  abundances because  its  abundance is
determined   rather  indirectly   \cite{14-lodders:2009};\\   
- increased  element
diffusion   rates   \cite{14-montalban:2004,   14-guzik:2005};\\  
- accretion   of
metal-poor  material  leading  to  a   `two-zone'  solar  model  in  terms  of
composition  \cite{14-guzik:2005,  14-castro:2007, 14-serenelli:2011}.\\   
Also solar models including some sort  of prescriptions to account for rotation and
other dynamical  effects have been  put forward, however their  performance is
quite poor.

So far, all  attempts of  finding a  solution to  all the manifestations  of 
the solar abundance problem have failed. In some cases $\ys$ can be brougth into
agreement  with helioseismology,  in other  cases $\rcz$  and the  sound speed
profile, but a simultaneous  solution to all the
problems has not yet being found.

The exceptions  are the two obvious  one: a) the {\em  low-Z} solar abundances
actually underestimate  the true metal content  of the Sun; b)  an increase of
radiative  opacities by the  right amount  (15\% to  20\% at  the base  of the
convective zone  down to about  3\% in the  solar core) to compensate  for the
decrease induced by  the {\em low-Z} abundances. The drawback  to this idea is
that current state-of-the-art radiative  opacity calculations differ by only 2
to 3\% at the  base of the convective envelope, much lower  from what would be
required by {\em low-Z} models. 

It  has  indeed been  shown that by increasing the radiative opacity  in  
{\rm  low-Z} SSMs  the
agreement  with helioseismology  can be  restored to  match results  from {\em
  high-Z} SSMs \cite{14-jcd:2009}. Addtionally, the $^7$Be and $^8$B fluxes of
a {\em  low-Z} SSMs  with increased  opacitiy coincide with  those from  a {\em
  high-Z} model \cite{14-serenelli:2010}.  As good  as this may seem, it shows
the intrinsic degeneracy between composition and opacities.

Recently, a novel approach, the Linear Solar Models (LSM), that 
relate changes in solar observables to modifications in the input physics by 
the calculation of kernels based on SSMs has been developed by 
\cite{14-villante:2010a}. LSMs\footnote{We remark that LSMs offer an efficient 
way of studying the response of the solar 
structure to changes in any of the physical ingredients entering solar model 
calculations that does not require the construction of solar models with the varied 
physics.} have been applied in particular to the solar 
abundance problem and the changes required in the radiative opacity to restore
the agreement between {\it low-Z} models and helioseismology 
\cite{14-villante:2010b}. Quantitively, results similar to those quoted above. 

By  using   $^8$B  and   now  $^7$Be  as   thermometers  of  the   solar  core
\cite{14-haxton:2008},  CNO neutrinos  represent a  unique way  to  break this
degeneracy and provide an independent  determination of the CNO abundances,
particularly the C+N  abundance in  the solar core.  Keeping in mind  the 
antagonism
between solar interior  and solar atmosphere models that  the solar abundance
problem  has established,  results from  CNO fluxes  will be  of  the outmost
relevance for solar, and by extension stellar, physics.

\subsection{The vacuum to matter transition?}

Solar neutrino experiments already measured the two extreme flavor conversion regimes, 
the vacuum term domination and matter term domination. There is no direct experimental evidence 
of the transition from one to the other. In fact, the lower energetic $^8$B neutrinos are sensitive 
to the rise of the spectrum from matter combination towards vacuum, but the data 
(still very uncertain) seems not to show it. More data coming from Super-Kamiokande, Borexino 
and SNO+ experiments will further explore the conversion in this regime.

The precise measurement of low energy neutrinos like pep, exploiting the fact that are more energetic than 
$^7$Be neutrinos, will also help to see small solar matter effects in the flavor conversion. This matter effects 
will be more precisely determined by the comparison of pep and pp neutrino measurements. In fact, the 
low energy neutrinos that are better suited to test matter effects are the CNO neutrinos. While the CNO neutrinos 
energy is around the pep neutrinos energy, the former are produced at higher temperatures and therefore 
at higher densities. The larger matter density where neutrinos are produced  
leads to larger matter effects 
for CNO neutrinos than for pep neutrinos. In fact, matter effects produce a significant spectral tilt of CNO 
neutrinos ($ \sim 10 \%$), which might be a good handle to separate the signal for background.

The determination of the vacuum to matter transition has a significant impact on the determination of the 
solar mass splitting derived by solar data, which adds to the implications of earth matter effects measured by 
comparing the neutrino fluxes during the day and night. The good match of the independently determined solar
 mass splitting by solar neutrino experiments and by reactor experiments leads to the best test on 
 non-standard neutrino physics to solar neutrinos. There are many possibilities but the two scenarios more 
 studied are the addition of new neutral current interactions \cite{14-Friedland:2004pp,14-Barranco:2007ej} which 
 modify the amplitude of matter effects and therefore 
 shift the effective mass splitting and the existence of a sterile neutrino which adds a new state with 
 the appropriate mass splitting \cite{14-deHolanda2010am}  to produce deviations of the flavor conversion 
 in the 1-3 MeV range.

%\subsection{Study of CNO neutrinos}\

\subsection{What else can we learn from CNO fluxes?}

The most fundamental information the CNO fluxes carry is the most obvious one:
that the CNO-bicycle operates in stars and it is a viable process for hydrogen
fusion. It must  not be forgotten that neutrinos are  the only direct evidence
of nuclear reactions being the  source of energy in solar (stellar) interiors.
For the Sun, models predict a marginal contribution to the total energy budget
from CNO  reactions, 0.8\% and  0.4\% for {\em  high-Z} and {\em  low-Z} solar
models. However, CNO  becomes the dominant mode for  hydrogen burning in stars
with masses right above the solar value.  Detection of CNO neutrinos will provide 
direct evidence that CNO reactions actually take place in nature, as originally 
envisioned by Hans Bethe \cite{14-bethe:1939}; it has been a long wait. 

The second important  aspect of CNO neutrinos is  the information they provide
about the abundance of metals in  the solar core. Knowing the abundance of CNO
elements in the solar core is  important by itself. In particular, a `perfect'
measurement  of   the  combined  $^{13}$N+$^{15}$O  flux   translates  into  a
determination  of  the  solar  C+N  abundance  with  $\sim  10\%$  uncertainty
\cite{14-haxton:2008},   and  the  dominant   sources  of   uncertainties  are
experimental  and can  be  potentially  reduced. Assuming  we  know the  solar
surface abundance of the same elements,  i.e. let us forget for the time being
about the solar abundance  problem, we can  then put constraints  on mixing
mechanisms that may have created composition gradients during the evolution of
the Sun. SSMs predict that the number  density of C+N is enhanced in the solar
core, at  present-day, by $\sim 16\%$ with  respect to the surface  due to the
effects of  microscopic diffusion. And, although helioseismology  shows that models
with  diffusion work  much  better than  models  without, there  is no  direct
evidence of how  efficient diffusion is. In fact,  there have been suggestions
that the  standard prescription \cite{14-thoul:1994}  may be too  efficient in
the Sun \cite{14-pinsonneault:2010} and that diffusion rates should be lowered
by $\sim 15  \%$. Solar CNO neutrinos could provide a  test for the efficiency
of diffusion.

There are other  possibilities that might create a  contrast between the solar
core and surface  composition. Recently, it has been shown that  the Sun has a
peculiar  composition when  compared to  `solar twins',  that is  stars almost
identical  to the  Sun  in their  surface properties  \cite{14-melendez:2009,
  14-ramirez:2010}.  The authors  found that  the Sun  is enhanced  in volatile
elements  with respect  to the  solar twins  that show  no sign  of harbouring
planets  by about  20\%.   In fact,  they  have associated  this  fact to  the
presence of rocky  planets in the Solar System,  where refractory elements are
locked,   and  the occurrence of   an   accretion   episode    of   volatile-enriched   
material
\cite{14-ramirez:2010} after rocky cores are formed in the protoplanetary disk.  If 
this  were true,  then the  Sun would  have an
envelope that  is richer  in CNO than  its interior.  If a measurement  of the
$^{13}$N+$^{15}$O flux  would yield as a  result a core  composition where the
abudance of  C+N would be comparable or  less than the surface  value, then we
would have an extremely exciting piece of evidence about the earlier phases of
planet formation in the solar system \cite{14-haxton:2008}.


\begin{thebibliography}{1}

\bibitem{14-Davis1968cp}
J.~N.~Bahcall, Phys.\ Rev.\ Lett.\  {\bf 12} (1964) 300; 
R.~J.~Davis, Phys.\ Rev.\ Lett.\  {\bf 12} (1964) 302; 
R.~J.~Davis, D.~S.~Harmer and K.~C.~Hoffman,
%``Search for neutrinos from the sun,''
Phys.\ Rev.\ Lett.\  {\bf 20} (1968) 1205.
%%CITATION = PRLTA,20,1205;%%
%
\bibitem{14-Pontecorvo}
B. Pontecorvo, J. Exptl. Theoret. Phys. 33 (1957) 549 [Sov. Phys. JETP 6 
(1958) 429]; B. Pontecorvo J. Exptl. Theoret. Phys. 34 (1958) 247 [Sov. Phys. 
JETP 7 (1958) 172].
%
\bibitem{14-Kamiokande89}
K.~S.~Hirata {\it et al.}  [KAMIOKANDE-II Collaboration],
%``Observation of B-8 Solar Neutrinos in the Kamiokande-II Detector,''
Phys.\ Rev.\ Lett.\  {\bf 63} (1989) 16.
%%CITATION = PRLTA,63,16;%%  
\bibitem{14-SK99}
Y.~Fukuda {\it et al.}  [Super-Kamiokande Collaboration],
%``Measurement of the solar neutrino energy spectrum using neutrino  electron
%scattering,''
Phys.\ Rev.\ Lett.\  {\bf 82} (1999) 2430.
%%CITATION = PRLTA,82,2430;%%
\bibitem{14-Hamp99} W. Hampel et al., Phys. Lett. B 447 (1999) 127.
\bibitem{14-Alt05} M. Altmann et al., Phys. Lett. B 616 (2005) 174.
\bibitem{14-Abd99} J.N. Abdurashitov et al., Phys. Rev. Lett. 83 (1999) 4686.
\bibitem{14-SNOES} 
%\cite{Ahmad:2001an}
%\bibitem{Ahmad:2001an}
Q.~R.~Ahmad {\it et al.}  [SNO Collaboration],
Phys.\ Rev.\ Lett.\  {\bf 87} (2001) 071301.
%%CITATION = PRLTA,87,071301;%%
%
\bibitem{14-SK2001}
S.~Fukuda {\it et al.}  [Super-Kamiokande Collaboration],
%``Solar 8B and hep Neutrino Measurements from 1258 Days of Super-Kamiokande
%Data,''
Phys.\ Rev.\ Lett.\  {\bf 86} (2001) 5651.
%%CITATION = PRLTA,86,5651;%%
\bibitem{14-Global2001}
%\cite{Bahcall:2001zu}
J.~N.~Bahcall, M.~C.~Gonzalez-Garcia and C.~Pena-Garay,
%``Global analysis of solar neutrino oscillations including SNO CC
%measurement,''
JHEP {\bf 0108} (2001) 014;
%%CITATION = JHEPA,0108,014;%%
%\cite{Fogli:2001vr}
%\bibitem{Fogli:2001vr}
G.~L.~Fogli, E.~Lisi, D.~Montanino and A.~Palazzo,
%``Model-dependent and independent implications of the first Sudbury  Neutrino
%Observatory results,''
Phys.\ Rev.\  D {\bf 64} (2001) 093007.
%%CITATION = PHRVA,D64,093007;%%
\bibitem{14-SNOI}
%\cite{Ahmad:2002jz}
%\bibitem{Ahmad:2002jz}
Q.~R.~Ahmad {\it et al.}  [SNO Collaboration],
Phys.\ Rev.\ Lett.\  {\bf 89} (2002) 011301.
%%CITATION = PRLTA,89,011301;%%
%
\bibitem{14-SNOII}
S.~N.~Ahmed {\it et al.}  [SNO Collaboration],
%``Measurement of the total active B-8 solar neutrino flux at the Sudbury
%Neutrino Observatory with enhanced neutral current sensitivity,''
Phys.\ Rev.\ Lett.\  {\bf 92} (2004) 181301.
%%CITATION = PRLTA,92,181301;%%
%
\bibitem{14-SNOIII} 
B.~Aharmim {\it et al.}  [SNO Collaboration],
%``An Independent Measurement of the Total Active 8B Solar Neutrino Flux Using
%an Array of 3He Proportional Counters at the Sudbury Neutrino Observatory,''
Phys.\ Rev.\ Lett.\  {\bf 101} (2008) 111301.
%%CITATION = PRLTA,101,111301;%%
%
\bibitem{14-Ahmad2002ka}
Q.~R.~Ahmad {\it et al.}  [SNO Collaboration],
Phys.\ Rev.\ Lett.\  {\bf 89} (2002)  011302.
%%CITATION = PRLTA,89,011302;%%

%
\bibitem{14-Davis2002fb}
R.~Davis,
%``Memories of a Nobel laureate,''
CERN Cour.\  {\bf 42N10} (2002) 15.
%%CITATION = CECOA,42N10,15;%%
%
\bibitem{14-firstKL}
K.~Eguchi {\it et al.}  [KamLAND Collaboration],
Phys.\ Rev.\ Letts.\  {\bf 90} (2003) 021802.
%%CITATION = PRLTA,90,021802;%%
%
\bibitem{14-bahcall:2001}  J.~N.   Bahcall,  M.  Pinsonneault   and  S.  Basu,
  Astroph. J. {\bf 555} (2001) 990. 

\bibitem{14-SK_KLpotentialities}
A.~Piepke  [KamLAND Collaboration],
Nucl.\ Phys.\ Proc.\ Suppl.\  {\bf 91} (2001) 99;
%%CITATION = NUPHZ,91,99;%%
%
A.~de Gouvea and C.~Pena-Garay,
%``Solving the solar neutrino puzzle with KamLAND and solar data,''
Phys.\ Rev.\  D {\bf 64} (2001) 113011;
%%CITATION = PHRVA,D64,113011;%%
P.~Aliani, V.~Antonelli, M.~Picariello and E.~Torrente-Lujan,
New J.\ Phys.\  {\bf 5} (2003) 2.
%%CITATION = NJOPF,5,2;%%
%
\bibitem{14-KL2004}
T.~Araki {\it et al.}  [KamLAND Collaboration],
Phys.\ Rev.\ Lett.\  {\bf 94} (2005) 081801.
%%CITATION = PRLTA,94,081801;%%
%
\bibitem{14-KLfollowing2}
S.~Abe {\it et al.}  [KamLAND Collaboration],
Phys.\ Rev.\ Lett.\  {\bf 100} (2008) 221803.
%%CITATION = PRLTA,100,221803;%%
%
% Reference non utilizzata
%\bibitem{14-Ackar}
%B.~Achkar {\it et al.},
%``Comparison of anti-neutrino reactor spectrum models with the Bugey-3
%measurements,''
%Phys.\ Lett.\  B {\bf 374} (1996) 243.
%%CITATION = PHLTA,B374,243;%%
\bibitem{14-KLfollowing3}
A.~Gando {\it et al.}  [The KamLAND Collaboration],
%``Constraints on $\theta_{13}$ from A Three-Flavor Oscillation Analysis of
%Reactor Antineutrinos at KamLAND,''
Phys.\ Rev.\  D {\bf 83} (2011) 052002. 
%%CITATION = PHRVA,D83,052002;%%
%%
%%%% References by Aldo

\bibitem{14-bahcall:1982} J.~N. Bahcall, W.~F. Huebner, S.~H. Lubow, P.~D.
 Parker and R.~K. Ulrich, Rev. of Modern Phys. {\bf 54} (1982) 767.

\bibitem{14-bahcall:1989} J.~N. Bahcall, {\it Neutrino Astrophysics}, 
(Cambridge Univ. Press, 1989). 

\bibitem{14-lodders:2009} K. Lodders, H. Palme and H.-P. Gail, in {\it
  Landolt-B{\"o}rnstein - Group VI Astronomy and Astrophysics Numerical Data 
  and Functional  Relationships in Science  and Technology} {\bf  44}, (2009),
  arXiv:0901.1149 

\bibitem{14-beeck:2012}  B. Beeck  {\it et  al.}, Astronom.  \&  Astroph. {\bf
  539} (2012) 121.

\bibitem{14-asplund:2009} M. Asplund, N. Grevesse, A.~J. Sauval and 
  P. Scott, Ann.\ Rev.\ Astron.\ Astroph. {\bf 47} (2009) 481.
 
\bibitem{14-asplund:2005}  M. Asplund,  N.  Grevesse and  J.  Sauval, in  {\it
  Cosmic  Abundances as  Records  of Stellar  Evolution and  Nucleosynthesis},
  eds. T.~G. Barnes III and F.~N. Bash, ASPC {\bf 336} (2005) 25.

\bibitem{14-allendeprieto:2001}   C.  Allende   Prieto,   D.~L.  Lambert   and
  M. Asplund, Astroph. J. Lett. {\bf 556}, L63 (2001)

\bibitem{14-grevesse:1993}  N.  Grevesse and  A.  Noels,  in  {\it Origin  and
  Evolution of Elements}, eds. N.  Prantzos, E. Vangioni-Flam and M. Casse 
 (1993) 15.

\bibitem{14-grevesse:1998} N.  Grevesse and  J. Sauval, Space  Science Reviews
  {\bf 85} (1998) 161.

\bibitem{14-caffau:2011} E.  Caffau, H.~G. Ludwig, M. Steffen,  B. Freytag and
  P. Bonifacio, Sol.\ Phys. {\bf 268} (2011) 255.

\bibitem{14-iglesias:1996} C.~A.  Iglesias and F.~J. Rogers,  Astroph. J. {\bf
  464} (1996) 943.

\bibitem{14-badnell:2005} N.~R. Badnell {\it et al.}, Monthly Notices of Royal
  Astron. Soc. {\bf 360} (2005) 458.
  
\bibitem{14-ferguson:2005} J.~W. Ferguson {\it et al.}, Astroph. J. {\bf 623} 
 (2005) 585.

\bibitem{14-adelberger:2011}     E.~G.     Adelberger     {\it    et     al.},
  Rev.\ Mod.\ Phys. {\bf 83} (2011) 195.

\bibitem{14-adelberger:1998} E.~G. Adelberger {\it et al.}, 
  Rev.\ Mod.\ Phys. {\bf 70} (1998) 1265.

\bibitem{14-jcd:2002} J.  Christensen Dalsgaard,  Rev.\ Mod.\ Phys.  {\bf 74}
 (2002) 1073. 

\bibitem{14-jcd:1996}  J. Christensen  Dalsgaard  {\it et  al.}, Science  {\bf
  272} (1996) 1286.

\bibitem{14-basu:1997a}  S. Basu  and H.  Antia, Monthly  Notices of  the Royal
  Astron. Soc. {\bf 287} (1997) 189.

\bibitem{14-basu:2004} S. Basu and H.  Antia, Astroph. J. Lett. {\bf 606}
  (2004)  85L.

\bibitem{14-basu:1997b} S. Basu {\it et al.},  Monthly  Notices of  the Royal
  Astron. Soc. {\bf 292} (1997) 243.

\bibitem{14-kosovichev:1997} A.~G.  Kosovichev {\it  et al.}, Sol.  Phys. {\bf
  170} (1997) 43.

\bibitem{14-basu:2009}  S.  Basu,  W.~J.  Chaplin,  Y. Elsworth,  R.  New  and
  A. M. Serenelli, Astroph. J. {\bf 699} (2009) 1403.

\bibitem{14-montalban:2004}   J.  Montalb\'an,   A.  Miglio,   A.   Noels  and
  N. Grevesse, in {\it SOHO 14 Helio- and Asteroseismology: Towards a Golden
    Future}, ESA Special Publication {\bf 559} (2004) 574.

\bibitem{14-turckchieze:2004}    S.      Turck-Chi\`eze    {\it    et    al.},
  Phys. Rev. Lett. {\bf 93} (2004)  211102. 

\bibitem{14-bahcall:2005a}  J.~N.  Bahcall, S.  Basu,  M.~H. Pinsonneault  and
  A. M. Serenelli, Astroph. J. {\bf 618} (2005)  1049. 

\bibitem{14-delahaye:2006} F.  Delahaye and M. Pinsonneault,  Astroph. J. {\bf
  649} (2006) 529.
\bibitem{14-serenelli:2011} A.~M. Serenelli, W.~C. Haxton and C. Pe\~na-Garay,
%``Solar models with accretion. I. Application to the solar abundance
%problem,'' 
Astrophys.\ J.\ {\bf 743} (2011) 24.
%%CITATION = ASJOA,743,24;%%
%
\bibitem{14-serenelli:2009}  A.~M.   Serenelli,  S.  Basu,   J.  Ferguson  and
  M. Asplund, Astroph. J. Lett. {\bf 705} (2009) L123.

\bibitem{14-antia:2006} H.~M. Antia  and S. Basu, Astroph. J.  {\bf 644} 
(2006) 1292. 

\bibitem{14-chaplin:2007}   W.~J.   Chaplin,   A.~M.   Serenelli,   S.   Basu,
  Y. Elsworth, R. New and G.~A. Verner, Astroph. J. {\bf 670} (2007) 872. 

\bibitem{14-roxburgh:2003}   I.~W.  Roxburgh,   I.~W.  and   S.~V.  Vorontsov,
  Astron. \& Astroph. {\bf 411} (2003) 215. 

\bibitem{14-clayton:1984}  D.~D.  Clayton,   in  {\em  Principles  of  Stellar
  Evolution and Nucleosynthesis}, University of Chicago Press (1984).


\bibitem{14-bahcall:2004}    J.~N.    Bahcall    and    M.~H.    Pinsonneault,
  Phys. Rev. Lett. {\bf 92} (2004) 121301. 


\bibitem{14-formicola:2004}  A. Formicola {\it  et al.}  (LUNA Collaboration),
  Phys.  Lett.  B {\bf 591} (2004) 61.

\bibitem{14-marta:2008}   M.  Marta   {\it  et   al.}   (LUNA  Collaboration),
  Phys. Rev. C {\bf 78} (2008) 022802. 

\bibitem{14-haxton:2008} W.~C.  Haxton and  A.~M. Serenelli, Astroph.  J. {\bf
  687} (2008) 678.

\bibitem{14-pepBX} G.~Bellini et al. (Borexino Collaboration), Phys. Rev. Lett. {\bf 
108} (2012) 051302.
%

\bibitem{14-bahcall:2006}  J.~N.   Bahcall,  A.~M.  Serenelli   and  S.  Basu,
  Astroph. J. Suppl. Series {\bf 165} (2006) 400.

\bibitem{14-bahcall:2005c} J.~N. Bahcall and A.~M. Serenelli, Astroph. J. {\bf
  626} (2005) 530.

%%%% End References by Aldo

\bibitem{14-afterSNO2002} 
P.~C.~de Holanda and A.~Y.~Smirnov,
%``Solar neutrinos: Global analysis with day and night spectra from SNO,''
Phys.\ Rev.\  D {\bf 66} (2002) 113005
%%CITATION = PHRVA,D66,113005;%%
and
%``LMA MSW solution of the solar neutrino problem and first KamLAND %results,''
JCAP {\bf 0302} (2003) 001;
%%CITATION = HEP-PH/0212270;%%
J.~N.~Bahcall, M.~C.~Gonzalez-Garcia and C.~Pena-Garay,
%``Before and after: How has the SNO neutral current measurement changed
%things?,''
JHEP {\bf 0207} (2002) 054 and 
%%CITATION = JHEPA,0207,054;%%
%``Solar neutrinos before and after KamLAND,''
JHEP {\bf 0302} (2003) 009;
%%CITATION = HEP-PH/0212147;%%; 
G.~L.~Fogli, E.~Lisi, A.~Marrone, D.~Montanino, A.~Palazzo and A.~M.~Rotunno,
%``Solar neutrino oscillation parameters after first KamLAND results,''
Phys.\ Rev.\ D {\bf 67} (2003) 073002;
%%CITATION = HEP-PH/0212127;%%
S.~Pascoli and S.~T.~Petcov,
%``The SNO solar neutrino data, neutrinoless double-beta decay and %neutrino
%mass spectrum,''
Phys.\ Lett.\  B {\bf 544} (2002) 239;
%%CITATION = PHLTA,B544,239;%%
M.~Maltoni, T.~Schwetz and J.~W.~F.~Valle,
%``Combining first KamLAND results with solar neutrino data,''
Phys.\ Rev.\ D {\bf 67} (2003) 093003;
%%CITATION = HEP-PH/0212129;%%
P.~Aliani, V.~Antonelli, M.~Picariello and E.~Torrente-Lujan,
%``Neutrino mass parameters from Kamland, SNO and other solar evidence,''
Phys.\ Rev.\ D {\bf 69} (2004) 013005;
%%CITATION = HEP-PH/0212212;%%
A.~Bandyopadhyay, S.~Choubey, S.~Goswami and D.~P.~Roy,
%``Implications of the first neutral current data from SNO for solar  %neutrino
%oscillation,''
Phys.\ Lett.\  B {\bf 540} (2002) 14;
%%CITATION = PHLTA,B540,14;%%bibitem{Bandyopadhyay:2002xj}
V.~Barger, D.~Marfatia, K.~Whisnant and B.~P.~Wood,
%``Imprint of SNO neutral current data on the solar neutrino problem,''
Phys.\ Lett.\  B {\bf 537} (2002) 179.
%%CITATION = PHLTA,B537,179;%%
%
\bibitem{14-SNOII_2005}
B.~Aharmim {\it et al.}  [SNO Collaboration],
%``Electron energy spectra, fluxes, and day-night asymmetries of B-8 solar
%neutrinos from the 391-day salt phase SNO data set,''
Phys.\ Rev.\  C {\bf 72} (2005) 055502.
%%CITATION = PHRVA,C72,055502;%%
%
\bibitem{14-SK2003SMY}
M.~B.~Smy {\it et al.}  [Super-Kamiokande Collaboration],
Phys.\ Rev.\ D {\bf 69} (2004) 011104.
%%CITATION = HEP-EX/0309011;%%
\bibitem{14-SKI-2005} 
J.~Hosaka {\it et al.}  [Super-Kamkiokande Collaboration],
%``Solar neutrino measurements in Super-Kamiokande-I,''
Phys.\ Rev.\  D {\bf 73} (2006) 112001.
%%CITATION = PHRVA,D73,112001;%%
%
\bibitem{14-Arpesella2008}
C.~Arpesella {\it et al.}  [The Borexino Collaboration],
%``Direct Measurement of the Be-7 Solar Neutrino Flux with 192 Days of
%Borexino Data,''
Phys.\ Rev.\ Lett.\  {\bf 101} (2008) 091302.
%%CITATION = PRLTA,101,091302;%%
%
\bibitem{14-LETA}
B.~Aharmim {\it et al.}  [SNO Collaboration],
%``Low Energy Threshold Analysis of the Phase I and Phase II Data Sets of the
%Sudbury Neutrino Observatory,''
Phys.\ Rev.\  C {\bf 81} (2010) 055504.
%%CITATION = PHRVA,C81,055504;%%
%
\bibitem{14-SNOphaseI_analisi2007}
B.~Aharmim {\it et al.}  [SNO Collaboration],
%``Measurement of the nu(e) and total B-8 solar neutrino fluxes with the Sudbury neutrino observatory phase I data set,''
Phys.\ Rev.\ C {\bf 75} (2007) 045502.
%
\bibitem{14-SKII}
J.~P.~Cravens {\it et al.}  [Super-Kamiokande Collaboration],
%``Solar neutrino measurements in Super-Kamiokande-II,''
Phys.\ Rev.\  D {\bf 78} (2008) 032002.
%%CITATION = PHRVA,D78,032002;%%
%
\bibitem{14-SKIII}
K.~Abe {\it et al.}  [Super-Kamiokande Collaboration],
%``Solar neutrino results in Super-Kamiokande-III,''
Phys.\ Rev.\ D {\bf 83} (2011) 052010.
%%CITATION = ARXIV:1010.0118;%%
%
\bibitem{14-SuperKamiokandeIV}
M.~Smy,
%``Low Energy Neutrino Astronomy in Super-Kamiokande,''
{\it arXiv:1110.0012 [hep-ex].} See also the ''Neutrino 2012'' Proceedings when they
will appear.
%%CITATION = ARXIV:1110.0012;%%
%
\bibitem{14-Bor09} 
G.~Alimonti {\it et al.}  [Borexino Collaboration],
%``The Borexino detector at the Laboratori Nazionali del Gran Sasso,''
Nucl.\ Instrum.\ Meth.\  A {\bf 600} (2009) 568, 
physics.ins-det/0806.2400.
%%CITATION = NUIMA,A600,568;%%
%Borexino collaboration, Nucl. Instrum. Meth. A 660, 568 (2009)
%
%
\bibitem{14-carlos} C.~Pe\~na-Garay, talk at the conference Neutrino Telescopes 2007", March 6-9, 2007, Venice, available 
online at {\tt neutrino.pd.infn.it/conference2007/}.
%%CITATION = NUCL-EX/0610020;%%
%
%
%Update
%Apparently this reference is never used
%\bibitem{Ali09} G.Alimonti et al, Nucl. Instrum. Methods A 609, 58 (2009).
%
\bibitem{14-bxfirstresults} 
C.~Arpesella {\it et al.}  [Borexino Collaboration],
%``First real time detection of Be7 solar neutrinos by Borexino,''
Phys.\ Lett.\  B {\bf 658} (2008) 101.
%%CITATION = PHLTA,B658,101;%%
%
\bibitem{14-ctf1} 
G.~Alimonti {\it et al.}  [Borexino Collaboration],
%``Ultra-low background measurements in a large volume underground detector,''
Astropart.\ Phys.\  {\bf 8} (1998) 141.
%%CITATION = APHYE,8,141;%%
%
\bibitem{14-ctf2} 
G.~Alimonti {\it et al.}  [Borexino Collaboration],
%``Science and Technology of BOREXINO: A Real Time Detector for Low Energy
%Solar Neutrinos SOLAR NEUTRINOS,''
Astropart.\ Phys.\  {\bf 16} (2002) 205.
%%CITATION = APHYE,16,205;%%
%
\bibitem{14-ctf3} 
C.~Arpesella {\it et al.}  [BOREXINO Collaboration],
%``Measurements of extremely low radioactivity levels in BOREXINO,''
Astropart.\ Phys.\  {\bf 18} (2002) 1.
%%CITATION = APHYE,18,1;%%
%
\bibitem{14-pmts1} 
A.~Ianni, P.~Lombardi, G.~Ranucci and O.~Smirnov,
%``The measurements of 2200 ETL9351 type photomultipliers for the Borexino
%experiment with the photomultiplier testing facility at LNGS,''
Nucl.\ Instrum.\ Meth.\  A {\bf 537} (2005) 683.
%%CITATION = NUIMA,A537,683;%%
%
\bibitem{14-pmts2} 
A.~Brigatti, A.~Ianni, P.~Lombardi, G.~Ranucci and O.~Smirnov,
%``The photomultiplier tube testing facility for the Borexino experiment  at
%LNGS,''
Nucl.\ Instrum.\ Meth.\  A {\bf 537} (2005) 521.
%%CITATION = NUIMA,A537,521;%%
%
\bibitem{14-cones} 
L.~Oberauer, C.~Grieb, F.~von Feilitzsch and I.~Manno,
%``Light concentrators for Borexino and CTF,''
Nucl.\ Instrum.\ Meth.\  A {\bf 530} (2004) 453.
%%CITATION = NUIMA,A530,453;%%
%
\bibitem{14-bxdetector} 
Borexino Collaboration, Nuclear Instruments and Methods in Physics Research 
A 600 (2009) 568-593.
%
\bibitem{14-lakn} H.~Simgen and G.~Zuzel, {\it Ultrapure gases - From the Production Plant to 
the Laboratory}, AIP Conference Proceedings Vol.~897, Topical Workshop on Low Radioactivity Techniques: LRT 2006, Aussois (France), pp. 45--50, ed. P.~Loaiza, Springer (2007).
%
\bibitem{14-SSM06}
J.~N.~Bahcall, A.~M.~Serenelli and S.~Basu,
%``10,000 Standard Solar Models: a Monte Carlo Simulation,''
Astrophys.\ J.\ Suppl.\  {\bf 165} (2006) 400.
%%CITATION = APJSA,165,400;%%
%
\bibitem{14-Bellini2011rx}
  G.~Bellini {\it et al.},
  %``Precision measurement of the 7Be solar neutrino interaction rate in
  %Borexino,''
  Phys.\ Rev.\ Lett.\  {\bf 107}, 141302 (2011)
  [arXiv:1104.1816 [hep-ex]].
  %%CITATION = PRLTA,107,141302;%%
\bibitem{14-metallicity} S.~Basu, ASP Conference Series {\bf 416} (2009) 193.
\bibitem{14-deutsch} M.~Deutsch, {\it ``Proposal for a Cosmic Ray Detection System for the Borexino Solar Neutrino Experiment''}, Massachusetts Institute of Technology, Cambridge, MA (1996).
\bibitem{14-pep-ctf} H.~Back et al. (Borexino Collaboration), Phys. Rev. C {\bf 74} (2006) 045805.
\bibitem{14-c11cris} C.~Galbiati, A.~Pocar, D.~Franco, A.~Ianni, L.~Cadonati, and S.~Sch\"onert, Phys. Rev. C {\bf 71} (2005) 055805.
\bibitem{14-bxmuon} G.~Bellini et al. (Borexino Collaboration), JINST {\bf 6}, P05005 (2011).
\bibitem{14-Davini}
S. Davini, {\it Measurement of the pep and CNO solar neutrino interaction rates in Borexino}, thesis submitted for the Ph.D. degree in Physics at the University of Genova, Academic Year 2011-12.
\bibitem{14-annihilation} Y.~Kino et al., Jour. Nucl. Radiochem. Sci {\bf 1} 
(2000) 63.
\bibitem{14-positronium} D.~Franco, G.~Consolati, and D.~Trezzi, Phys. Rev. C {\bf 83} (2011) 015504.
\bibitem{14-tmva} TMVA Users Guide, http://tmva.sourceforge.net/docu/ TMVAUsersGuide.pdf.
%
\bibitem{14-maneschg} W.~Maneschg et al., 
%{arXiv:1110.1217}.
{\it  ``Production and characterization of a custom-made $^{228} Th$ source with reduced neutron source strength for the Borexino experiment''}, in 
{\it Nuclear Instruments and Methods in Physics Research Section A: Accelerators,
Spectrometers, Detectors and Associated Equipment}, Vol. 680, 11 July 2012, Pgg.161-167.
%
\bibitem{14-pdg2010} Review of Particle Physics, K. Nakamura et al. (Particle Data Group), J. Phys. G {\bf 37} (2010) 075021. 
%
\bibitem{14-SNOI+II+III}
B.~Aharmin et al. (SNO Collaboration), 
%``Combined Analysis of all Three Phases of Solar Neutrino Data from the Sudbury Neutrino Observatory,''
{\it arXiv:1109.0763 (nucl-ex)}.
%%CITATION = ARXIV:1109.0763;%%
\bibitem{14-BahcallRadiativeCorrection} J.N.~Bahcall, M.~Kamionkowski and A.~Sirlin, Phys. Rev. D {\bf 51} (1995) 6146.
\bibitem{14-erlerRadCorr} J.~Erler and M.J.~Ramsey-Musolf, Phys. Rev. D {\bf 72} (2005) 073003.
%
% Non usato
%\bibitem{14-bxliquid} G.~Alimonti et al. (Borexino Collaboration), Nucl. 
%Instr. and Meth. A {\bf 609}, 58 (2009).
%
\bibitem{14-Gallium09}
J. N. Abdurashitov et al. (SAGE Collaboration), Phys. Rev. C 80 (2009) 015807,
contains combined analysis with the following references: M. Altmann et al. 
(GNO Collaboration), Physics Letters B 616 (2005) 174; F. Kaether, 
{\it Datenanalyse der Sonnenneutrinoexperiments Gallex}, Ph.D. thesis, Heidelberg (2007).
%
\bibitem{14-Cleveland1998}
B.~T.~Cleveland {\it et al.},
%``Measurement of the solar electron neutrino flux with the Homestake
%chlorine
Astrophys.\ J.\  {\bf 496} (1998) 505.
%%CITATION = ASJOA,496,505;%%
%
\bibitem{14-Bellini2008mr}
Borexino Collaboration, G.~Bellini {\em et~al.},
\newblock Phys. Rev. {\bf D82} (2010) 033006.
%, arXiv:0808.2868.
%
\bibitem{14-T2K2011}
K.~Abe {\it et al.}  [T2K Collaboration],
%``Indication of Electron Neutrino Appearance from an Accelerator-produced
%Off-axis Muon Neutrino Beam,''
Phys.\ Rev.\ Lett.\  {\bf 107} (2011) 041801.
%%CITATION = PRLTA,107,041801;%%
%
\bibitem{14-MINOS2011}
P.~Adamson {\it et al.}  [MINOS Collaboration],
%``Improved search for muon-neutrino to electron-neutrino oscillations in
%MINOS,''
Phys.\ Rev.\ Lett.\  {\bf 107} (2011) 181802.
%%CITATION = PRLTA,107,181802;%%
%
\bibitem{14-FogliLisi2011}
G.~L.~Fogli, E.~Lisi, A.~Marrone, A.~Palazzo and A.~M.~Rotunno,
%``Evidence of theta(13)>0 from global neutrino data analysis,''
Phys.\ Rev.\  D {\bf 84} (2011) 053007.
%%CITATION = PHRVA,D84,053007;%%
%
\bibitem{14-CHOOZ03}
M.~Apollonio {\it et al.}  [CHOOZ Collaboration],
%``Search for neutrino oscillations on a long base-line at the CHOOZ nuclear
%power station,''
Eur.\ Phys.\ J.\  C {\bf 27} (2003) 331.
%%CITATION = EPHJA,C27,331;%%
%
\bibitem{14-DoubleChooz}
Y.~Abe {\it et al.}  [DOUBLE-CHOOZ Collaboration],
%``Indication for the disappearance of reactor electron antineutrinos in the Double Chooz experiment,''
Phys.\ Rev.\ Lett.\  {\bf 108} (2012) 131801.
%%CITATION = ARXIV:1112.6353;%%
On the possible implications of Double Chooz results see also:
C.~Giunti and M.~Laveder,
%``Effect of the reactor antineutrino anomaly on the first Double-Chooz results,''
Phys.\ Rev.\ D {\bf 85} (2012) 031301.
%%CITATION = ARXIV:1111.5211;%%
%
\bibitem{14-DayaBay}
F.~P.~An {\it et al.}  [DAYA-BAY Collaboration],
%``Observation of electron-antineutrino disappearance at Daya Bay,''
Phys.\ Rev.\ Lett.\  {\bf 108} (2012) 171803.
%%CITATION = ARXIV:1203.1669;%%
\bibitem{14-RENO}
J.~K.~Ahn {\it et al.}  [RENO Collaboration],
%``Observation of Reactor Electron Antineutrino Disappearance in the RENO Experiment,''
Phys.\ Rev.\ Lett.\  {\bf 108} (2012) 191802.
%%CITATION = ARXIV:1204.0626;%%
\bibitem{14-Daya_Bay_neu2012} 
D. Dwyer [for the Daya-Bay Collaboration], {\it talk at Neutrino 2012}.  
%
\bibitem{14-Fogli2012ua}
G.~L.~Fogli, E.~Lisi, A.~Marrone, D.~Montanino, A.~Palazzo and A.~M.~Rotunno,
%``Global analysis of neutrino masses, mixings and phases: entering the era % of leptonic CP violation searches,''
{\it arXiv:1205.5254 [hep-ph]}.
%CITATION = ARXIV:1205.5254;%%
%
\bibitem{14-Tortola2012te}
D.~V.~Forero, M.~Tortola and J.~W.~F.~Valle,
%``Global status of neutrino oscillation parameters after recent reactor % measurements,''
{\it arXiv:1205.4018 [hep-ph]}.
%%CITATION = ARXIV:1205.4018;%%
\bibitem{14-Schwetz2012}
T. Schwetz, {\it talk at NuTURN 2012, Workshop on ``Neutrino at the Turning Point'' 
(Laboratori Nazionali del Gran Sasso, Italy, 2012), available at agenda.infn.it/
conferenceDisplay.py?confId=4722; talk at ``What is NU ?'', Workshop at the Galileo 
Galilei Institute (Florence, Italy, 2012); available at www.ggi.fi.infn.it }.
%
\bibitem{14-Serenelli2009yc}
A.~Serenelli, S.~Basu, J.~W.~Ferguson and M.~Asplund,
%``New Solar Composition: The Problem With Solar Models Revisited,''
Astrophys.\ J.\  {\bf 705} (2009) L123.
%%CITATION = ASJOA,705,L123;%%
%
\bibitem{14-Friedland:2004pp} 
A.~Friedland, C.~Lunardini and C.~Pena-Garay,
%``Solar neutrinos as probes of neutrino matter interactions,''
Phys.\ Lett.\ B {\bf 594} (2004) 347.
%%CITATION = HEP-PH/0402266;%%
%
\bibitem{14-deHolanda2010am}
P.~C.~de Holanda and A.~Y.~.Smirnov,
%``Solar neutrino spectrum, sterile neutrinos and additional radiation in %the Universe,''
Phys.\ Rev.\ D {\bf 83} (2011) 113011.
%%CITATION = ARXIV:1012.5627;%%

\bibitem{14-spirovignaud}M. Spiro and D. Vignaud, Phys. Lett. {\bf B 242} 
(1990) 279.

\bibitem{14-luminosity}J.N. Bahcall, Phys. Rev. {\bf C 65} (2002) 025801. 

\bibitem{14-roadmap} 
  J.~N.~Bahcall and C.~Pena-Garay,  JHEP {\bf 0311} (2003) 004.

\bibitem{14-concha:2010} Gonz\'alez-Garc\'ia, M.~C., Maltoni and M. Salvado, 
Journal of High Energy Physics, 72 (2010) 5.

\bibitem{14-Chavarria2012sd}
A.~Chavarria,
%``Solar Neutrinos in 2011,''
arXiv:1201.6311 [astro-ph.SR].
%%CITATION = ARXIV:1201.6311;%%
%
\bibitem{14-Boger1999bb}
J.~Boger {\it et al.}  [SNO Collaboration],
%``The Sudbury neutrino observatory,''
Nucl.\ Instrum.\ Meth.\ A {\bf 449} (2000) 172.
%%CITATION = NUCL-EX/9910016;%%
%
\bibitem{14-Fukuda2003s} S. Fukuda {\it et al.} [Super-Kamiokande 
Collaboration], 
Nucl.\ Instrum.\ Meth.\ A {\bf 501} (2003) 418.
%
\bibitem{14-Abdurashitov1999zd}
J.~N.~Abdurashitov {\it et al.}  [SAGE Collaboration],
%``Measurement of the solar neutrino capture rate with gallium metal,''
Phys.\ Rev.\ C {\bf 60} (1999) 055801.
%%CITATION = ASTRO-PH/9907113;%%
%
\bibitem{14-Kraus2010zzb}
C.~Kraus {\it et al.}  [SNO+ Collaboration],
%``The rich neutrino programme of the SNO+ experiment,''
Prog.\ Part.\ Nucl.\ Phys.\  {\bf 64} (2010) 273.
%%CITATION = PPNPD,64,273;%%
%

\bibitem{14-CLEAN2004}
D.~N.~McKinsey and K.~J.~Coakley,
%``Neutrino detection with CLEAN,''
Astropart.\ Phys.\  {\bf 22} (2005) 355.
%%CITATION = ASTRO-PH/0402007;%%
%
\bibitem{14-Aprile2009dv} 
E.~Aprile and T.~Doke,
%``Liquid Xenon Detectors for Particle Physics and Astrophysics,''
Rev.\ Mod.\ Phys.\  {\bf 82} (2010) 2053.
%%CITATION = ARXIV:0910.4956;%%
%
\bibitem{14-MEG}  
Baldini, A. et al., 2002, {\it Research Proposal to INFN, The
MEG experiment, search for the $\mu \to e \gamma$ decay at PSI}.
%
\bibitem{14-Moriyama2011zz}
S.~Moriyama [XMASS Collaboration],
%``Status of XMASS experiment,''
PoS IDM {\bf 2010} (2011) 057.
%%CITATION = POSCI,IDM2010,057;%%
%
\bibitem{14-XMASStalk}
K.~Kobayashi [XMASS Collaboration], 
{\it XMASS experiment}, talk given at the ''TeV Particle Astrophysics 2010'', Paris, July 2010.
%
\bibitem{14-DARWIN}
L.~Baudis,
%``DARWIN: dark matter WIMP search with noble liquids,''
PoS IDM {\bf 2010} (2011) 122 [arXiv:1012.4764 [astro-ph.IM]].
%%CITATION = ARXIV:1012.4764;%%
L.~Baudis [DARWIN Consortium Collaboration],
%``DARWIN: dark matter WIMP search with noble liquids,''
{\it arXiv:1201.2402 [astro-ph.IM]}.
%%CITATION = ARXIV:1201.2402;%%
%
\bibitem{14-LENA}
The next-generation liquid-scintillator neutrino observatory LENA\\
Astroparticle Physics, In Press, Uncorrected Proof, Available online 10 March 2012
%
\bibitem{14-PXE}
  Borexino Collab.,
{\it ``Study of phenylxylylethane (PXE) as scintillator for low energy neutrino experiments''}, in
{\it Nuclear Instruments and Methods in Physics Research Section A: Accelerators, Spectrometers, 
Detectors and Associated Equipment}, Vol. 585, Issues 12, 21 January 2008, 
pgg. 48-60
\bibitem{14-Arpesella2008mt}
Borexino Collaboration, C.~Arpesella {\em et~al.},
\newblock Phys. Rev. Lett. {\bf 101}, 091302 (2008), arXiv:0805.3843.

\bibitem{14-wur09phd}
M.~Wurm,
\newblock {\em Cosmic Background Discrimination for the Rare Neutrino Event
  Search in Borexino and LENA},
\newblock PhD thesis, Technische Universit\"at M\"unchen, 2009.

\bibitem{14-Dangelo2006}
D.~D'Angelo,
\newblock {\em {Towards the detection of low energy solar neutrinos in
  BOREXino: data readout, data reconstruction and background identification}},
\newblock Phd thesis, Technische Universit\"at M\"unchen, 2006.

\bibitem{14-Ianni2005ki}
A.~Ianni, D.~Montanino, and F.~L. Villante,
\newblock Phys. Lett. {\bf B627} (2005) 38.
%physics/0506171.
%
\bibitem{14-Raghavan1976yc}
R.~S.~Raghavan,
%``Inverse beta decay of 115-In to 115-Sn*: a new possibility for detecting 
% solar neutrinos from the proton-proton reaction,''
Phys.\ Rev.\ Lett.\  {\bf 37} (1976) 259.
%%CITATION = PRLTA,37,259;%%
\bibitem{14-Raghavan2004}
R. S. Raghavan, 
{\it ``Discovery potential of low energy solar neutrino experiments''}, Notes for APS-SAWG, Mar 15 2004.
%

\bibitem{14-antia:2005} H.  Antia and  S. Basu, Astroph.  J. Lett.  {\bf 620},
  L129 (2005)

\bibitem{14-bahcall:2005}   J.N.  Bahcall,  S.   Basu  and   A.~M.  Serenelli,
  Astroph. J. {\bf 631} (2005) 1281.

\bibitem{14-guzik:2005}   J.~A.   Guzik,   L.~S.   Watson   and   A.~N.   Cox,
  Astroph. J. {\bf 627} (2005) 1049.

\bibitem{14-castro:2007}  M. Castro, S.  Vauclair and  P. Richard,  Astron. \&
  Astroph. {\bf 463} (2007) 755.

\bibitem{14-jcd:2009} J.  Christensen-Dalsgaard, M.P. di Mauro,  G. Houdex and
  F. Pijpers, Astron. \& Astroph. {\bf 494} (2009) 205.

\bibitem{14-serenelli:2010} A. M. Serenelli, Astroph. Space Sci. {\bf 328}
  (2010) 13.

\bibitem{14-villante:2010a} F.~L. Villante, F.~L. and B. Ricci, Astroph. J. 
{\bf 714} (2010) 944.

\bibitem{14-villante:2010b} F.~L. Villante, Astroph. J. {\bf 724} (2010) 98. 

\bibitem{14-Barranco:2007ej} 
  J.~Barranco, O.~G.~Miranda, C.~A.~Moura and J.~W.~F.~Valle,
  Phys.\ Rev.\ D {\bf 77} (2008) 093014.

\bibitem{14-bethe:1939} H.~A. Bethe, Phys. Rev. {\bf 55} (1939) 434.

\bibitem{14-thoul:1994}   A.~A.   Thoul,    J.~N.   Bahcall   and   A.   Loeb,
  Astroph. J. {\bf 412} (1994) 828.

\bibitem{14-pinsonneault:2010}     F.     Delahaye,    M.~H.     Pinsonneault,
  L. Pinsonneault and C.~J. Zeippen, arXiv:1005.0423 (2010)

\bibitem{14-melendez:2009}  J.  Mel\'endez,  M.  Asplund,  B.   Gustafsson and
  D. Yong, Astroph. J. Lett. {\bf 704}, L66 (2009)

\bibitem{14-ramirez:2010}   I.   Ram\'{\i}rez,   M.   Asplund,   P.   Baumann,
  J. Mel\'endez and T. Bensby, Astron. \& Astroph. {\bf 512} (2010) 33.


\end{thebibliography}
\end{document}